\documentclass[hidelinks, 12pt]{article}

\usepackage[hidelinks]{hyperref}
\usepackage{amssymb}
\usepackage{amsmath}
\usepackage{amsfonts, amsthm}
\usepackage{amsbsy}
\allowdisplaybreaks
\usepackage{thmtools}
\renewcommand\thmcontinues[1]{Continued}
\usepackage{mathtools}
\usepackage{dsfont}
\usepackage{titling}
\usepackage{tikz}
\usetikzlibrary{chains, fit, positioning}
\usepackage{graphicx}
\usepackage{pdflscape}
\usepackage{afterpage}
\usepackage{tabularx}
\usepackage{array}
\usetikzlibrary{arrows.meta}
\usepackage{longtable}
\usepackage{caption}
\usepackage{subcaption}
\usepackage{rotating}
\usepackage{cleveref}
\usepackage{fancyhdr}

\newcommand{\Var}{\mathrm{Var}}
\newcommand{\Cov}{\mathrm{Cov}}
\newcommand{\Eqref}[1]{Eq.~\eqref{#1}}
\newcommand\blfootnote[1]{%
  \begingroup
  \renewcommand\thefootnote{}\footnote{#1}%
  \addtocounter{footnote}{-1}%
  \endgroup
}
\theoremstyle{remark}

\newtheorem{theorem}{Theorem}[section]

\newtheorem{lemma}[theorem]{\bf{Lemma}}
\newtheorem{definition}[theorem]{\bf{Definition}}

\newtheorem{example}[theorem]{\bf{Example}}
\usepackage{array}
\usepackage{subcaption}
\usepackage{comment}
\newcolumntype{P}[1]{>{\centering\arraybackslash}p{#1}}
\usepackage{graphicx}
\usepackage{natbib}
\usepackage{url} 

\newcommand{\blind}{0}

\begin{document}

\def\spacingset#1{\renewcommand{\baselinestretch}%
{#1}\small\normalsize} \spacingset{1}

\if0\blind
{
  \title{\bf The Double Emulator}
  \author{Conor Crilly\thanks{School of Mathematics, University of Bristol, Fry Building, Woodland Road, Bristol, BS8 1UG, UK} 
  \and Oliver Johnson\protect\footnotemark[1] \thanks{Corresponding author: \texttt{O.Johnson@bristol.ac.uk}}
  \and Alexander Lewis\thanks{AWE, Aldermaston, Berkshire, RG7 4PR, UK}
  \and Jonathan Rougier\protect\footnotemark[1] \protect\footnotemark[3]
}
  \date{}
  \maketitle
  \blfootnote{\copyright \, UK Ministry of Defence Crown Owned Copyright 2024/AWE}
} \fi

\if1\blind
{
    \title{\bf The Double Emulator}
    \medskip
    \date{}
    \maketitle
} \fi

\bigskip
\begin{abstract}
Computer models (simulators) are vital tools for investigating physical processes. Despite their utility, the prohibitive run-time of simulators hinders their direct application for uncertainty quantification. Gaussian process emulators (GPEs) have been used extensively to circumvent the cost of the simulator and are known to perform well on simulators with smooth, stationary output. In reality, many simulators violate these assumptions. Motivated by a finite element simulator which models early stage corrosion of uranium in water vapor, we propose an adaption of the GPE, called the double emulator, specifically for simulators which `ground' in a considerable volume of their input space. Grounding is the process by which a simulator attains its minimum and can result in violation of the stationarity and smoothness assumptions used in the conventional GPE. We perform numerical experiments comparing the performance of the GPE and double emulator on both the corrosion simulator and synthetic examples.
\end{abstract}

\noindent%
\vfill

\newpage
\spacingset{1} 

\section{Introduction}
\label{Introduction}
Computer models (simulators) have become essential for our understanding of physical systems, particularly when conducting the physical experiment is constrained ethically, financially or legally \citep{Santner03, Gramacy2020}. In this article, we are interested in high-fidelity deterministic simulators, meaning that evaluating the simulator at a given input will always return the same output. Despite this, attaining complete knowledge of the corresponding input-output relationship is limited by available computing resources; this is known as code uncertainty \citep{Kennedy01}. To circumvent the simulator expense a cheaper surrogate model is often used in place of the expensive simulator, with Gaussian process emulators (GPEs) receiving considerable attention within the computer experiments community, beginning with \citet{Sacks89}.

One way of defining a GPE is as the sum of two components; the first a regression component capturing the systematic behavior of the simulator and the second a \textit{separable stationary} Gaussian process capturing the residual behavior \citep[e.g.][]{Goldstein04}. When the simulator behavior is similar across the entire input space, the conventional treatment of the emulator residual as smooth and stationary can perform well out-of-the-box. However, \textit{both} of these assumptions can be challenged in simulators with outputs which attain their minimum over a substantial volume of the input space. In particular, GPEs incorporate a mean square smoothness assumption regarding the sample paths of the Gaussian process via the covariance kernel, which is determined by the kernel's behavior around zero \citep{Rasmussen05, Santner03}. When a simulator attains its minimum, perhaps with a discontinuity in its first derivative, the typical choice of either a squared exponential (Gaussian) kernel, or a kernel from the M\'atern family, may be restricting the emulator's performance.

To deal with non-stationarity and, more generally, simulators which undergo extreme change in shape, increasingly flexible methods for function approximation including deep Gaussian processes \citep{Damianou13, dunlop18, Sauer23}, non-stationary kernels \citep{Higdon98, paciorek04, paciorekthesis}, input warping \citep{Sampson92}, output warping \citep{snelson03}, kernel mixtures \citep{volodina20} and local approximations \citep{gramacy2015local} have been proposed. We are interested specifically in emulating simulators which challenge the stationarity assumption by attaining their minimum, which we call \textit{grounding}, in a large volume of the input space, which we call the \textit{grounded region}, \textit{and} which potentially challenge typical smoothness assumptions by grounding with a discontinuity in their first derivative. We refer to the process of grounding with a discontinuity as a \textit{hard landing}; if the simulator attains its minimum smoothly it is said to land \textit{softly}. Grounding simulators arise naturally in scenarios where one quantity can be eliminated e.g. infectious disease modeling, fossil fuel extraction or the estimation of glacial melt-water runoff. A specific example motivating this work, for which the behavior around the grounded region is not well understood, is a simulator modeling the oxidation of uranium \citep{Natchiar20, Natchiar21}. Improved emulation and increased understanding of this oxidation process has direct implications for improving safety in any industry which involves handling, storage, or disposal of uranium related products \citep{govreport}. 

To improve emulation near the grounding line separating the grounded and non-grounded regions, we introduce the double emulator; a framework which combines a probabilistic classifier with a GPE. Using variations of synthetic simulators in \citet{Dette_Pepelyshev_2010} and \citet{rosenbrock60} our analysis focuses on the following two questions related to the shape of the simulator output. First, for a given simulator, what is the effect of the volume of the grounded region on the emulator's performance? Second, what is the effect of the hardness of the landing? The different synthetic simulators allow us to investigate each of these properties under varied input dimension. We also investigate the effect of a varied number of training runs on the oxidation simulator. Note, it is not necessary to restrict the GPE and double emulator to use the same covariance kernel. However, our comparisons between any given GPE and double emulator will always use the same kernel. Our results suggest the following:
\begin{enumerate}
    \item On examples with a grounded region of moderate volume, the double emulator regularly outperforms the GPE. The improvement provided by using the double emulator over the GPE increases as the change in the simulator's derivative across the grounding line increases, see Figure \ref{fig:results} where $b < 1$.
    \item On examples where the volume of the grounded region is very large the GPE can outperform the double emulator as the GPE within the double emulator struggles due to a lack of training data.  
    \item The influence of the classifier within the double emulator on root mean squared error (RMSE) is less clear than with the continuous ranked probability score (CRPS) \citep{Gneiting07}, where it is almost always the case that a better classifier provides a better model.
\end{enumerate}

We note that classifiers have been used previously in the literature to improve the performance of GPEs, although for different purposes than we consider here. For example, \citet{Gramacy08} split the input space using trees and independent GPEs are fit in each region of the space, whilst \cite{Isberg_Welch_2022} use trees to target regions of the input space where the simulator response is most complex. Furthermore, we highlight the existence of similar, yet ultimately different, applications in which GPEs, contours and expensive simulators form the main elements of the statistical analysis. For example, the estimation of contours and percentiles of expensive simulators using GPEs has been studied from a Bayesian optimization perspective \citep{ranjan08, roy14}, where extensions of the well known expected improvement acquisition functions are derived \citep{mockus78, schonlau97, schonlau98}. In particular, our approach differs from existing approaches in that our aim is to improve the predictive performance of the GPE, with emphasis on the region around the grounding line, as opposed to estimating the volume of a failure region \citep{Cole23}, contour location \citep{Marques18}, or optimization of an expensive black-box function. Our method also differs from the existing techniques for modeling non-stationary simulator output mentioned previously. 

The remainder of the article is as follows. Section \ref{components} motivates our contribution, reviews the basic components required for our model, and discusses the limitations of the conventional emulator for emulating simulators which ground. Section \ref{DoubleEmulator} presents the mathematical details of the double emulator. Section \ref{Experiments} compares the GPE with the double emulator on both a family of synthetic examples and the oxidation simulator, and includes a derivation of the CRPS score used to assess the double emulator. Section \ref{Discussion} discusses future research directions.

\section{Review}
\label{components}
We begin with a pedagogical example to clarify grounding. We then review the basics of GPEs, their limitations for grounding simulators, and outline our proposed solution. 
\subsection{Motivation}
\begin{figure}[ht!]
    \centering
    \begin{subfigure}{0.3\textwidth}
        \centering
        \includegraphics[width=\linewidth]{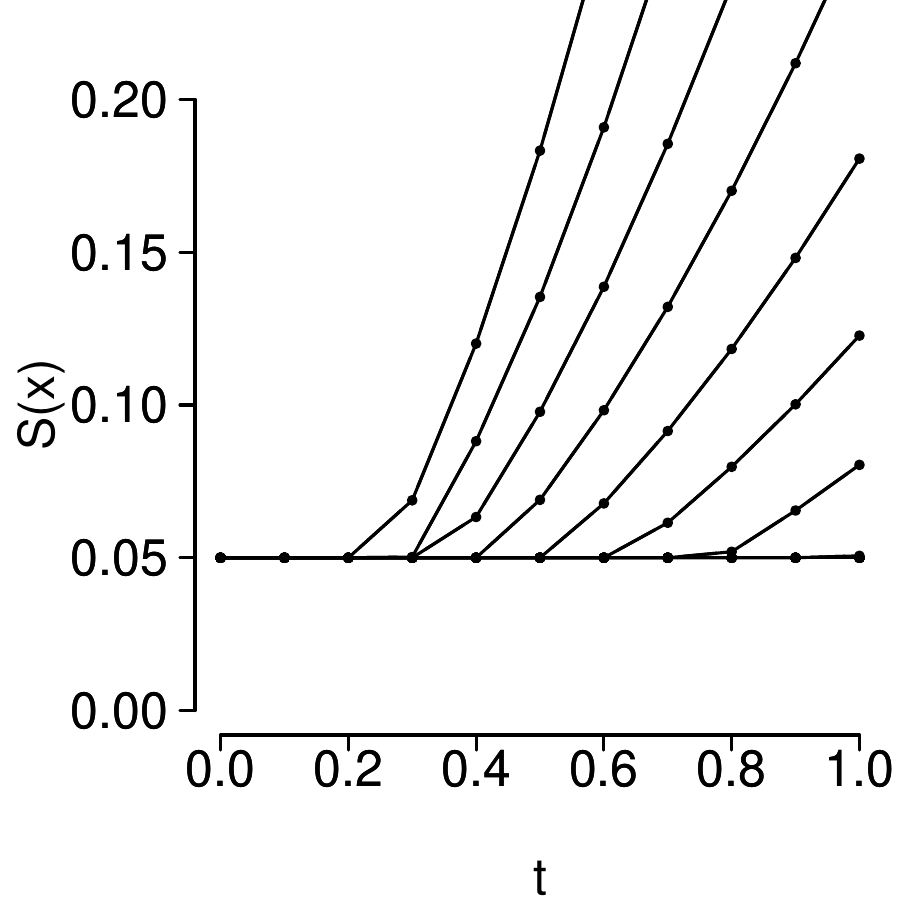}
        \caption{}
        \label{fig:motivatingruns}
    \end{subfigure}
    \begin{subfigure}{0.3\textwidth}
        \centering
        \includegraphics[width=\linewidth]{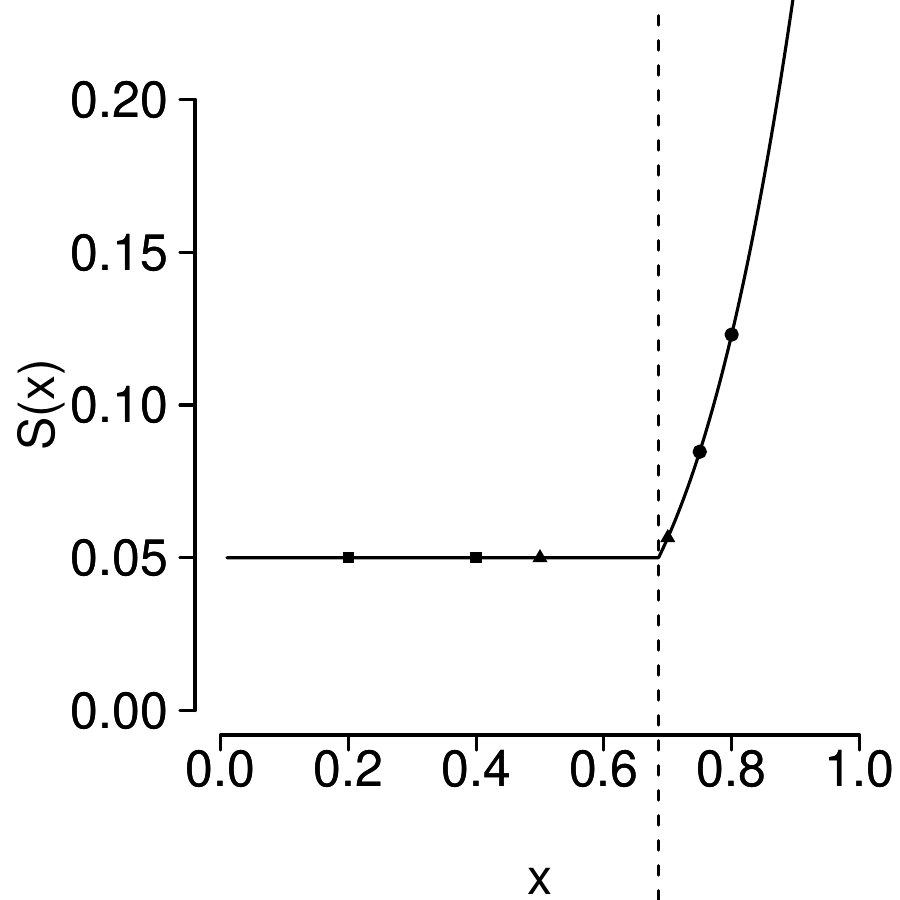}
        \caption{}
        \label{fig:motivatinggrounding}
    \end{subfigure}
    \begin{subfigure}{0.3\textwidth}
        \centering
        \includegraphics[width=\linewidth]{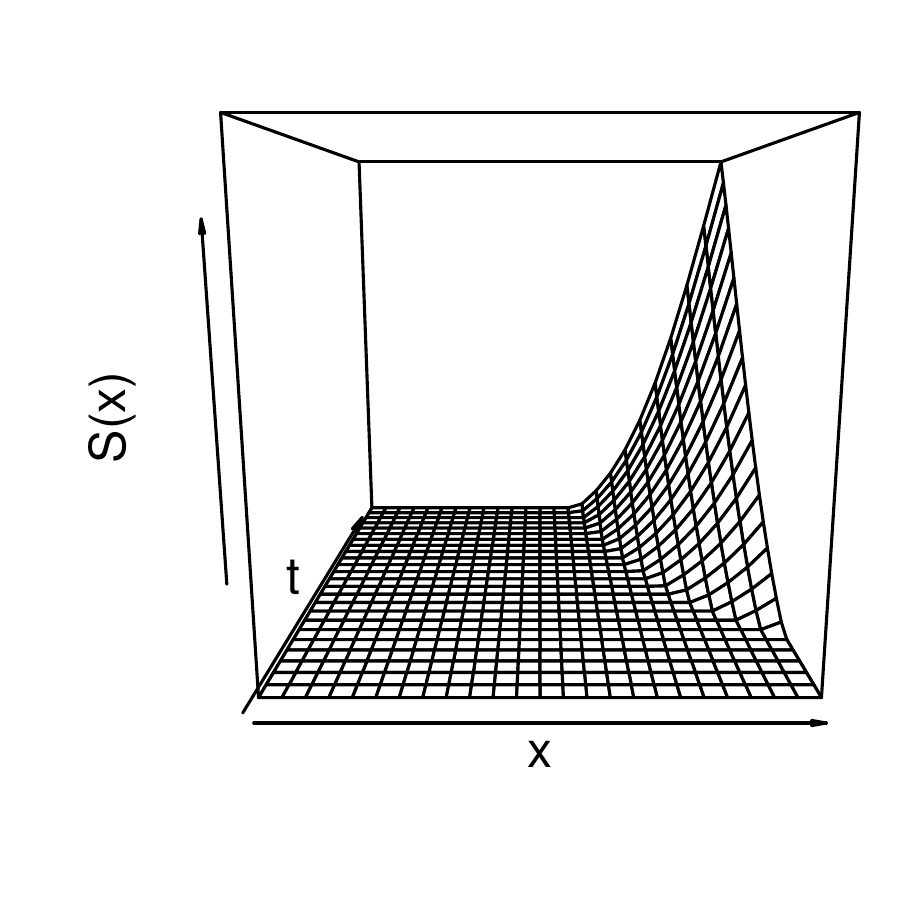}
        \caption{}
        \label{fig:motivatingsurface}
    \end{subfigure}
    \caption{An illustration of grounding for the simulator in \eqref{motivatingexample}. Three input pairs are highlighted: the solid triangles straddle the grounding point, whilst the solid squares and circles do not. Straddling and non-straddling pairs affect hyperparameter estimation differently, see Section \ref{Existing}.}
    \label{fig:Grounding}
\end{figure}

Consider the following synthetic simulator:
\begin{equation}
    \label{motivatingexample}
    S(x) = \max\left\{\sin^2\left(x^3t\right), \, 0.05 \right\}, \; t \in \left\{0, 0.1, \dots, 0.9, 1 \right\}.
\end{equation}
Evaluating $S$ at any value of the input variable, $x$, returns a time series of eleven points, indexed by the output index, $t$. This is an example of a simulator which returns regular output as the output is restricted to lie on the same grid for each evaluation of $S$. We are interested in scalar emulators for simulators which \textit{ground}. Grounding is the process by which the simulator attains its minimum as a function of the \textit{input} variables, $x$, for some \textit{fixed} output index, $t$. Figure \ref{fig:Grounding} depicts the grounding process for the simulator in \eqref{motivatingexample}. In particular, Figure \ref{fig:motivatingruns} contains 20 runs of the simulator for different $x$ values; for each run, the solid black dots represent the simulator output at $t \in \{0, 0.1, \dots, 0.9, 1\}$. Note, for the chosen values of $x$ many of the runs equal the grounded value of $0.05$ for every $t$ value. Figure \ref{fig:motivatinggrounding} then analyses, specifically at $t = 0.7$, the variation in the simulator output as a function of $x$, with the vertical dashed line denoting the grounding point; \textit{this is grounding}. 

In a computer experiment, the training data consists of a finite number of simulator evaluations at a set of $d$ dimensional inputs typically chosen to fill the input space, for example using a Latin hypercube design \citep{McKay_Beckman_Conover_1979}. As a result, the position of the grounding line, which is a manifold of dimension $d-1$ separating the grounded and non-grounded regions, is an unknown quantity; accurately estimating this is a crucial component of the double emulator. In this simple example, the data is linearly separable. However, in higher dimensional input spaces, which is common for computer experiments, this is very unlikely and locating the grounding line will require a suitable classification algorithm with the ability to detect extremely nonlinear, non-axially-aligned grounding lines, see Section \ref{ClassifierChoice}. For illustrative purposes, Figure \ref{fig:motivatingsurface} contains a visualization of the simulator on a dense grid of $x$ and $t$ values to show that, for different $t$ values, the shape of the grounding in $x$ will be different; we emphasize that emulating the simulator response as a function of $t$ is not the focus of this article, though efficient methods for tackling this problem do exist \citep[e.g.][]{Rougier08}. Instead, we extract the simulator response at a fixed value of $t$, emulating the scalar-valued response as a function of $x$.
\subsection{Background on GPEs}
\label{Existing}
A common approach for emulating a simulator which returns functional or multivariate output involves building independent scalar emulators for the response at a specific output index \citep[e.g.][]{bayarri2009using}; this is a scalar emulation task. We are interested in simulators for which many of the simulator runs return the known minimum simulator response at the chosen output index. Using a GPE is a tried and tested technique for emulating expensive scalar valued deterministic simulators \citep[e.g.][]{Sacks89, currin91}. In this paradigm, an $n$-vector of training data $\boldsymbol{y}^T = (S(\boldsymbol{x}^{(1)}), \dots, S(\boldsymbol{x}^{(n)}))$ corresponds to $n$ runs of the simulator at $n$ different combinations of the input parameters $\boldsymbol{x}^T = (x_1, \dots, x_d)$, with the runs contained in the $n \times d$ design matrix $\boldsymbol{X}$ with entries $X_{ij} = x^{(i)}_j$. The simulator response, $S$, is treated as a sample path of the stochastic process
\begin{equation}
    S(\boldsymbol{x}) = \mu(\boldsymbol{x}) + Z(\boldsymbol{x}).
\end{equation}
Global behavior of the simulator response is modeled using the regression $\mu(\boldsymbol{x})$, whilst $Z(\boldsymbol{x})$ is a zero-mean Gaussian process with covariance $\Cov(S(\boldsymbol{x}), S(\boldsymbol{x}')) = \sigma^2 r(\boldsymbol{x}, \boldsymbol{x}')$ modeling the local deviation away from the regressors. The positive-definite function $r(\boldsymbol{x}, \boldsymbol{x}')$ represents the correlation between the simulator response value at any two inputs, with $\sigma^2$ the process variance. 

There is considerable flexibility in the specification of both the form of $\mu(\boldsymbol{x})$ and the correlation function, $r(\boldsymbol{x}, \boldsymbol{x}')$. In this article, we assume 
\begin{equation}
\label{assumptions}
    \mu(\boldsymbol{x}) = g^T(\boldsymbol{x})\boldsymbol{\beta} \;\; \text{and} \;\;     r(\boldsymbol{x},\boldsymbol{x}') = \prod_{j=1}^d r_j(\left|x_j - x'_j\right|;\lambda_j),
\end{equation}
where $g^T(\cdot) = (g_1(\cdot), \dots, g_q(\cdot))$ are $q$ known regression functions. For emulating simulators of physical processes, where monotonicity is often present in the output, fixing $q = d+1$ and defining $g_1(\boldsymbol{x}) = 1$ and $g_j(\boldsymbol{x}) = x_{j-1}$ for $j \in \{2,\dots, q\}$ is a natural choice, as higher order regressors can result in wild extrapolation outside of the convex hull of the training data. In addition, it is almost always the case for computer experiments that the separable stationary product correlation function in \eqref{assumptions} is used \citep{Sacks89, Kennedy01, Higdon04}. In turn, specifying $r$ reduces to the specification of the univariate correlation functions, $r_j$, the most common choice being the squared exponential, which produces sample paths which are infinitely differentiable in mean-square. This smoothness assumption can be overly restrictive, prompting the popularization of the Mat\'ern class of kernels for modeling environmental data \citep{stein99} and simulator output \citep{Rougier09}. In particular, the choice of covariance kernel should reflect the modeler's beliefs about the smoothness of the simulator response, with a poor choice potentially resulting in poor emulation.

With $\mu$ and $r$ specified, the next task involves estimating the unknown parameters $\boldsymbol{\beta} = (\beta_1, \dots, \beta_q)^T$ and $\sigma^2$, as well as the hyperparameters $\boldsymbol{\lambda} = (\lambda_1, \dots, \lambda_d)^T$. Freedom of choice with regards to the estimation procedure admits further flexibility, with MLE, REML and Bayesian approaches all permeating the literature. Prohibitive expense of fully Bayesian estimation has rendered MLE, REML and marginal posterior mode estimation the canonical out-of-the-box methods. The likelihood of the data is given by
\begin{equation}
    \label{lik}
    \mathcal{L}(\boldsymbol{\beta}, \sigma, \boldsymbol{\lambda}) = p(\boldsymbol{y} \lvert \boldsymbol{\beta}, \sigma, \boldsymbol{\lambda}) \propto \left| \sigma^2 \boldsymbol{R} \right|^{-1/2} \exp\left\{-\frac{1}{2\sigma^2}(\boldsymbol{y}-\boldsymbol{G}\boldsymbol{\beta})^T\boldsymbol{R}^{-1}(\boldsymbol{y}-\boldsymbol{G}\boldsymbol{\beta})\right\},
\end{equation}
where $\boldsymbol{G}$ is the $n \times q$ matrix with entries $G_{ij} = g_{j}(\boldsymbol{x}^{(i)})$, and $\boldsymbol{R}$ denotes the $n \times n$ Gram-matrix of correlation function evaluations with entries $R_{ij} = r(\boldsymbol{x}^{(i)}, \boldsymbol{x}^{(j)})$. Under MLE, REML, and posterior mode estimation, estimates of $\boldsymbol{\beta}$ and $\sigma^2$ can be expressed in closed-form as functions of $\boldsymbol{\lambda}$. Denoting the MLE of $\boldsymbol{\beta}$ and $\sigma^2$ by $\boldsymbol{\hat{\beta}}$ and $\hat{\sigma}^2$, respectively, and the REML estimate of $\sigma^2$ by $\tilde{\sigma}^2$, in each case $\boldsymbol{\lambda}$ is chosen as the minimizer of the following:
\begin{equation}
    \label{REMLLambda}
    \begin{aligned}
        &\text{(MLE)} \quad &&n\ln\left( \hat{\sigma}^2(\boldsymbol{\lambda})\right) + \ln \left| 
\boldsymbol{R}(\boldsymbol{\lambda}) \right|, \\
&\text{(REML)} \quad &&(n-q)\ln\left( \tilde{\sigma}^2 \right) + \ln \left| 
 \boldsymbol{R}(\boldsymbol{\lambda}) \right| + \ln \left| 
 \boldsymbol{G}^T \boldsymbol{R}^{-1}(\boldsymbol{\lambda}) \boldsymbol{G} \right|.
    \end{aligned}
\end{equation}
In \eqref{REMLLambda}, the explicit dependence of the Gram matrix $\boldsymbol{R}$ on $\boldsymbol{\lambda}$ emphasizes the role that $\boldsymbol{R}$ plays in the estimation of $\boldsymbol{\lambda}$. The same is true under posterior mode estimation, which simply augments the REML objective function with the reference prior for $\boldsymbol{\lambda}$, which also depends on $\boldsymbol{R}$ \citep{Gu18}. 

The Gram matrix is an encoding of the pairwise correlations for the training data entirely in terms of \textit{pairwise} distances. Hence, each objective function is a highly complicated function of the set of pairwise distances. In relation to grounding, see Figure \ref{fig:motivatinggrounding}, there are two distinct types of `pair' influencing the estimation of the hyperparameter vector $\boldsymbol{\lambda}$ when $S$ grounds. A non-straddling pair will have both points on one each side of the grounding point, or grounding line for higher dimensional $x$, with both points corresponding to either grounded or non-grounded values of the simulator. A straddling pair will have one point on either side of the grounding line. Due to the difference in simulator behavior across the grounding line compared with simulator behavior on either side of the grounding line, each type of pair will affect the parameter estimation differently. Pairs straddling the grounding line will encourage a smaller estimate of $\boldsymbol{\lambda}$, favoring wiggliness, whilst non-straddling pairs with both points on one side of the grounding line will encourage larger $\boldsymbol{\lambda}$, favoring smoothness. We suspect that the complex interaction between straddling and non-straddling pairs in the training data, and their conflicting influence on the estimation of $\boldsymbol{\lambda}$, through the Gram matrix, is likely to be harming the performance of the GPE on grounding simulators.  

\subsection{Solution Outline}
\begin{figure}[ht!]
    \centering
    \begin{subfigure}{0.42\textwidth}
        \centering
        \includegraphics[width=\linewidth, trim={0 0 0 0},clip]{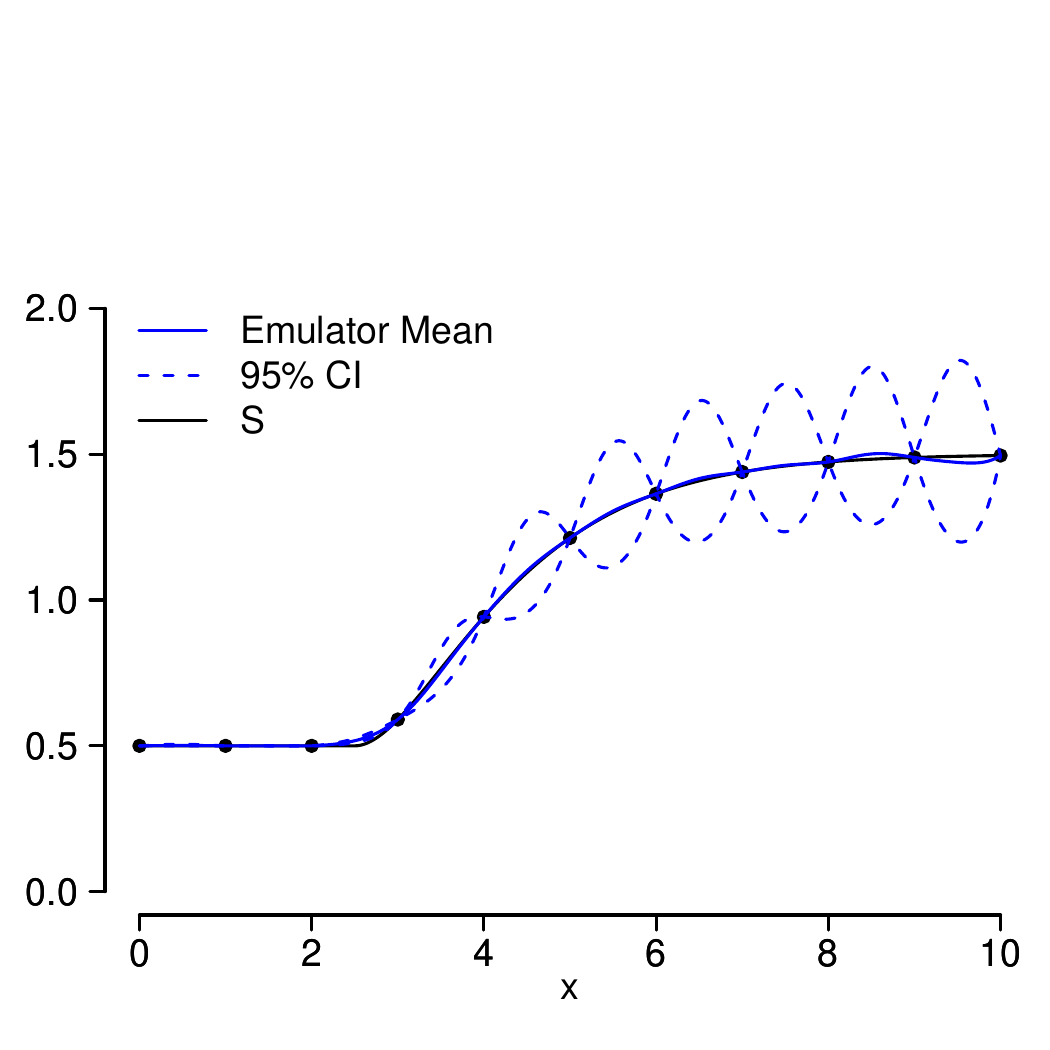}
    \end{subfigure}
    \begin{subfigure}{0.42\textwidth}
        \centering
        \includegraphics[width=\linewidth, trim={0 0 0 0},clip]{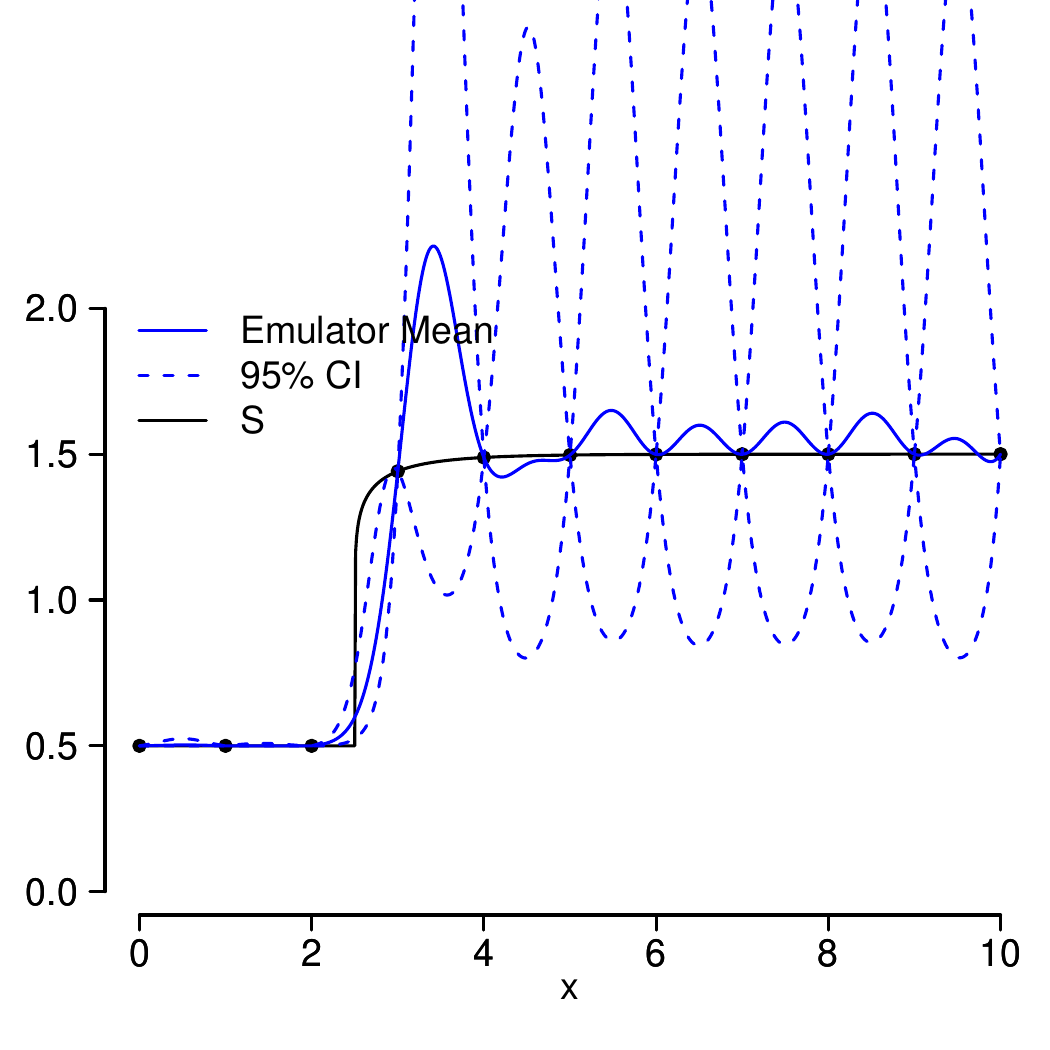}
    \end{subfigure}
    \begin{subfigure}{0.42\textwidth}
        \centering
        \includegraphics[width=\linewidth, trim={0 0 0 4cm},clip]{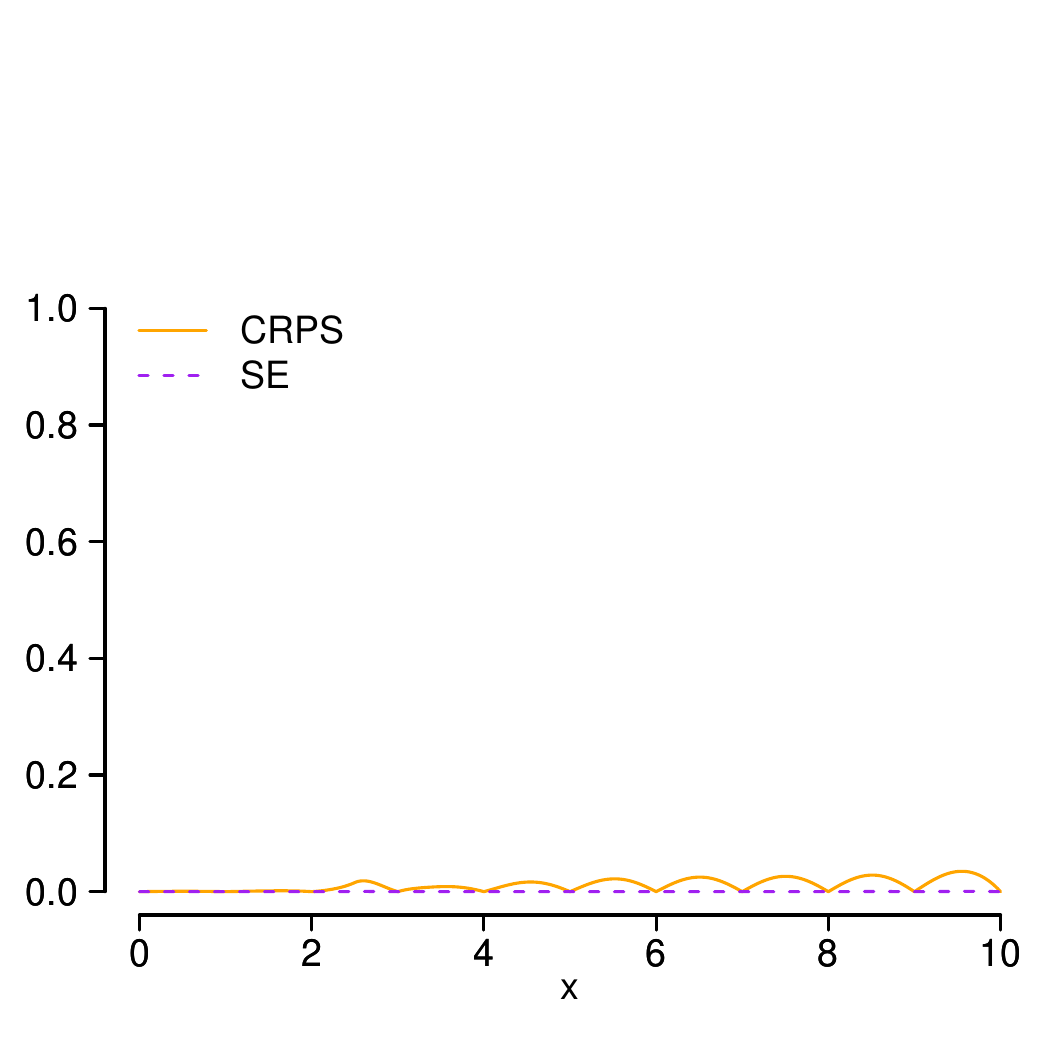}
    \end{subfigure}
    \begin{subfigure}{0.42\textwidth}
        \centering
        \includegraphics[width=\linewidth, trim={0 0 0 4cm},clip]{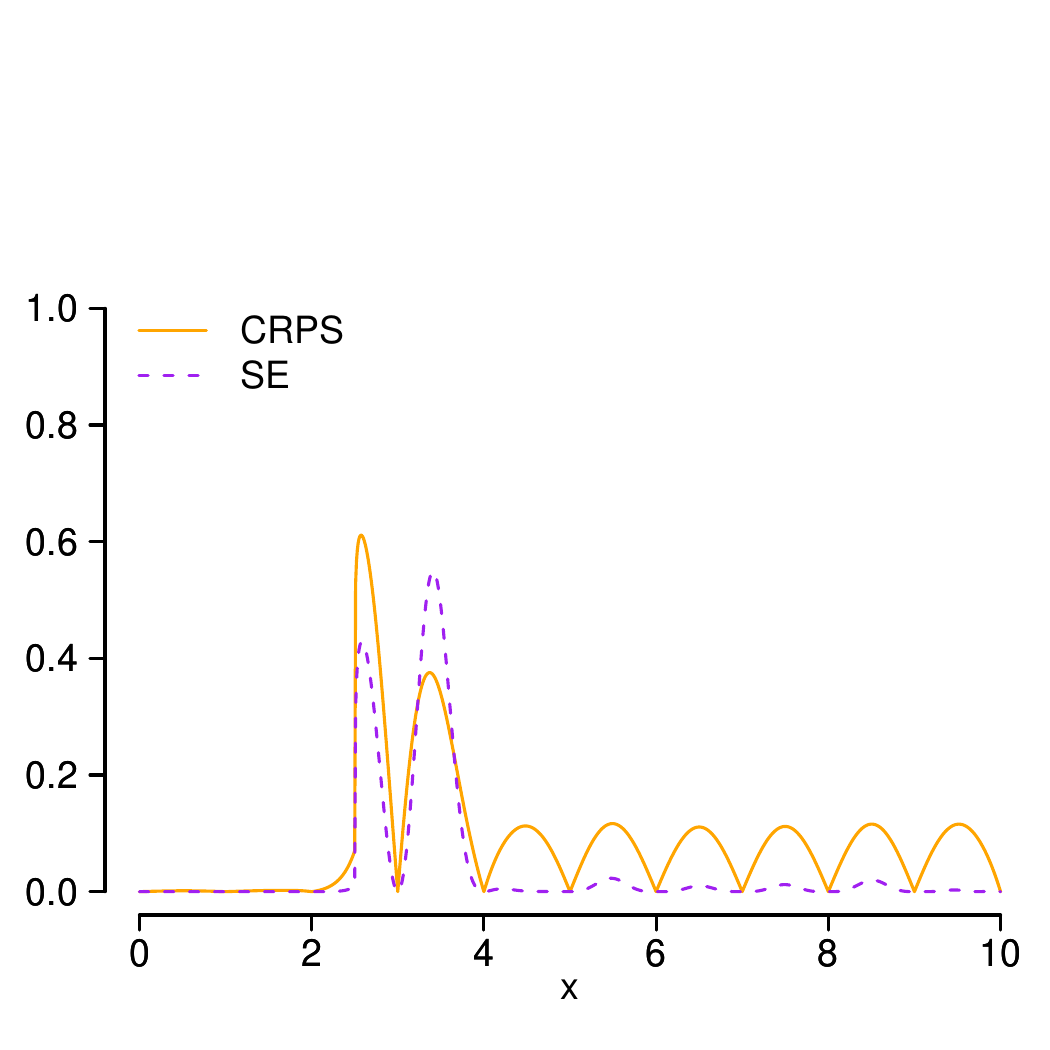}
    \end{subfigure}
    \caption{The fit of the GPE (top row) on the simulator in \eqref{1Dsimulator}. The left column is the soft landing, $\alpha = 2$, and the right column is the hard landing, $\alpha = 0.1$. The $95\%$ CI is asymmetrical as the GPE is fit using a log transform, see Section \ref{DoubleEmulator} for details. We also display the squared error (SE) and CRPS scores as functions of $x$ (bottom row).}
    \label{fig:1DSim}
\end{figure}

We outline our solution using an example.
\begin{example}[label=exa:cont]
Consider the synthetic simulator, based on the Gamma distribution function, with one-dimensional input and one-dimensional output, defined by
\begin{equation}
\label{1Dsimulator}
    S(x) = \begin{cases}
        F(x-s;\alpha, \sigma) + g, \, &(x-s) \geq 0, \\
        0, &(x-s) < 0,
    \end{cases}
\end{equation}
where $F$ is the distribution function of a Gamma random variable with shape and scale parameters $\alpha$ and $\sigma$. Suppose we are interested in emulating $S$ for $x \in [0,10]$. The triple $(s, \alpha, g) \in \mathbb {R}^3$ facilitates control over the shape of the simulator as follows:
\begin{enumerate}
    \item The volume of the grounded space, $[0,s]$, is controlled through $s$.
    \item The magnitude of the derivative at the point of grounding is controlled through the shape, $\alpha$.
    \item The grounding value of the simulator is determined by the value of $g$.
\end{enumerate}

Figure \ref{fig:1DSim} (top left) analyses the fit of the GPE for $s = 2.5$ and $\alpha = 2$; a `soft' landing. Figure \ref{fig:1DSim} (top right) does the same for $s = 2.5$ and $\alpha = 0.1$; a `hard' landing. Specifically, a hard landing corresponds to the simulator having a discontinuity in its first derivative when attaining its minimum. This provides a visual cue as to how the GPE struggles to emulate a simulator which changes shape rapidly when attaining its minimum. With the hard landing, as the GPE approaches $x = 2.5$ from the right, the posterior mean of the GPE considerably underestimates the simulator output, whilst from the left the posterior mean overestimates the simulator output, leading to a spike in both the CRPS and squared error (SE) near the grounding point at $x = 2.5$. In addition, close to the grounding point, the simulator output does not even lie within the $95\%$ posterior prediction interval of the GPE, which is denoted by the dashed lines, supporting the heuristic Gram matrix argument in Section \ref{Existing}. In effect, the GPE is over confident about predictions which are poor. Under a calibrated performance metric such as the CRPS score, this is worse than making predictions equally as far from the true value of the simulator but with larger uncertainty. Visually, the GPE performance does not degrade so obviously when the simulator lands softly, approaching its minimum smoothly without a discontinuity in its first derivative, and where the smoothness assumptions of the GPE are valid. In each case linear basis functions were used for the GPE's regression component, and a Mat\'ern-$5/2$ kernel was used for the residual. With the hard landing, the smoothness assumptions imposed by the Mat\'ern-$5/2$ kernel are possibly limiting the emulator's ability to capture the behavior of the simulator around the landing.
\end{example}
This analysis leads to the following observations:
\begin{enumerate}
    \item A GPE struggles to estimate functions displaying rapid shape change when attaining their minimum. 
    \item A GPE should not be fit to the data points lying in the grounded region; an emulator should simply return the known minimum output value with zero uncertainty at any point in this region.
    \item Fitting a GPE only to data points outside the grounded region will lead to increased variance across the grounded region as the GPE extrapolates either out of the convex hull of its training data, or between pairs of points in its training data.
    \item The region on which the simulator attains its minimum is unknown and the probability that a given input lies in this region can be estimated using a probabilistic classifier, such as a random forest (RF) \citep{breiman01} or kernel support vector machine (SVM) \citep{cortes95}.
\end{enumerate}

\section{The Double Emulator}
\label{DoubleEmulator}
We now present the mathematical details of our novel approach; a mixture of a Gaussian process and a point mass with component weights determined using a probabilistic classifier. 
\subsection{The Model}
We consider a more general scenario than the one-dimensional example in \eqref{1Dsimulator} by defining a statistical model for any simulator $S: \mathcal{X} \xrightarrow[]{} \mathcal{Y}$ which grounds at $g$, where $\mathcal{X} \subseteq \mathbb{R}^d$ and $\mathcal{Y} \subset \mathbb {R}$ i.e. $S(\boldsymbol{x}) \geq g$ for all $\boldsymbol{x} \in \mathcal{X}.$ Our statistical model for the simulator at any $\boldsymbol{x}$ is defined as follows:
\begin{definition}
\label{DEDefinition}
\begin{equation}
\begin{aligned}
        F(y;\boldsymbol{x}) &\coloneqq  \mathbb{P}(S(\boldsymbol{x})\leq y), \\
        &= 
    \begin{cases} 
      0, & y < g, \\
      1 - p(\boldsymbol{x}), & y = g, \\
      1- p(\boldsymbol{x}) + p(\boldsymbol{x}) \cdot L(y-g+\gamma;m(\boldsymbol{x}), v(\boldsymbol{x})), & y > g,
\end{cases}
\end{aligned}
\label{DEDef}
\end{equation}
where $L(\cdot;m(\boldsymbol{x}),v(\boldsymbol{x}))$ is the Lognormal distribution function, with $m(\boldsymbol{x})$ and $v(\boldsymbol{x})$ its `meanlog' and `varlog' parameters, respectively, $\gamma$ is a small positive constant improving the stability of the log scale emulation, and $p(\boldsymbol{x}) \coloneqq \mathbb{P}(S(\boldsymbol{x}) > g)$ is an estimated probability that the simulator is above ground. Supplement \ref{lognormaldetails} contains details on the Lognormal distribution.
\end{definition}
Specifying that a random variable $Y$ has a Lognormal distribution function with meanlog $m$ and varlog $v$ is equivalent to specifying that $\ln Y$ is normally distributed with mean $m$ and variance $v$. Furthermore, the closed form expressions for the mean and variance of the Lognormal distribution, in terms of $m$ and $v$, are well known, see, for example, \citet{casella02} and Supplement \ref{lognormaldetails}. Our main reason for using a log transformation is because the simulators of interest are bounded below; na\"ively fitting a GPE without transforming the output would place non-zero probability on values below the simulator minimum.

Define the indicator function $\mathds{1}_g: \mathcal{Y} \xrightarrow[]{} \{0,1\}$ as
\begin{equation}
    \begin{aligned}
        \mathds{1}_g(y) \coloneqq \begin{cases} 
      1, & y > g, \\
      0, & y \leq g.
   \end{cases}
    \end{aligned}
\end{equation}
Letting $\{X,Y\}$ denote the entire ensemble of simulator runs, estimation of the model parameters $p(\cdot), \, m(\cdot)$ and $v(\cdot)$, can be performed as follows:
\begin{enumerate}
    \item Estimate $p(\boldsymbol{x}) \coloneqq \mathbb{P}(S(\boldsymbol{x}) > g)$ by training a probabilistic classifier on $\{X,\mathds{1}_g(Y)\}$.
    \item Estimate $m(\boldsymbol{x})$ and $v(\boldsymbol{x})$ by training a GPE on the data $\{X, \ln (Y-g+\gamma)\}$ using only the subset of $\{X, Y\}$ for which $y > g$. 
\end{enumerate}
Once $p(\cdot), \, m(\cdot)$ and $v(\cdot)$ are estimated, the distribution function at any $\boldsymbol{x}$ is obtained via \eqref{DEDef}. Due to the presence of an atom at $g$, the random variable defined by \eqref{DEDef} does not admit a density. However, expressions for the mean, $\mathbb{E} \, S(\boldsymbol{x})$, and variance, $\Var \, S(\boldsymbol{x})$, can be calculated using the Law of Iterated Expectation and Law of Total Variance, respectively, by conditioning on the indicator $I(\boldsymbol{x}) \coloneqq \mathds{1}_g(S(\boldsymbol{x}))$. This is encompassed in the following lemma:
\begin{lemma}
Since $Y-g+\gamma \sim L(m(\boldsymbol{x}), v(\boldsymbol{x}))$, the expressions for the mean and variance of $Y$, denoted by $M(\boldsymbol{x})$ and $V(\boldsymbol{x})$, respectively, are given by
\begin{equation}
    \label{DEMom}
    \begin{aligned}
        M(\boldsymbol{x}) &= e^{m(\boldsymbol{x}) + v(\boldsymbol{x})/2} + g - \gamma, \\
        V(\boldsymbol{x}) &= e^{2m(\boldsymbol{x}) + v(\boldsymbol{x})}\,(e^{v(\boldsymbol{x})}-1).
    \end{aligned}
\end{equation}
The double emulator mean function is then given by
\begin{equation}
    \label{DEMean}
    \begin{aligned}
        \mathbb{E} \, S(\boldsymbol{x}) &= \mathbb{E} \, \mathbb{E} \, (S(\boldsymbol{x}) \lvert I(\boldsymbol{x})), \\
        &= (1-p(\boldsymbol{x}))g + p(x)M(\boldsymbol{x}).
    \end{aligned}
\end{equation}
Similarly, the double emulator variance function is given by
\begin{equation}
    \label{DEVar}
    \begin{aligned}
        \Var \, S(\boldsymbol{x}) &= \Var \, \mathbb{E} \, (S(\boldsymbol{x}) \lvert I(\boldsymbol{x})) + \mathbb{E} \, \Var \, (S(\boldsymbol{x}) \lvert I(\boldsymbol{x})), \\
        &= (1-p(\boldsymbol{x}))g^2 + p(\boldsymbol{x})M(\boldsymbol{x})^2 - (\mathbb{E} \, S(\boldsymbol{x}))^2 + p(\boldsymbol{x}) V(\boldsymbol{x}), \\
        &= p(\boldsymbol{x})(1-p(\boldsymbol{x}))(g-M(\boldsymbol{x}))^2 + p(\boldsymbol{x})V(\boldsymbol{x}).
    \end{aligned}
\end{equation}
\Eqref{DEMom}--\eqref{DEVar} define the mean and variance functions of the double emulator.
\end{lemma}
\subsection{Classifier Choice}
\label{ClassifierChoice}
In principle any probabilistic classifier could be used to estimate the probability of a non-grounding, $p(\boldsymbol{x})$, for the double emulator. Despite this, the logical approach would involve using any information relating to the shape of the grounding line to tailor the choice of classifier to the simulator under investigation. In high dimensional input spaces this is extremely difficult, if not impossible, to intuit, so using a sufficiently flexible method is paramount. With this in mind, we consider two well known classifiers; the SVM the RF. The arguments for and against these classifiers amount to the trade-off between flexibility and interpretability. For benchmarking, we also use an artificial omniscient classifier which always returns the correct class membership; this isolates the gain in performance solely by improving the classifier. 

As discussed by \citet{hastie09}, there are several ways of presenting and interpreting the SVM. For the double emulator the main strength of the SVM is its ability to estimate extremely curved, wiggly grounding lines. In particular the SVM has this ability because it exploits the `kernel trick'. Contrary to the SVM, which has the ability to fit an extremely flexible non-axially-aligned decision boundary, the decision boundaries of the constituent trees in a RF are composed of axially-aligned cuts corresponding to binary splits on the input variables. Combining the trees as a RF splits the entire input region into smaller rectangular regions, the boundaries of which remain axially-aligned. Hence, for a non-axially-aligned grounding line, such as a diagonal line in a two-dimensional input space, using a RF will only approximately `tile over' the non-axially-aligned sections of the grounding line. 

Despite the greater flexibility of the SVM, a compelling argument in favor of the RF is its interpretability. Specifically, with a small number of interpretable tuning parameters, unexpected behavior in the RF can be more easily diagnosed compared with the SVM, which is more of a black-box. With this in mind, using a RF within the double emulator may be preferable from the practitioner's perspective. We emphasize that, although each classifier can be improved by tuning, bearing in mind the importance of ease of implementation, in this article we choose \textit{not to tune} the classifiers, only comparing the classifiers out-of-the-box. Hence, default settings for each classifier are used. In addition, for these very reasons we do not use classifiers such as neural networks in this article, although we acknowledge the appeal of their power and flexibility; we prefer to use two popular, simple classifiers, and benchmark them with the perfect classifier. 

For the SVM, class probabilities are estimated using Platt scaling \citep{platt99} and the default kernel is Gaussian. Class probabilities for the RF are approximated using the proportion of votes of the individual trees. For our experiments in Section \ref{Experiments} we use the implementation contained in the \texttt{kernlab} package \citep{kernlab} for the SVM, and the \texttt{randomForest} package \citep{Liaw02} for the RF. All experiments are conducted in \textsf{R} \citep{rcompenv}. Full details on the classifiers and their default settings can be found in \citet{kernlab} and \citet{Liaw02}. 
\section{Experiments}
\label{Experiments}
This section compares the double emulator with the conventional emulator on a range of synthetic examples and a real-world example. The synthetic examples are particularly useful as we can control the aspects of the simulator output which we anticipate will challenge the GPE and influence the performance of the double emulator. In particular, we expect the volume of the grounded region, the complexity of the grounding line and the magnitude of the simulator's derivative upon grounding all to play a crucial role in the analysis. First, we detail the performance metrics used in the analysis.

We use two performance metrics; RMSE and CRPS. The definition of RMSE, see Supplement \ref{rmsesection}, is straightforward; the crucial point is that it only assesses the quality of the emulator's posterior mean function as a point estimate for the simulator. As an emulator is a full probabilistic representation of the simulator output as a function of the inputs, a comprehensive assessment of the emulator requires a metric for comparing a forecasted distribution with a single realization. We can use the (negatively oriented) CRPS score \citep{Gneiting07}, which for a forecasted distribution, $F$, and observation, $y$, is defined as
\begin{equation}
\label{crpsdef}
    \text{CRPS}(F,y) = \int_{\mathbb{R}} (F(z) - H(z \geq y))^2 \; dz,
\end{equation}
where $H$ is the Heaviside step function.

When the distribution function $F$ can be evaluated, the CRPS can be approximated by numerically integrating \eqref{crpsdef}. For `straightforward' distributions, closed form expressions for the CRPS exist, though in our case the existence of an exact expression is not immediately obvious. In fact, an exact formula, for which the proof can be found in Supplement \ref{crpssection}, is as follows:
\begin{lemma}
\label{crpslemma}
For $u \geq g$ the CRPS score for the model, $F$, in Definition \ref{DEDefinition} is given by
\begin{align}
\text{CRPS}(F, u) &= (u-g) + 2 p \int_\gamma^{u-g+\gamma} (L(z) - 1) dz + p^2 \int_\gamma^\infty (L(z)-1)^2 dz.
\end{align}
Explicit dependence of $p, \, m$ and $v$ on $\boldsymbol{x}$ is omitted for brevity, whilst exact expressions for the integrals are available in Supplement \ref{crpssection}.
\end{lemma}
\subsection{Synthetic Simulators}
To assess the performance of the double emulator on simulators with inputs of dimension $d>1$, we use the `curved' (DP) function \citep{Dette_Pepelyshev_2010}, as well as Rosenbrock's `Banana' function \citep{rosenbrock60}. Variants `A' and `B' \citep{Kok09} of the Banana are common choices for testing methods such as kriging \citep{picheny13}, whilst Gaussian process emulation also formed the basis of the analysis by \citet{Dette_Pepelyshev_2010}, albeit from a design perspective. The DP function is defined as 
\begin{equation}
    \label{DPFunction}
    f(\boldsymbol{x}) = 4(x_1 - 2 + 8x_2 -8x_2^2)^2 + (3-4x_2)^2 + 16\sqrt{(x_3+1)}(2x_3-1)^2.
 \end{equation}
With the Banana, we focus on variant `B', defined for $\boldsymbol{x} \in \mathbb{R}^d, \; d \geq 2$ as 
\begin{equation}
\label{banana}
    f(\boldsymbol{x}) = \sum_{i=1}^{d-1} \left[ (1 - x_i)^2 + 100 \cdot (x_{i+1} - x_i^2)^2 \right].
\end{equation}
For $d = 2$ we have 
\begin{equation}
\label{banana2}
    f(\boldsymbol{x}) = (1-x_1)^2 + 100 \cdot (x_2-x_1)^2,
\end{equation}
which is the function in \citet{rosenbrock60}. Though each function is defined on $\mathbb{R}^d$, for the purposes of our analysis we restrict the domain to $[0,1]^d$, fixing $d = 8$ for the Banana. We adopt a strategy similar to that in the one dimensional case in \eqref{1Dsimulator}, where we introduce two further parameters $(a,b) \in \mathbb{R}$, analogous to $s$ and $\alpha$ in \eqref{1Dsimulator}, allowing us to control the size of the grounded region and the behavior of the derivative around the grounding line, respectively. For each $f$, a family of simulators is then defined by 
\begin{equation}
\label{bananasim}
    S(\boldsymbol{x}) \coloneqq \max\Bigg\{ 0, \bigg|\frac{f(\boldsymbol{x}) - a}{m}\bigg|^b \Bigg\},
\end{equation}
where $m \coloneqq \max \{f : {\boldsymbol{x} \in [0,1]^d}\}$ is used to scale $S(\boldsymbol{x})$ onto $[0,1]$. 

\begin{figure}[ht!]
    \centering
    \begin{subfigure}{0.3\textwidth}
        \centering
        \includegraphics[width=\linewidth]{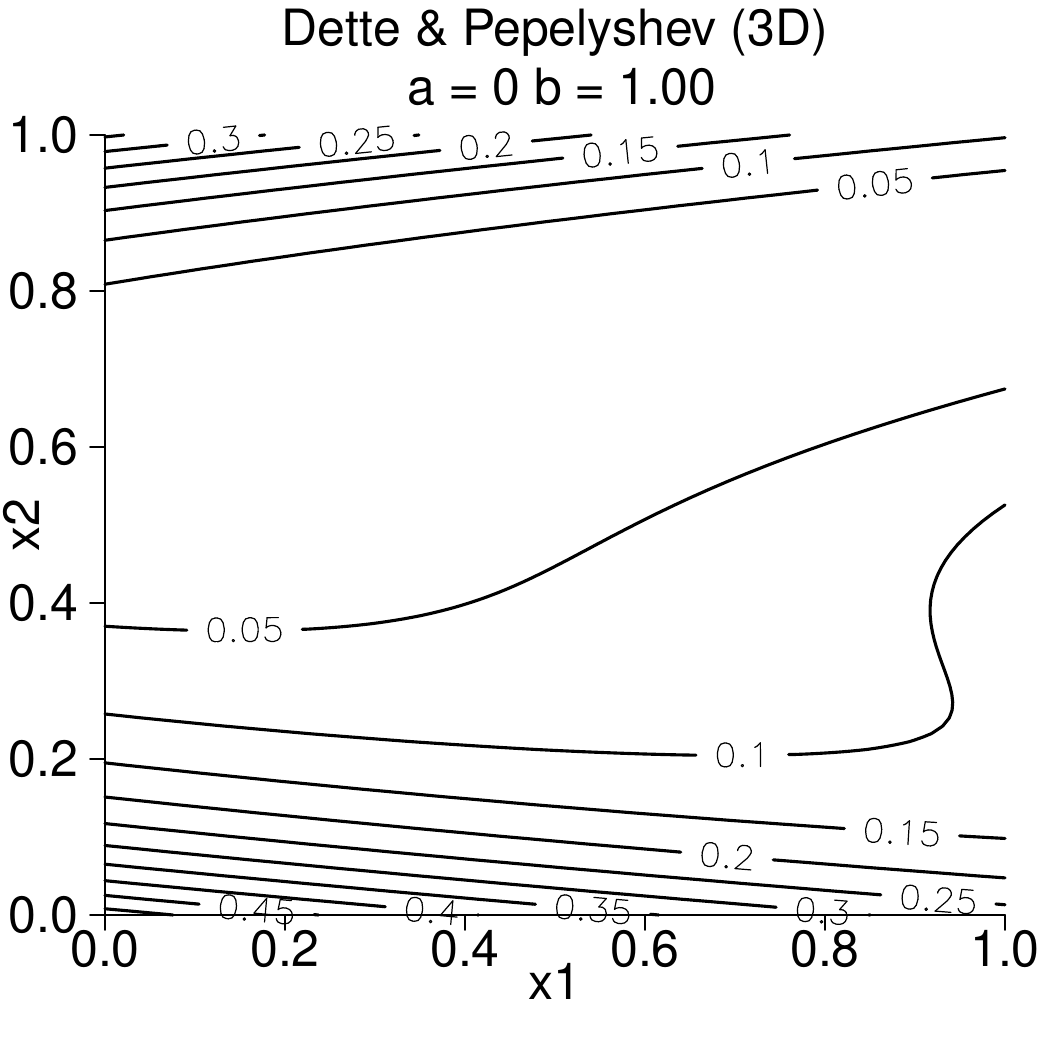}
    \end{subfigure}
    \begin{subfigure}{0.3\textwidth}
        \centering
        \includegraphics[width=\linewidth]{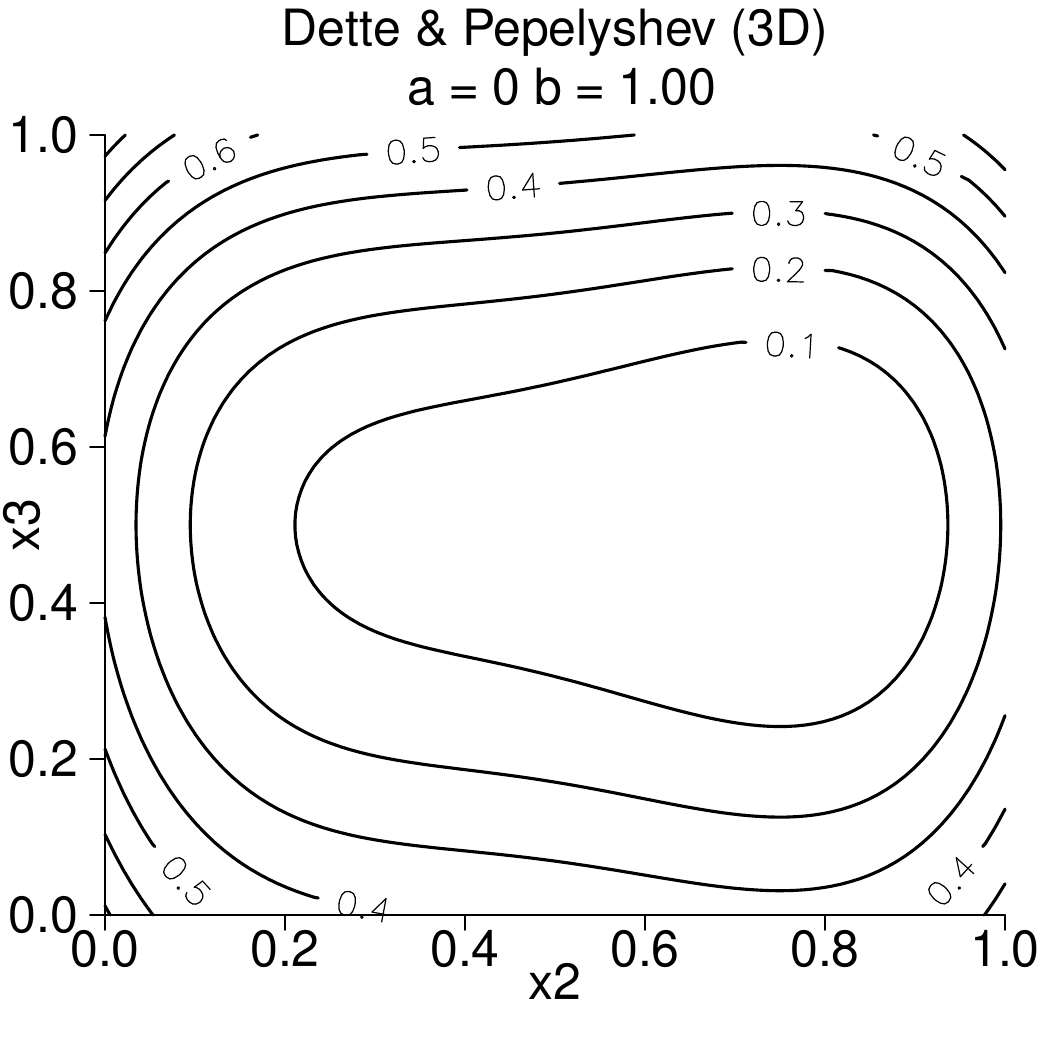}
    \end{subfigure}
    \begin{subfigure}{0.3\textwidth}
        \centering
        \includegraphics[width=\linewidth]{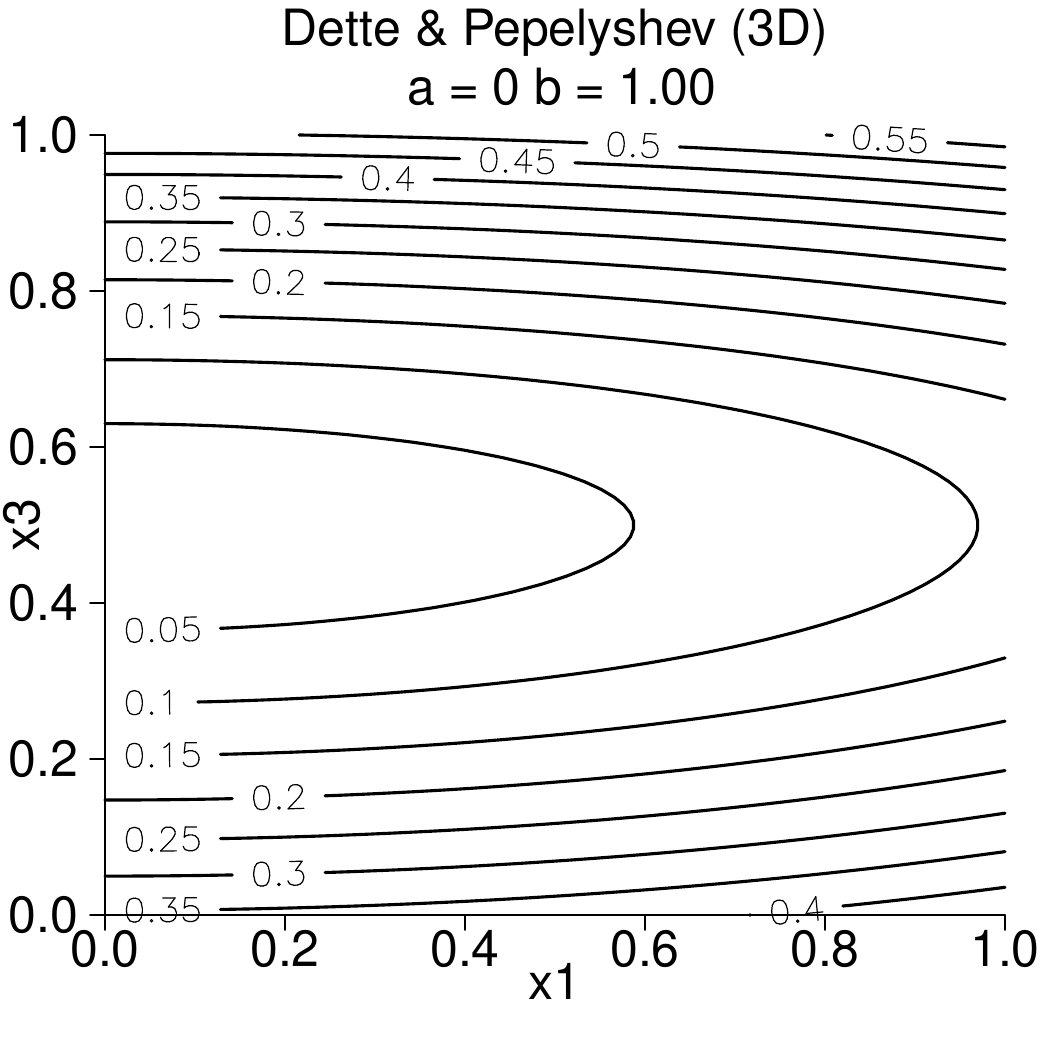}
    \end{subfigure}
    \begin{subfigure}{0.3\textwidth}
        \centering
        \includegraphics[width=\linewidth]{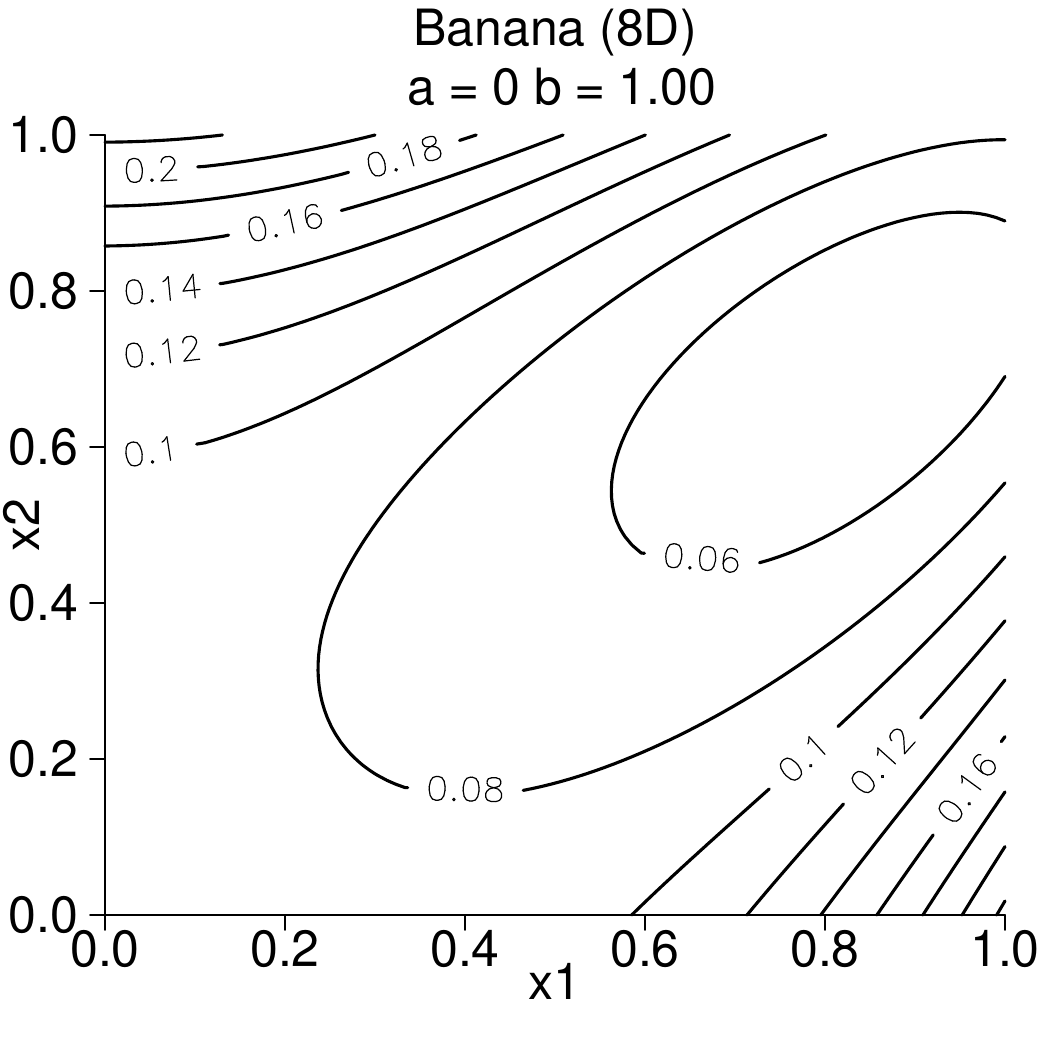}
    \end{subfigure}
    \begin{subfigure}{0.3\textwidth}
        \centering
        \includegraphics[width=\linewidth]{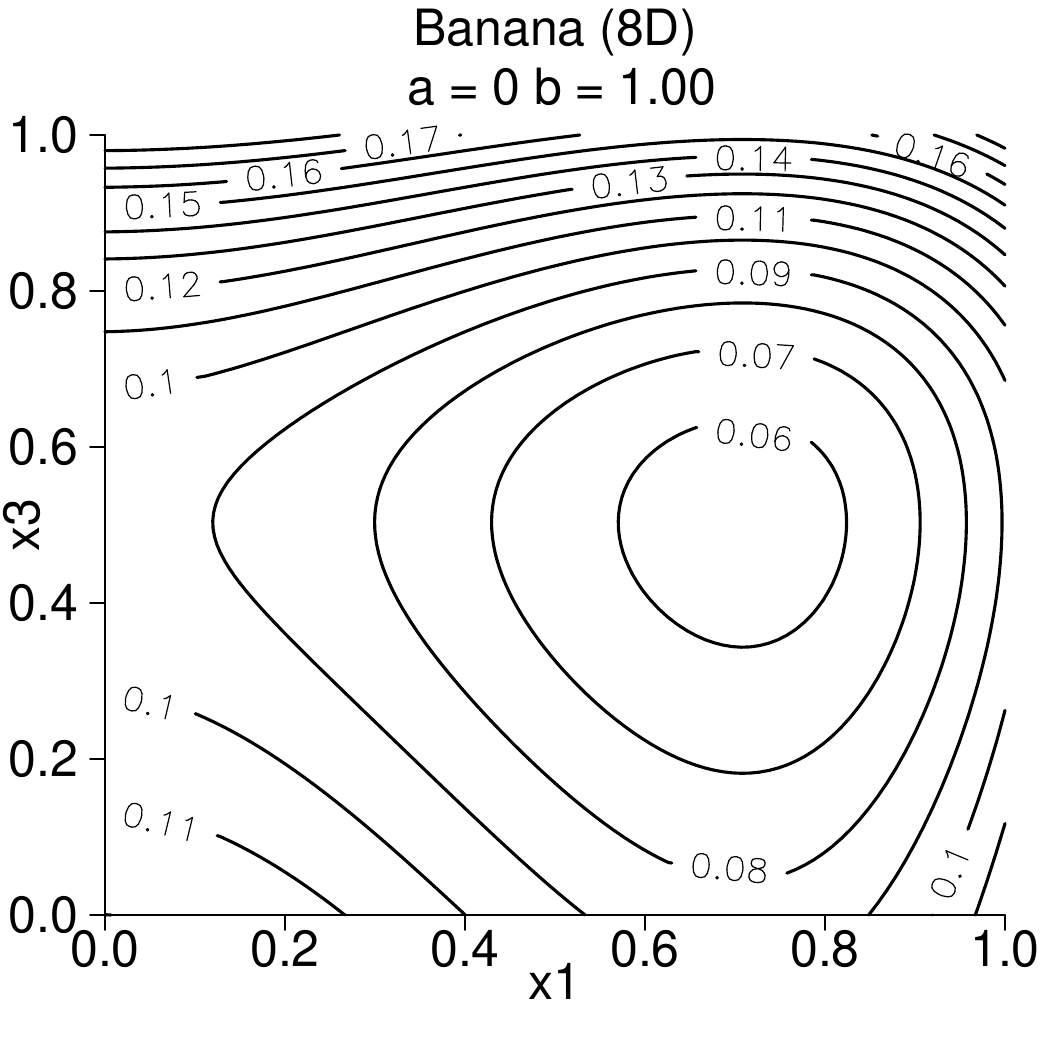}
    \end{subfigure}
    \begin{subfigure}{0.3\textwidth}
        \centering
        \includegraphics[width=\linewidth]{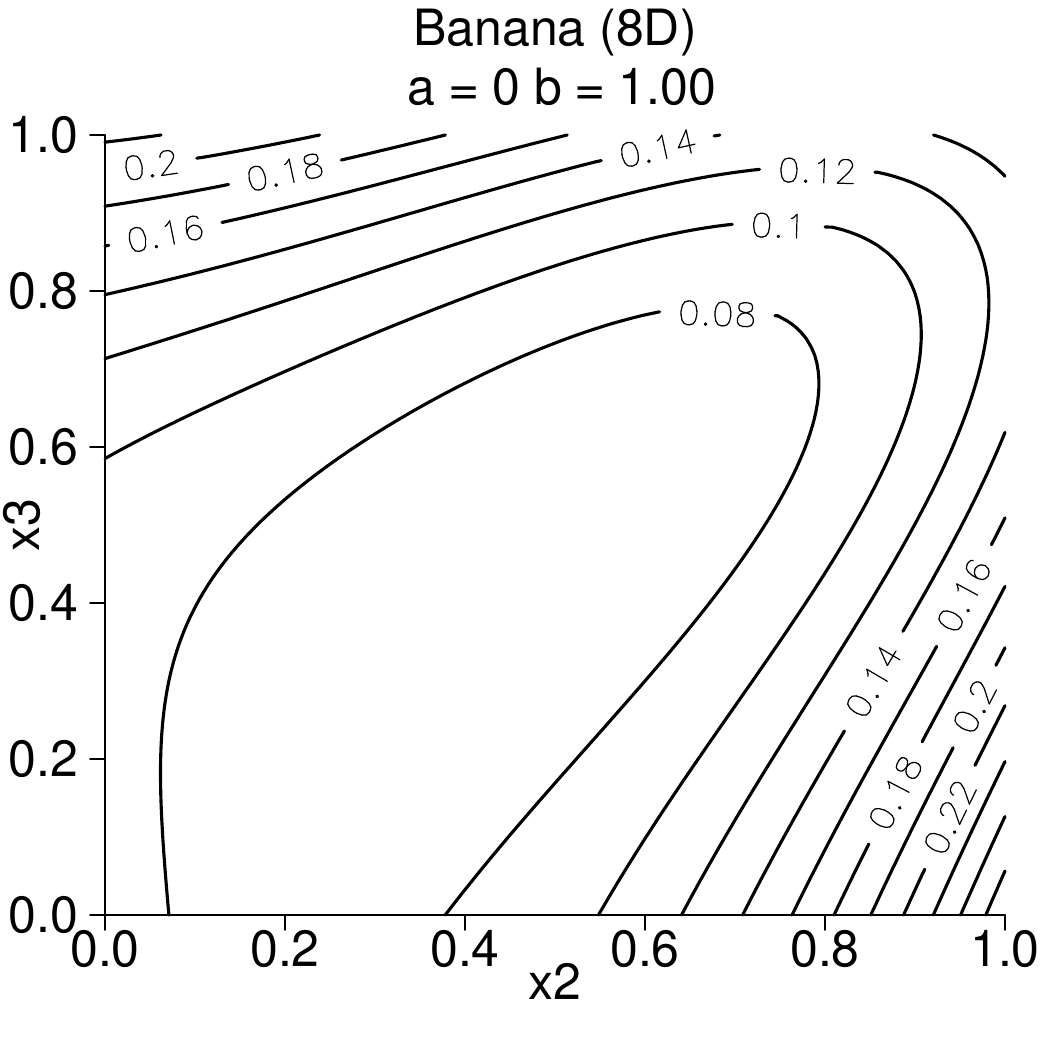}
    \end{subfigure}
    \caption{Nominal plots of Dettel \& Pepelyshev's 3D function (above) and Rosenbrock's eight dimensional variant `B' Banana (below). In both cases, the rich shape of the grounding region that will be introduced by varying the threshold parameter $a$ in \eqref{bananasim} is reflected through the richness of the level sets. Note, the numbers on the level sets correspond to the simulator output value.}
    \label{fig:NomBans}
\end{figure}

A subset of the `nominal runs' of each synthetic simulator, generated by varying two of the inputs over a dense grid on $[0,1]^2$, with all other inputs fixed at nominal values of 0.5, are included in Figure \ref{fig:NomBans}. In each case, this reveals the richness of the shape of the grounding line generated as a result of varying the threshold parameter, $a$. In particular, it is now evident that using a classifier which can accurately detect non-axially-aligned grounding lines will be crucial for the method to function well.

\begin{figure}[ht!]
    \centering
    \begin{subfigure}{0.3\textwidth}
        \centering
        \includegraphics[width=\linewidth]{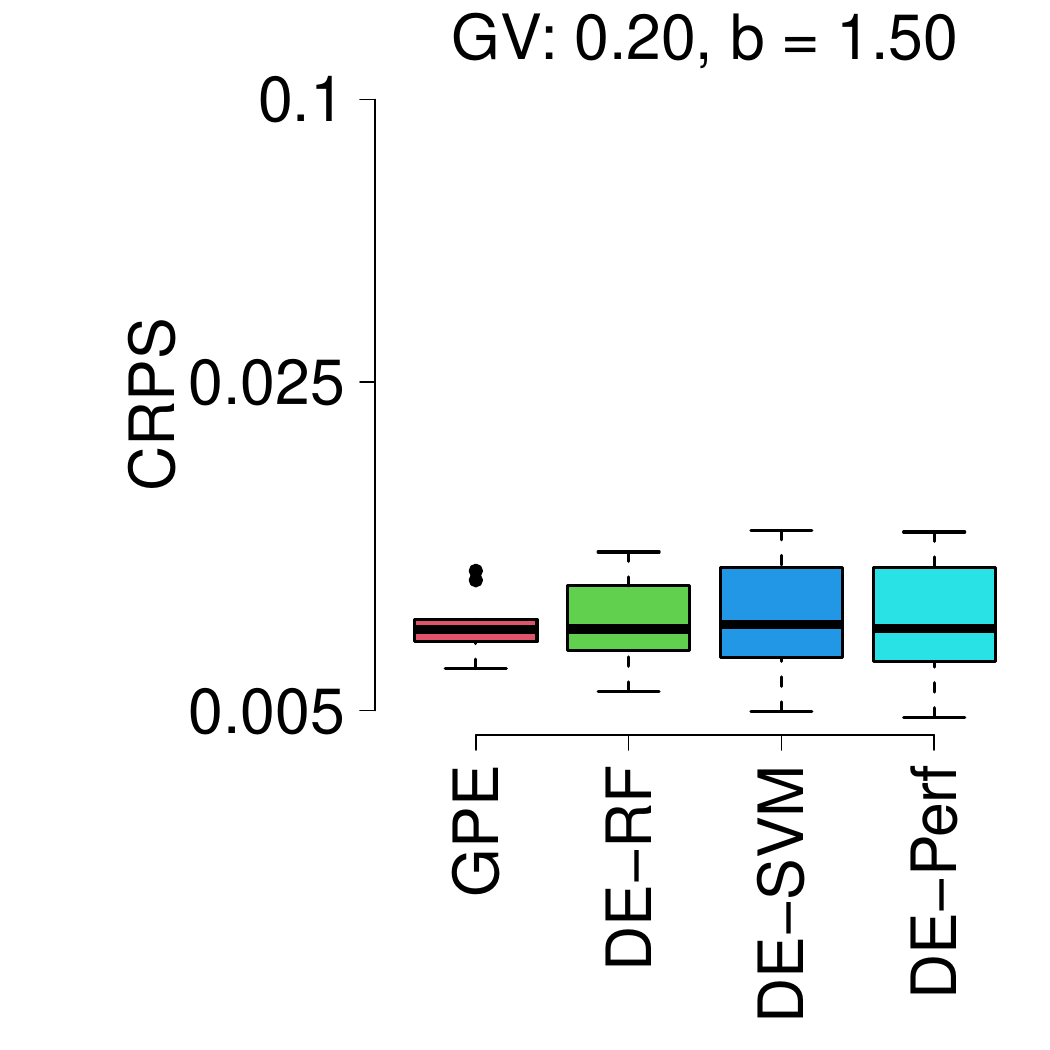}
    \end{subfigure}
    \begin{subfigure}{0.3\textwidth}
        \centering
        \includegraphics[width=\linewidth]{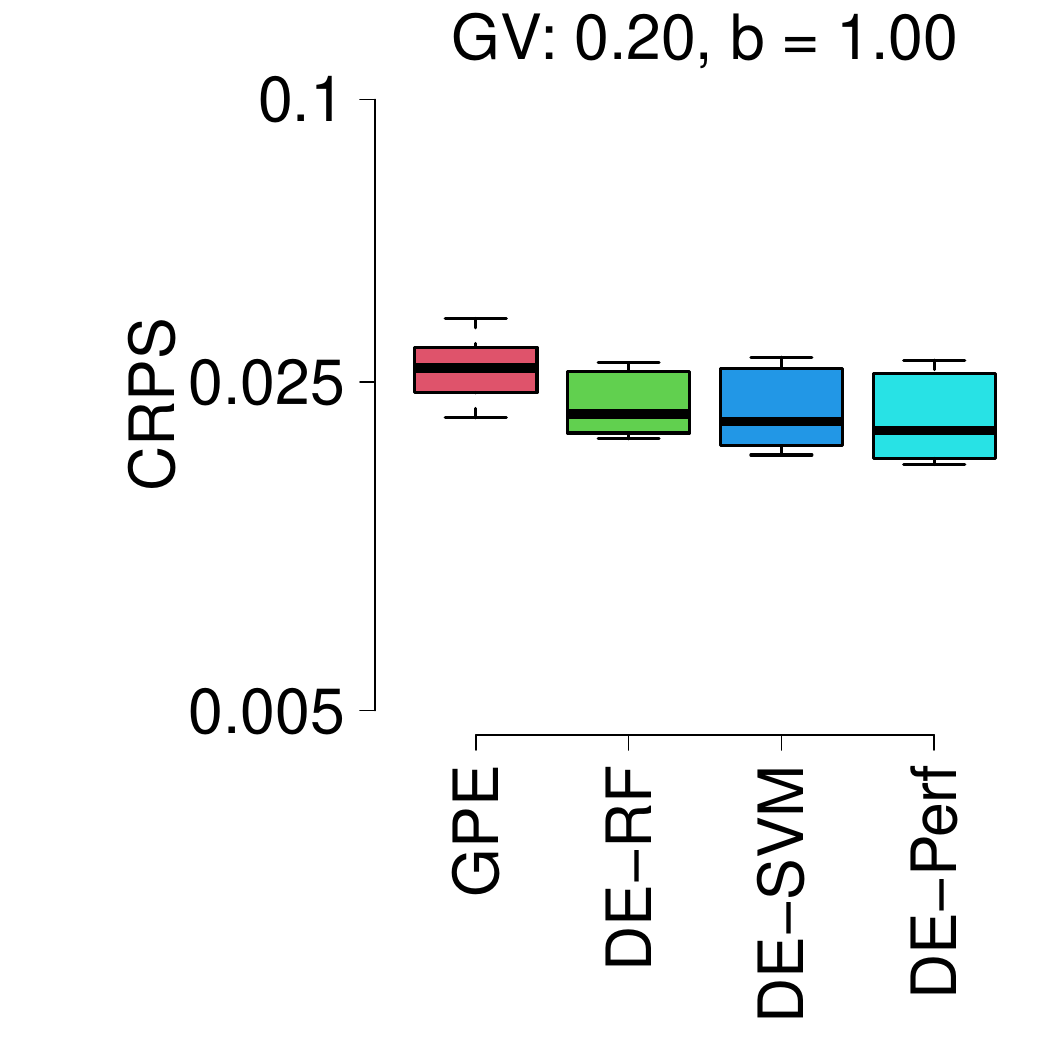}
    \end{subfigure}
    \begin{subfigure}{0.3\textwidth}
        \centering
        \includegraphics[width=\linewidth]{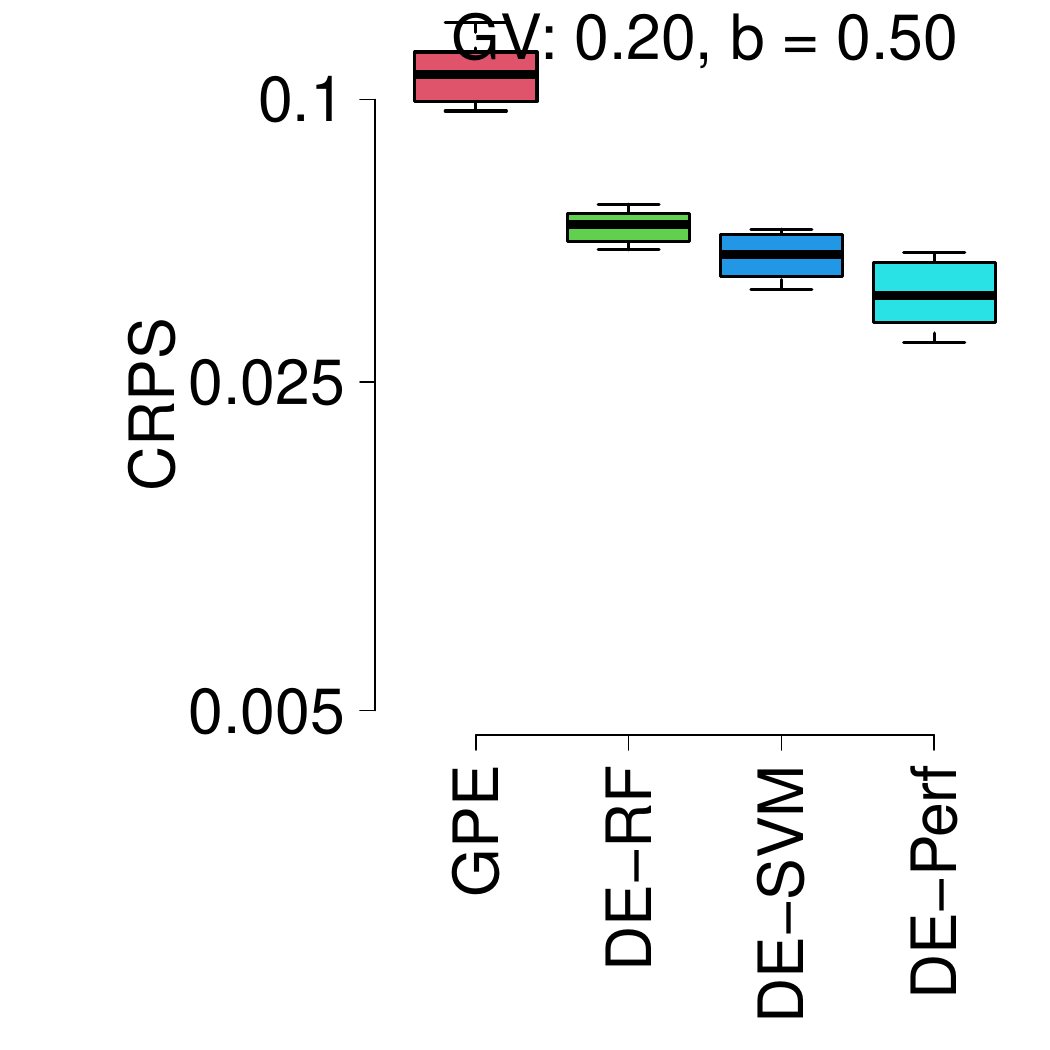}
    \end{subfigure}
    \begin{subfigure}{0.3\textwidth}
        \centering
        \includegraphics[width=\linewidth]{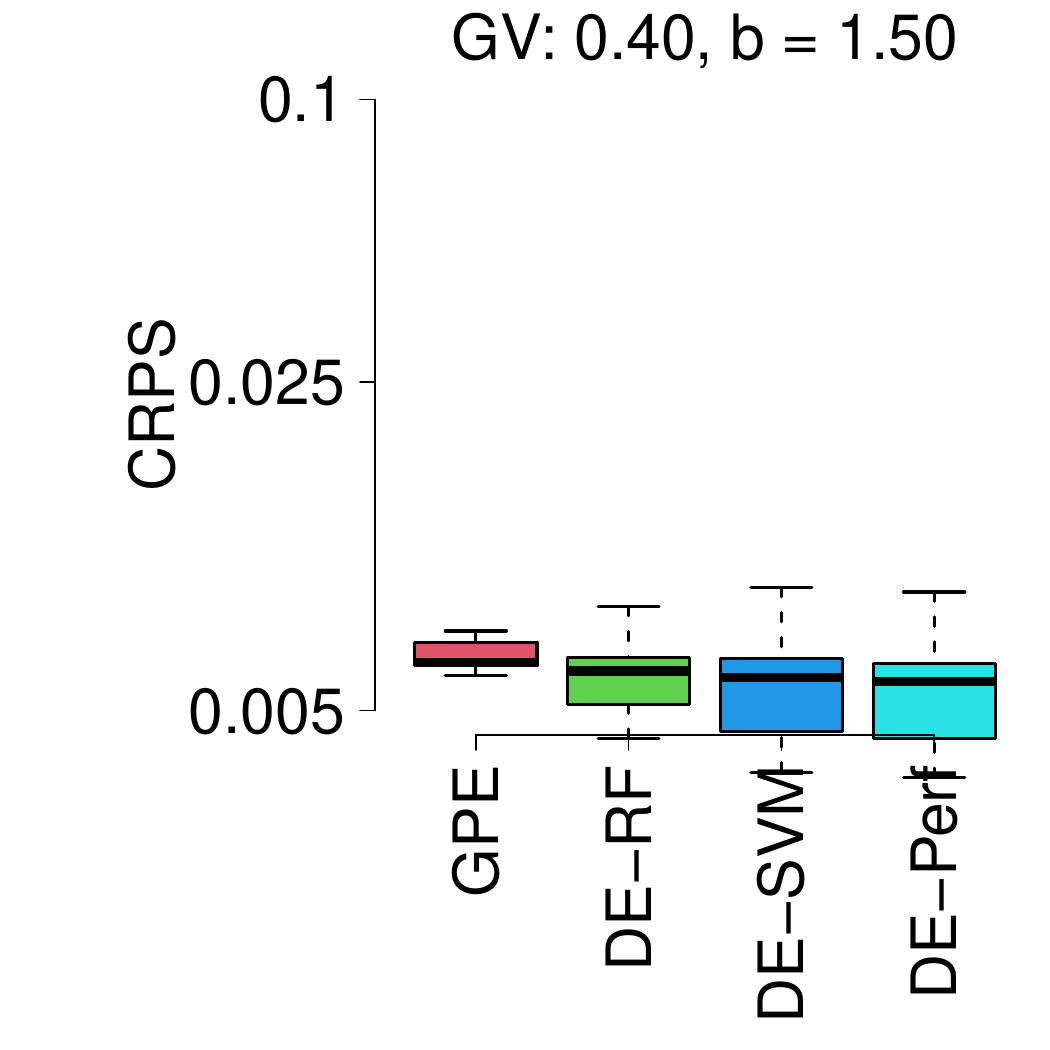}
    \end{subfigure}
    \begin{subfigure}{0.3\textwidth}
        \centering
        \includegraphics[width=\linewidth]{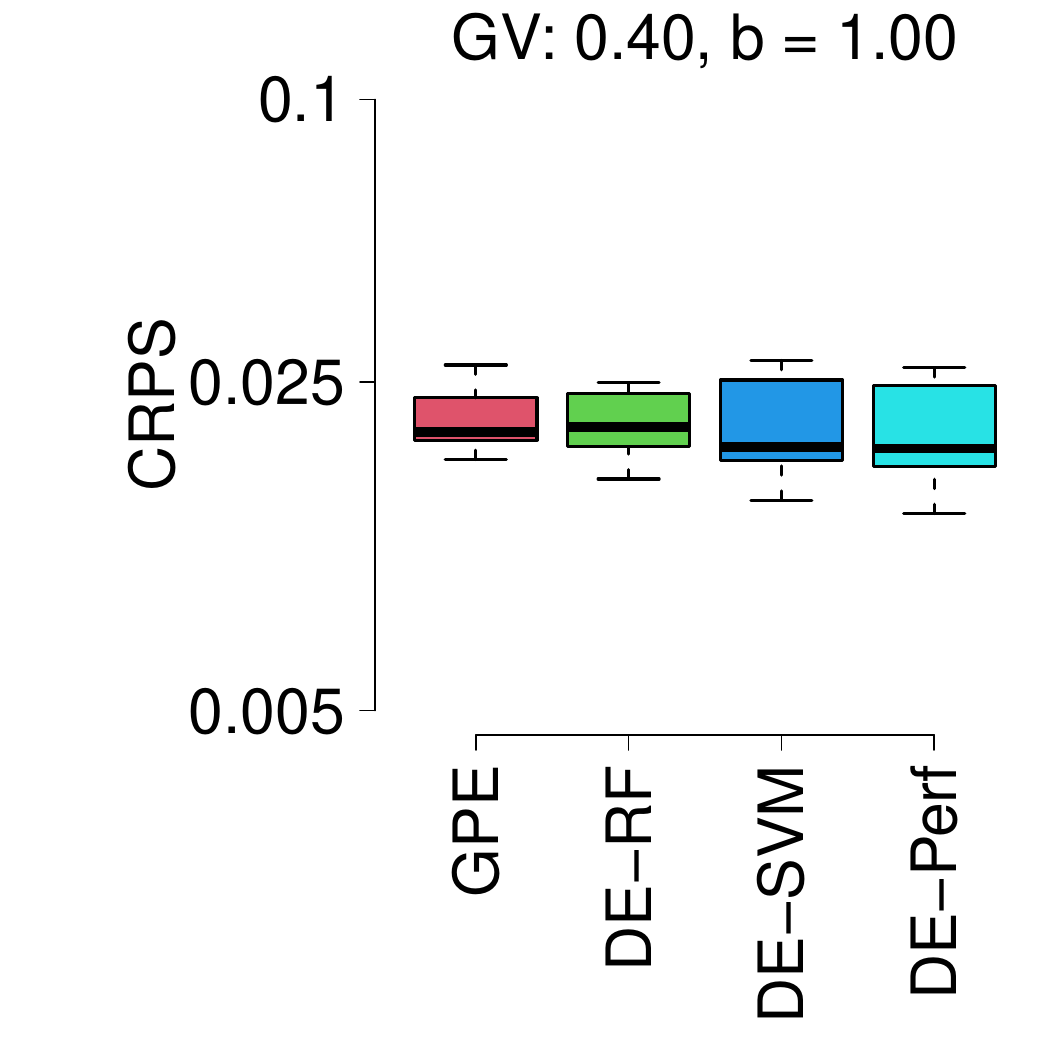}
    \end{subfigure}
    \begin{subfigure}{0.3\textwidth}
        \centering
        \includegraphics[width=\linewidth]{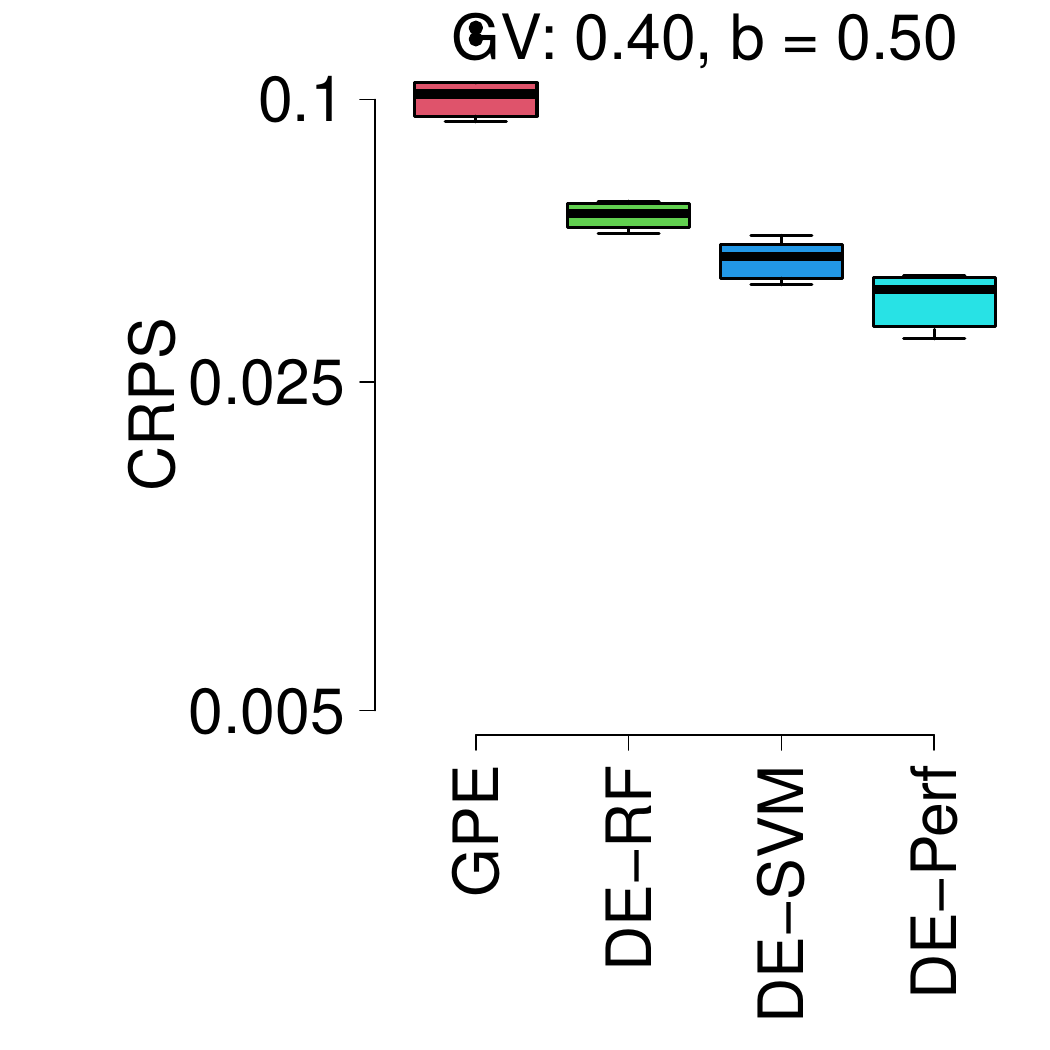}
    \end{subfigure}
    \begin{subfigure}{0.3\textwidth}
        \centering
        \includegraphics[width=\linewidth]{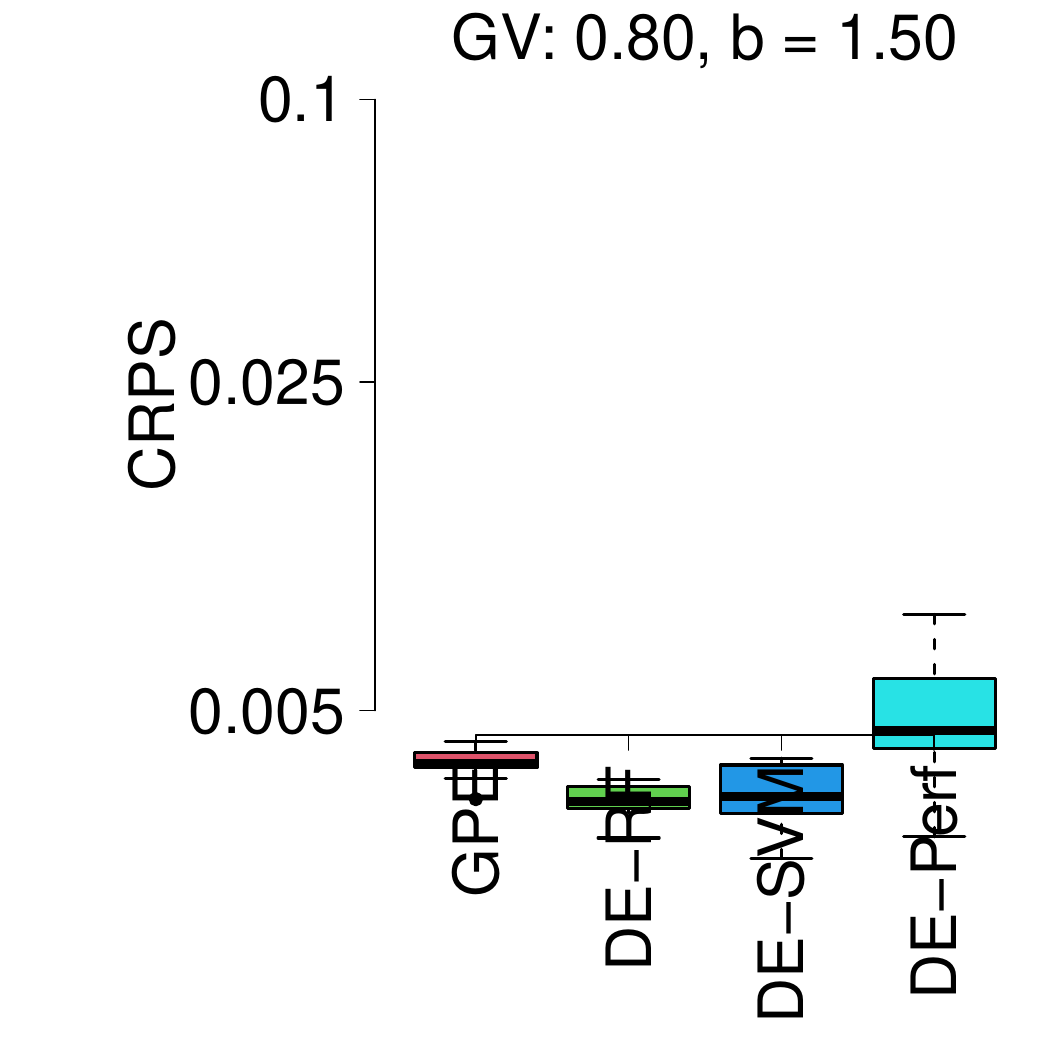}
    \end{subfigure}
    \begin{subfigure}{0.3\textwidth}
        \centering
        \includegraphics[width=\linewidth]{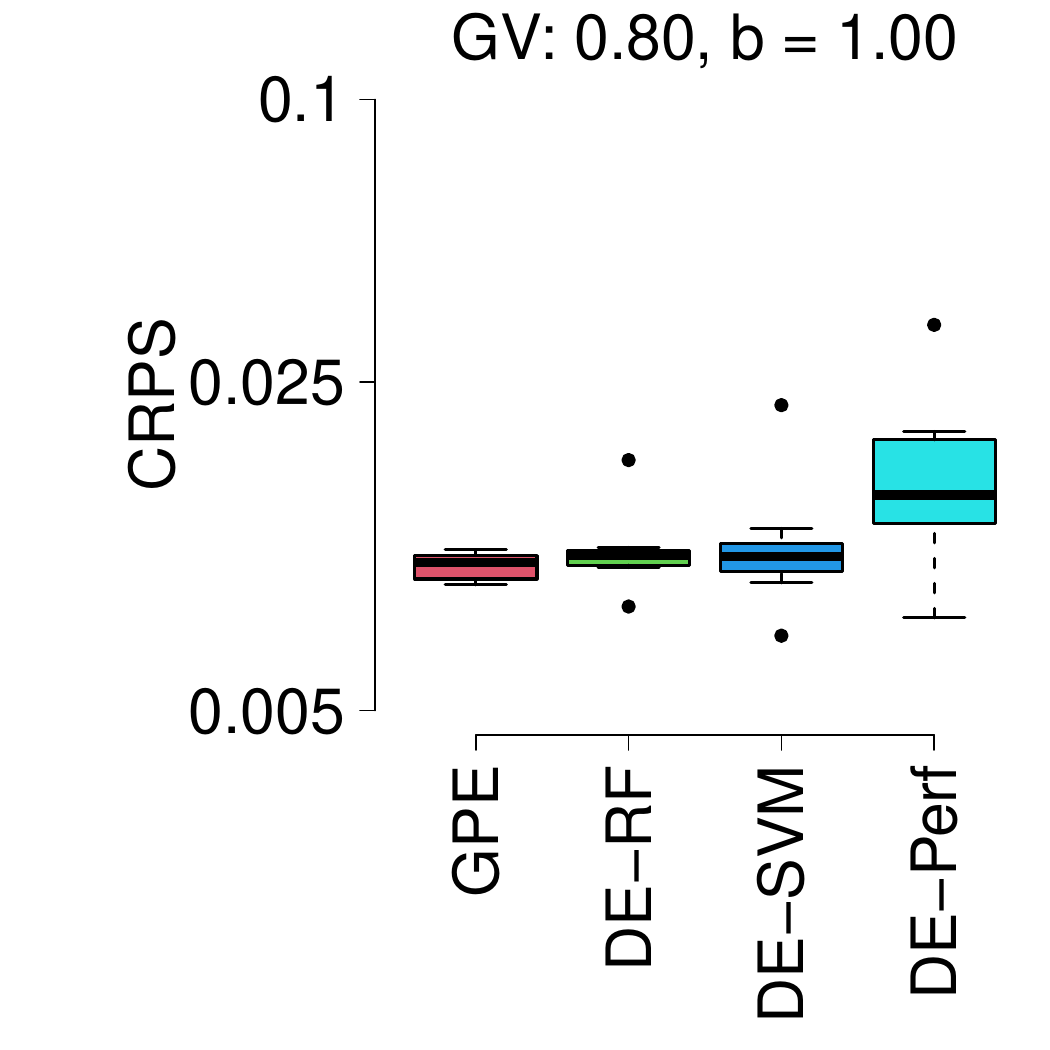}
    \end{subfigure}
    \begin{subfigure}{0.3\textwidth}
        \centering
        \includegraphics[width=\linewidth]{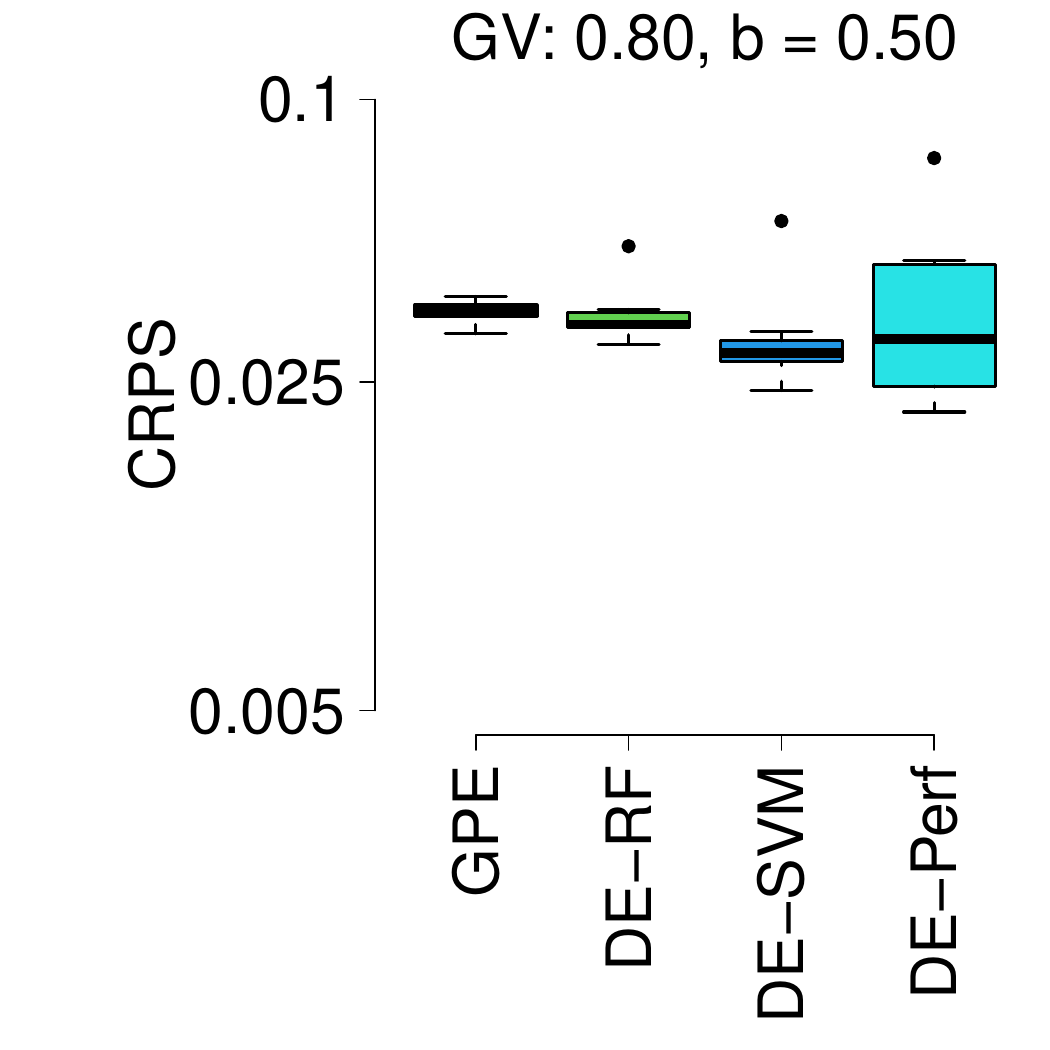}
    \end{subfigure}
    \caption{Comparing the performance of the GPE against three instances of the double emulator, fit using a RF (DE-RF), SVM (DE-SVM) and `perfect' classifier (DE-Perf), on the Banana simulator in \eqref{banana} using the CRPS score. The grounded volume (GV) is controlled via the offset, $a$, which is chosen so that the grounded volume increases from top to bottom from $20\%$ to $80\%$ of the input space. The hardness of the landing is increased from left to right via the exponent, $b$, in \eqref{bananasim}. The top left figure shows that little is gained by using a double emulator when the landing is very soft. Surprisingly, nothing is gained by having an omniscient classifier in that case. However, as the hardness of the landing increases the improvement gained by using a double emulator becomes apparent. When GV $=0.8$, the lack of training data for the GPE in the double emulator causes the model to struggle. Additional experiments analyzing the RMSE for the Banana, as well as the CRPS and RMSE for the simulator in \eqref{DPFunction} are available in supplement \ref{furtherresults}.}
    \label{fig:results}
\end{figure}

Using \eqref{bananasim}, we conducted a series of Monte Carlo experiments for various values of $a$ and $b$ in \eqref{bananasim} to investigate the effect of the volume of the grounded region and the behavior of the derivative around the grounding line on emulation performance, using both the curved simulator in \eqref{DPFunction} and the $8$-dimensional Banana in \eqref{banana} as test cases. For given values of $a$ and $b$ in \eqref{bananasim}, our empirical assessment of each statistical model was aggregated over ten Monte Carlo iterations, the experimental set-up for each iteration was as follows: (a) Training set of size 200 - the same for each model; (b) Testing set of size 1000 - the same for each model; (c) A Mat\'ern $5/2$ kernel along with constant and linear basis functions for the GPE component of each model; (d) Kernel hyperparameters estimated using the default settings in \texttt{RobustGaSP} \citep{Gu22}; (e) Classifiers fit using default settings. In each iteration, the training and testing datasets were each chosen as the best by maxi-min distance from 30 randomly generated LHDs \citep[e.g.][Chapter~17]{dean15}. The results of the series of experiments on the Banana are included in Figure \ref{fig:results}.

We choose specifically to discuss the results on the Banana, as they are more indicative of the strengths and weaknesses of the double emulator; the results for the DP function are included in Supplement \ref{furtherresults} and are more favorable for the double emulator, which bests the conventional GPE according to CRPS in every experiment, and according to RMSE in nearly all experiments. 

With the Banana, when the landing is soft (when $b \geq 1$ in Figure \ref{fig:results}) and the grounded volume is relatively small, the double emulator provides little, if any, advantage over the GPE. When the grounded volume is extremely large (GV $= 0.8$), the performance of the double emulator degrades further. Since the GPE component of the double emulator is only fit using the non-grounded data, the GPE within the double emulator is being deprived of training data. This is compounded by the extremely imbalanced data on which the classifier is being trained. It appears that accurately predicting the grounded region is not enough to compensate for the reduction in training data available to the GPE. Specifically, the perfect classifier does not consistently provide an advantage over the SVM or RF in these specific examples. It is worth noting that out-of-the-box application of the SVM to small datasets with highly imbalanced classes, for instance as a result of either a very small or very large grounded region, can throw error messages. Thus, our results correspond to cases where we can compare all four models out-of-the-box avoiding edge cases.

In contrast, the results generated by altering the hardness of the landing by varying the exponent, $b$, in \eqref{banana2}, reveal an interesting insight in favor of the double emulator. As the exponent decreases and the derivative correspondingly increases around the grounding line, the double emulator provides a performance advantage over the GPE in terms of both RMSE and CRPS. This is the case both with the curved function and the Banana. Moreover, for both test cases, when the landing is the hardest, corresponding to $b=0.5$, the gain in performance when using the double emulator is the greatest. In this case, the accuracy of the classifier appears to be more important, with the perfect classifier double emulator outperforming both the corresponding RF and SVM models.

In summary, these results suggest, empirically, that for soft landings, especially when the grounded volume is extremely large, the GPE could perform more favorably. However, when the landing is hard, the GPE will struggle and the double emulator provides a performance advantage over a range of grounded volumes, both with respect to the accuracy of the double emulator mean function as a point estimate of the simulator output, and the corresponding uncertainty quantification provided by the double emulator variance function. Finally, we reiterate that the results on the DP function considerably favored the double emulator. However, the training size ($200$) was the same for both simulators, despite the difference in the dimension of their input. Thus, we expect the shortcomings of the double emulator on simulators with higher dimensional input could be mitigated with a larger or more cleverly chosen design \cite[e.g.][]{Isberg_Welch_2022}.
\subsection{The Oxidation Simulator}
\label{Oxidation}
The corrosion simulator models the oxidation of uranium in water vapor environment, for which the chemical reaction is given by
\begin{equation}
\label{Corrosion}
    \begin{aligned}
        \textnormal{U(s)} + 2\textnormal{H}_2\textnormal{O(g)} \xrightarrow[]{} \textnormal{UO}_2\textnormal{(s)} + 2\textnormal{H}_2\textnormal{(g)}.
    \end{aligned}
\end{equation}
As opposed to proceeding directly as in \eqref{Corrosion}, a detailed reaction-diffusion scheme was proposed in \citet{Natchiar21} following experimental observations. A rigorous mathematical derivation and asymptotic analysis of the system of partial differential equations (PDEs), as well as a simulator solving the corresponding system was derived in \citet{Natchiar21} and \citet{Natchiar20}. For the purposes of emulation, we highlight that non-dimensionalization of the mathematical model occurs within the simulator, and the input is the eight dimensional parameter vector
\begin{equation}
\label{SimulatorInputs}
   x = (k_1, k_2, D_h^{H}, D_h^{M}, D_h^{O}, D_c^{H}, D_c^{M}, D_c^{O}),
\end{equation}
where $k_1$ and $k_2$ are elementary reaction rates, and the final six inputs correspond to the diffusivity of each diffusing species in each static component. In particular, the subscripts $c$ and $h$ refer to $\textnormal{OH}^-$ and $\textnormal{H}^{\boldsymbol{\cdot}}$, respectively, whilst the superscripts $H,M$ and $O$ denote the hydride, metal and oxide components. 

Each run of the simulator solves the system of coupled PDEs and a high resolution spatio-temporal grid of (non-dimensional) concentration values is returned for each of the five species; $\text{OH}^{-}, \text{H}^{\cdot}, \text{UO}_2, \text{UH}_3$ and $\text{U}$. The numerical solution scheme employs an adaptive time stepper, therefore varying the inputs causes the temporal indices of the simulator outputs to change; the simulator output is not regular. To circumvent this issue, we interpolate the simulator output onto a fixed set of knots in the temporal (t) domain using splines. By applying the quantity of interest functional $\mathcal{Q}_{(z,t,i)}$, extracting the concentration of species $i$ at the space-time point $(z,t)$, we can assess our model on data from the oxidation simulator. Hence, we are interested in emulating the map $\mathcal{Q}_{(z,t, i)}[S] : \mathcal{X} \xrightarrow[]{} \mathbb{R}_{\geq 0}$
for \textit{fixed} $(z,t,i)$, where $\mathcal{X}$ is the eight dimensional input parameter space and $S$ is the raw simulator. Strictly speaking, accounting for all sources of uncertainty necessitates the acknowledgment of uncertainty arising due to interpolation. However, we choose to ignore this for the purposes of this work as we are interested only in analyzing emulator performance. Further, emulating each output coordinate independently is not uncommon, though emulation of functional output has been explored extensively \citep{Higdon08, Rougier08}.

\begin{figure}[ht!]
    \centering
    \begin{subfigure}[t]{0.42\textwidth}
        \centering
        \includegraphics[width=\linewidth, trim = {0cm 0.5cm, 0cm, 0.5cm}, clip]{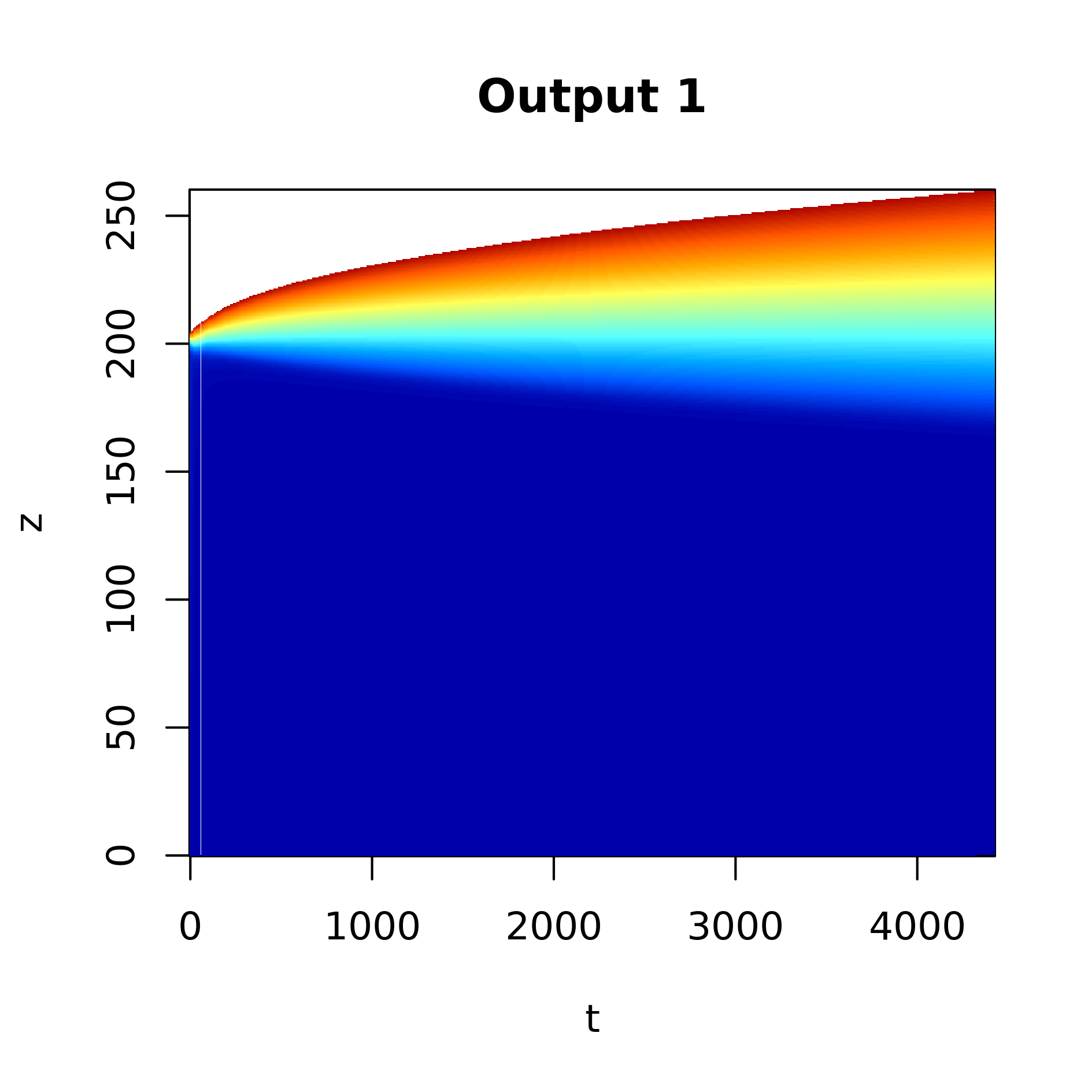}
    \end{subfigure}
    \begin{subfigure}[t]{0.42\textwidth}
        \centering
        \includegraphics[width=\linewidth, trim = {0cm 0.5cm, 0cm, 0.5cm}, clip]{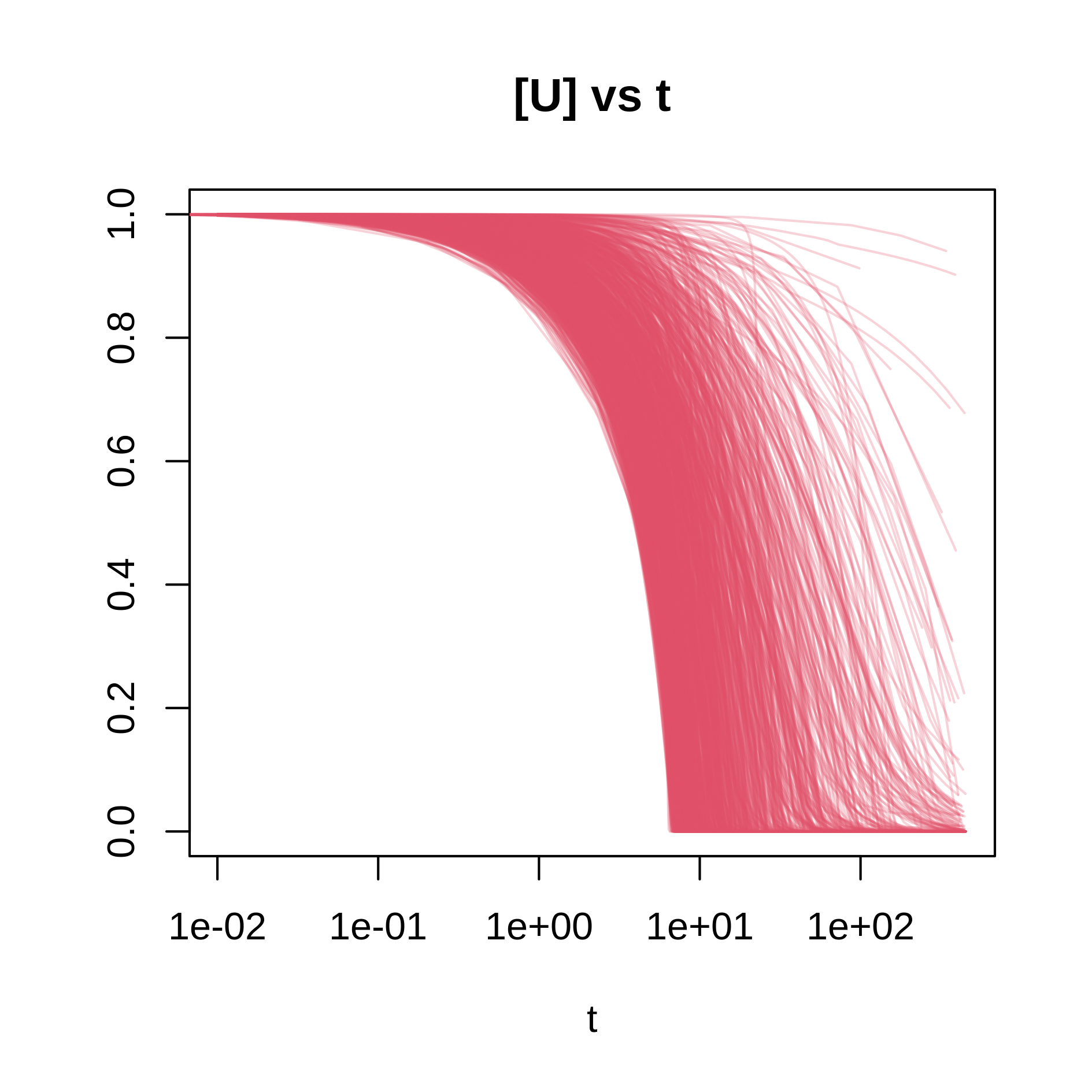}
    \end{subfigure}
    \caption{\textit{Left}: One of the five space time grids corresponding to one example simulator run. \textit{Right}: Traces of metal concentration over time for 1000 different runs of the simulator. Some traces are shorter than others due to the time limit of the submission scripts.}
    \label{fig:tracesanddomain}
\end{figure}

The data available from the oxidation simulator consisted of a 1000 run randomly generated LHD over parameter ranges suggested in \cite{Natchiar21}. We chose to emulate the concentration of metal, U, close to the hydride-metal interface where there would likely exist a non-trivial grounded region. Figure \ref{fig:tracesanddomain} contains example output over the spatio-temporal domain for one input combination and also illustrates the variation, as a function of time, in the concentration of metal at a fixed spatial location; each trace corresponds to one of the 1000 runs. The scalar valued quantity is then extracted from the 1000 runs by selecting a specific value of $t$, requiring interpolation as explained above; we chose $t = 25$.

Unlike the synthetic experiments, for the oxidation simulator we do not artificially vary the volume of the grounded region or derivative near the grounding line. The grounded region, using a grounding value equal to the square root of machine precision, occupied approximately $65\%$ of the input space. Instead, our experiment here simply compares the four statistical models according to RMSE and CRPS for training datasets of size $N \in \{300,400,500\}$. Similar to the synthetic experiments, our empirical assessment of each statistical model was aggregated over ten Monte-Carlo iterations. For a given $N$, the experimental set-up for each Monte Carlo iteration was as follows: (a) Sample a testing set of 500 runs from the 1000 total runs at random; (b) From the remaining 500 runs, randomly sample a training set of size $N$; (c) A Mat\'ern $3/2$ kernel along with constant and linear basis functions for the GPE component of each model; (d) Kernel hyperparameters estimated using the default settings in \texttt{RobustGaSP}; (e) Classifiers fit using default settings.

\begin{figure}[ht!]
    \centering
    \begin{subfigure}[t]{0.45\textwidth}
        \centering
        \includegraphics[width=\linewidth]{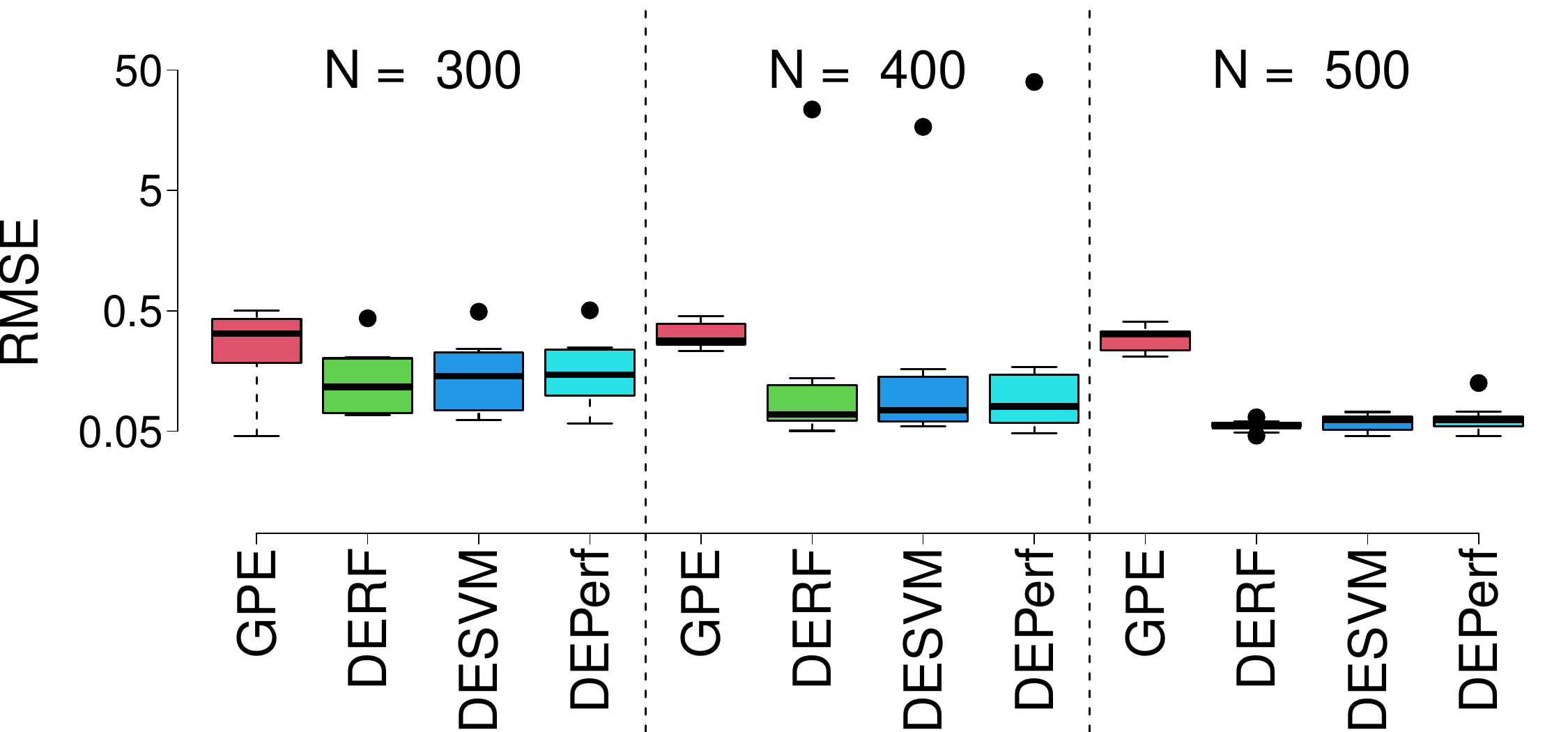}
    \end{subfigure}
    \begin{subfigure}[t]{0.45\textwidth}
        \centering
        \includegraphics[width=\linewidth]{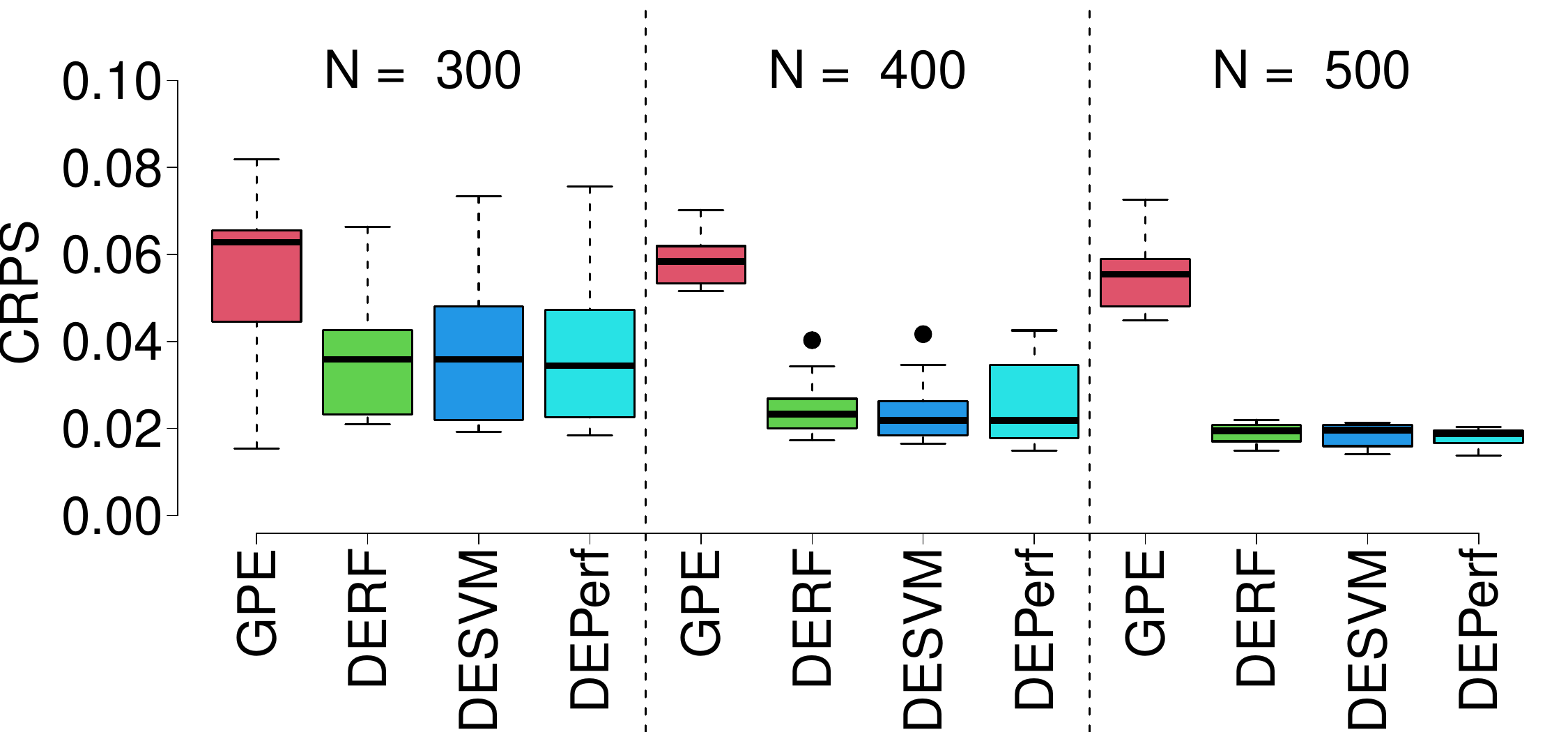}
    \end{subfigure}
    \caption{Analyzing the effect of varying training dataset size on the performance of the GPE and double emulators on the oxidation simulator in Section \ref{Oxidation}. Note, RMSE is on a log-scale to aid the visualization of the outliers.}
    \label{fig:staticoxres}
\end{figure}

The results are given in Figure \ref{fig:staticoxres}, showing that, according to CRPS, the double emulators outperform the GPE for each dataset. In terms of RMSE, we suspect, again, since the grounded volume is large, the double emulator will be struggling for the same reasons as described in the synthetic examples, in some cases resulting in large outliers. In fact, experiments for small $N$ demonstrated this aspect of the double emulator. 

\section{Discussion}
\label{Discussion}
Typical application of the GPE for emulation of simulators involves using a stationary, separable GP residual \citep{Santner03, Gramacy2020}. We challenged this model formulation in the context of simulators that attain their minimum value on a considerable volume of their input domain, potentially experiencing discontinuities in their derivative whilst doing so. We introduced an emulator which combines the GPE with a probabilistic classifier in a novel manner, where the role of the classifier was to estimate whether or not the simulator had `grounded'. Using pathological synthetic examples, as well as a real simulator modeling the corrosion of uranium, we demonstrated that our double emulator regularly outperformed the basic GPE on examples where the grounded region was of moderate size. Furthermore, the gain in performance increased as the change in the derivative of the simulator across the grounding line increased.

Despite our emulator regularly outperforming the GPE on examples in this article, we highlight several areas for future improvement. One of the main limitations of the double emulator is its application to simulators with grounded regions occupying an extremely large volume of the input space. Theoretically, with a perfect classifier, we would expect, as the grounded volume tends to the volume of the input domain, the double emulator should be unbeatable. However, in practice, this was not the case, partially due to instability of the GPE within the double emulator. In particular, the GPE within the double emulator is trained only on the non-grounded data points and GPEs are known to struggle when extrapolating outside of the convex hull of the training data. When the grounded volume was very large, the accurate prediction of the grounded region was not enough to counteract the instability of the GPE. We therefore advise either careful implementation of the GPE component of the double emulator, or simply applying the basic GPE in these limiting cases. For a similar reason imbalanced classes can cause instability of the out-of-the-box performance of the classifier; tuning the classifiers is one potential remedy. Experimentation with alternative classifiers such as artificial neural networks as well as alternative implementations of the GPE pave the way for further enhancement of the method.

Our final comment relates to design. We are currently investigating a novel sequential design scheme for improving the double emulator by targeting two sources of uncertainty; the classifier uncertainty, measured according to (binary) entropy, and the double emulator variance. Preliminary results are promising. An interesting extension involves finding an optimal strategy for splitting computational resources between each source of uncertainty. Identification of metrics other than entropy is also of interest.
\clearpage

\if0\blind
{
\subsection*{Acknowledgments}
We would like to thank Rich Hewitt (University of Manchester) and Phil Monks (AWE) for providing the code for the oxidation simulator, as well as their helpful discussions on running the simulator and their advice on appropriate ranges for the input parameters. CC would like to thank the EPSRC Centre for Doctoral Training in Computational Statistics and Data Science, grant number EP/S023569/1, and AWE for financial support in the form of a PhD studentship.} \fi

\begin{center}
{\large\bf SUPPLEMENTARY MATERIAL}
\end{center}
The supplementary material includes additional figures and results for the synthetic experiments, as well as the relevant \textsc{R} code and datasets.

\bibliographystyle{jasa}
\bibliography{bibliography}

\newpage 

\appendix
\begin{center}
{\large\bf SUPPLEMENTARY MATERIAL}
\end{center}
\appendix
\section{Emulating the Log}
\label{lognormaldetails}
\renewcommand{\thesubsection}{\Alph{section}.\arabic{subsection}}
A random variable $Z$ is log-normally distributed if $\ln(Z) \sim \mathcal{N}(m, v)$; we use the notation $Z \sim \text{Lognormal}(m,v)$ to express this. The distribution function, mean and variance of $Z$ are defined by
\begin{equation}
    \begin{aligned}
        L(z;m,v) &\coloneqq \frac{1}{2}\left[1 + \text{erf}\left(\frac{\ln{z} - m}{\sqrt{2v}}\right) \right], \\
        M &= \exp{\left(m + v/2\right)}, \\
        V &= \exp{\left(2m + v\right)}\,[\exp\left({v}\right)-1].
    \end{aligned}
\end{equation}
Note that $m$ and $v$ are referred to as the `meanlog' and `varlog' parameters of the Lognormal distribution (mean and variance of the corresponding Normal distribution for $\ln{Z}$), whilst $M$ and $V$ are the mean and variance of the Lognormal distribution.

\section{Performance Metrics}
\label{app2}
\subsection{RMSE}
\label{rmsesection}
For a test set $\{X,Y\}$ consisting of $n$ simulator runs, the RMSE is defined as 
\begin{equation}
    \begin{aligned}
        \textnormal{RMSE} &\coloneqq \sqrt{ \frac{1}{n} \sum_{i=1}^{n} \left(Y^{(i)}-\hat{Y}^{(i)}\right)^2},
    \end{aligned}
\end{equation}
where $\hat{Y}^{(i)}$ is a point estimate of the emulator at the $i^{th}$ run.

\subsection{CRPS}
\label{crpssection}
\begin{figure}[ht!]
    \centering
    \begin{subfigure}[t]{0.42\textwidth}
        \centering
        \includegraphics[width=\linewidth]{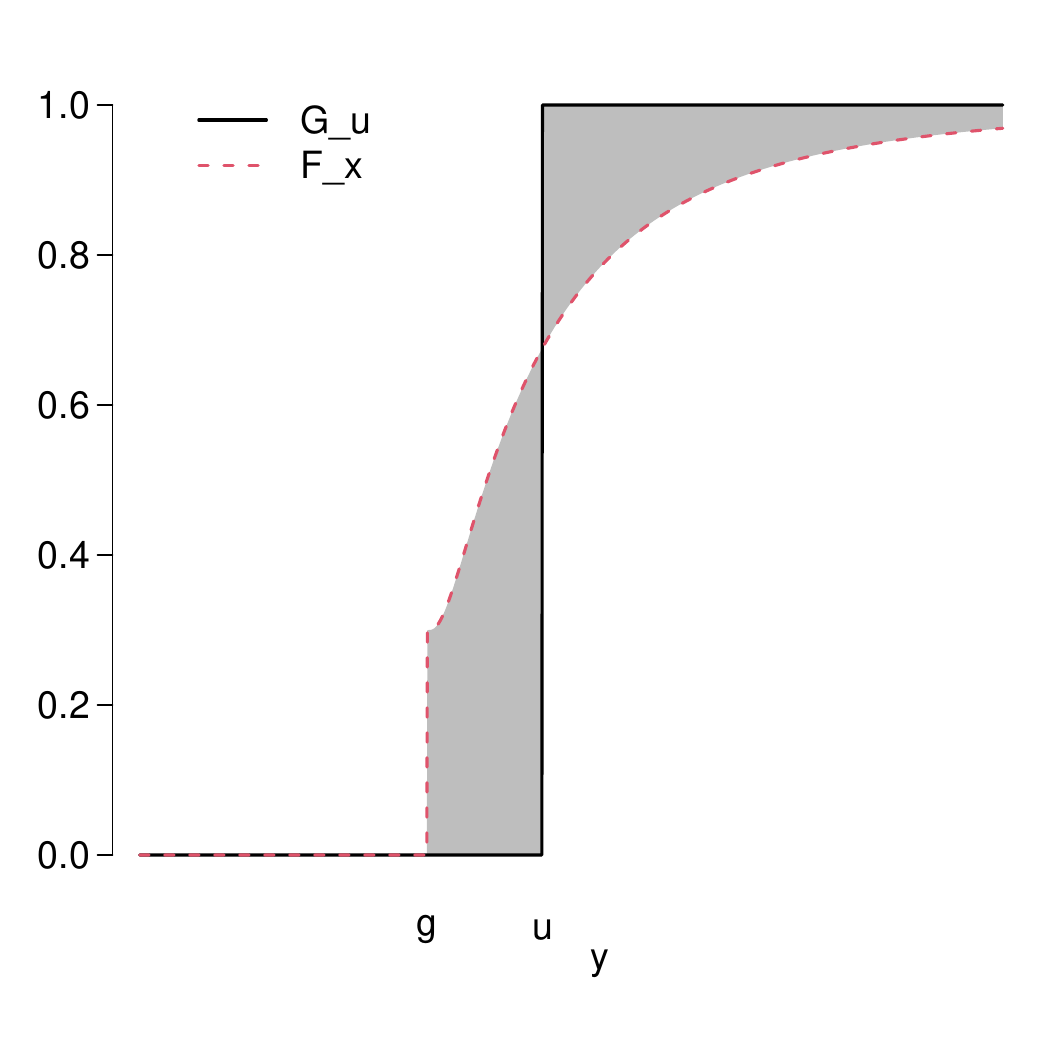}
    \end{subfigure}
    \caption{A schematic diagram of the distribution function (dashed line) of the double emulator, and the distribution function of a point mass at the observed value $y=u$ (solid line). Here, the observed value $u>g$, and $p=0.7$. The gray shaded area represents the CRPS score, which can be either be estimated using numerical integration or, in our case, exactly.}
    \label{fig:crpsschematic}
\end{figure}

The proof of Lemma \ref{crpslemma} is as follows:
\begin{proof}
Writing $G_u(y) = \mathbb{I}( y \geq u)$ for the distribution function of a point mass at $u \geq g$ we can write the (negatively oriented) CRPS score $\text{CRPS}( F, u)$ as 
\begin{align}
\MoveEqLeft[2] \int_{-\infty}^\infty (F(y) - G_u(y))^2 dy \nonumber \\
&= \int_{g}^u (F(y) - G_u(y))^2 dy + \int_{u}^\infty (F(y) - G_u(y))^2 dy \nonumber \\
&= \int_g^u \left[ 1+ 2 p(L(y-g+\gamma) - 1) + p^2 (L(y-g+\gamma)-1)^2 \right] dy \nonumber \\
&\quad + \int_{u}^\infty p^2 (L(y-g+\gamma)-1)^2 dy \nonumber \\
&= (u-g) + 2 p \int_g^{u} (L(y-g+\gamma) - 1) dy + p^2 \int_g^\infty (L(y-g+\gamma)-1)^2 dy \nonumber \\
&= (u-g) + 2 p \int_\gamma^{u-g+\gamma} (L(z) - 1) dz + p^2 \int_\gamma^\infty (L(z)-1)^2 dz \label{eq:CRPStot}
\end{align}
(writing $1-p + p L(z) = 1 +p(L(z)-1)$ and
using the fact that $F(y) = G_u(y) = 0$ for $y \leq g$). We can use integration by parts to write
\begin{align}
\MoveEqLeft[2] \int_\gamma^{u-g+\gamma} (L(z) - 1) dz \nonumber \\
&= \left[ (L(z) - 1) z \right]_\gamma^{u-g+\gamma} - \int_\gamma^{u-g+\gamma} \ell(z) z dz \nonumber \\
&= (L(u-g+\gamma) - 1) (u-g+\gamma) - (L(\gamma) - 1) \gamma  \nonumber \\
&\quad - \int_\gamma^{u-g+\gamma} \frac{1}{\sqrt{2 \pi v}} \exp \left( - \frac{ (\ln z - m)^2}{2 v} \right) dz \nonumber \\
&= (L(u-g+\gamma) - 1) (u-g+\gamma) - (L(\gamma) - 1) \gamma \nonumber \\ 
&\quad - \exp(m + v/2) \left\{ \Phi \left( \frac{\log(u-g + \gamma) - m -v }{\sqrt{v}} \right) - \Phi \left( \frac{\log(\gamma) - m -v }{\sqrt{v}} \right) \right\} \label{eq:CRPSterm1}
\end{align}
where we write $\ell(z) = \frac{d}{dz} L(z)$ and use the exact expression for $\ell$. Using the formula for the log-normal CRPS score in
\cite[P.2293]{baran2015log} we can express:
\begin{align}
\label{eq:CRPSterm2}
  \int_\gamma^\infty (L(z)-1)^2 dz &= \text{CRPS}(L,\gamma) \nonumber \\ 
  &= \gamma \left( 2\Phi\left(\frac{\log \gamma - m}{\sqrt{v}}\right) - 1 \right) \nonumber \\
  &\quad - 2 \exp( m + v/2) \left\{ \Phi\left(\frac{\log \gamma - m}{\sqrt{v}} - \sqrt{v}\right) + \Phi\left(\sqrt{v/2}\right) - 1\right\}.
\end{align}
 The result follows on substituting \eqref{eq:CRPSterm1} and \eqref{eq:CRPSterm2} in
\eqref{eq:CRPStot}.
\end{proof}
A schematic of the CRPS is included in Figure \ref{fig:crpsschematic}.

\section{1D Gamma Example}
\begin{figure}[ht!]
    \centering
    \begin{subfigure}{0.42\textwidth}
        \centering
        \includegraphics[width=\linewidth, trim={0 0 0 1.5cm},clip]{Figures/1DSimulator/Soft/GPESoftLanding.pdf}
    \end{subfigure}
    \begin{subfigure}{0.42\textwidth}
        \centering
        \includegraphics[width=\linewidth, trim={0 0 0 1.5cm},clip]{Figures/1DSimulator/Soft/GPESoftLandingMetrics.pdf}
    \end{subfigure}
    \begin{subfigure}{0.42\textwidth}
        \centering
        \includegraphics[width=\linewidth, trim={0 0 0 1.5cm},clip]{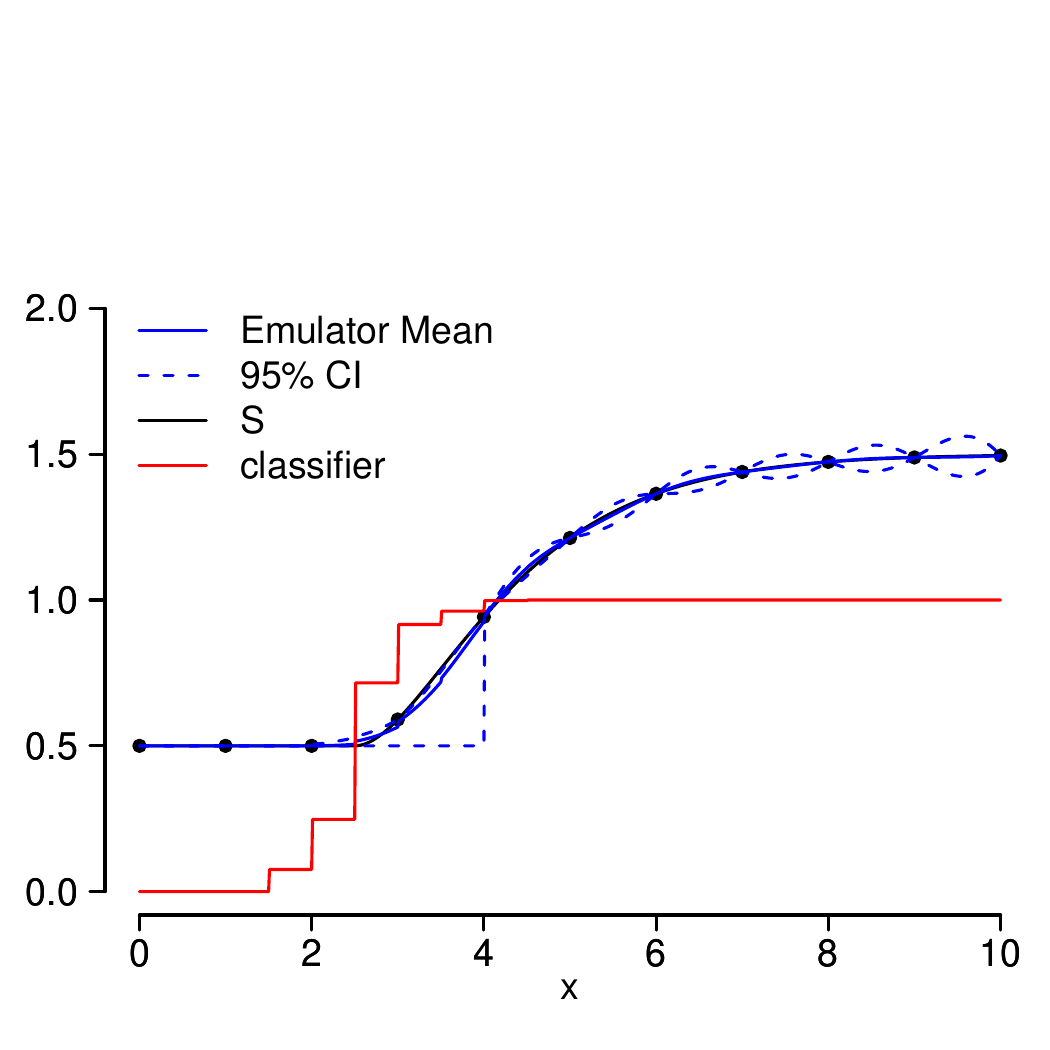}
    \end{subfigure}
    \begin{subfigure}{0.42\textwidth}
        \centering
        \includegraphics[width=\linewidth, trim={0 0 0 1.5cm},clip]{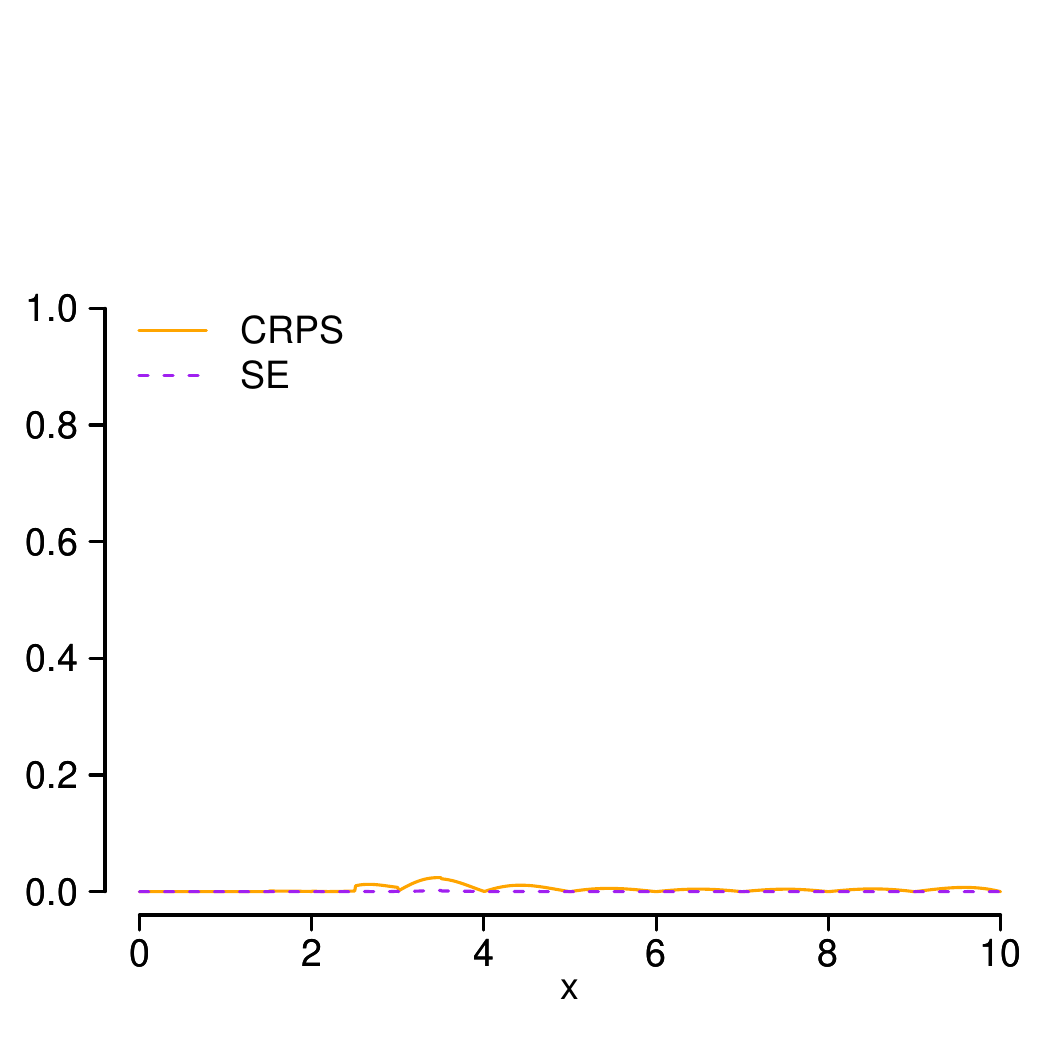}
    \end{subfigure}
    \begin{subfigure}{0.42\textwidth}
        \centering
        \includegraphics[width=\linewidth, trim={0 0 0 1.5cm},clip]{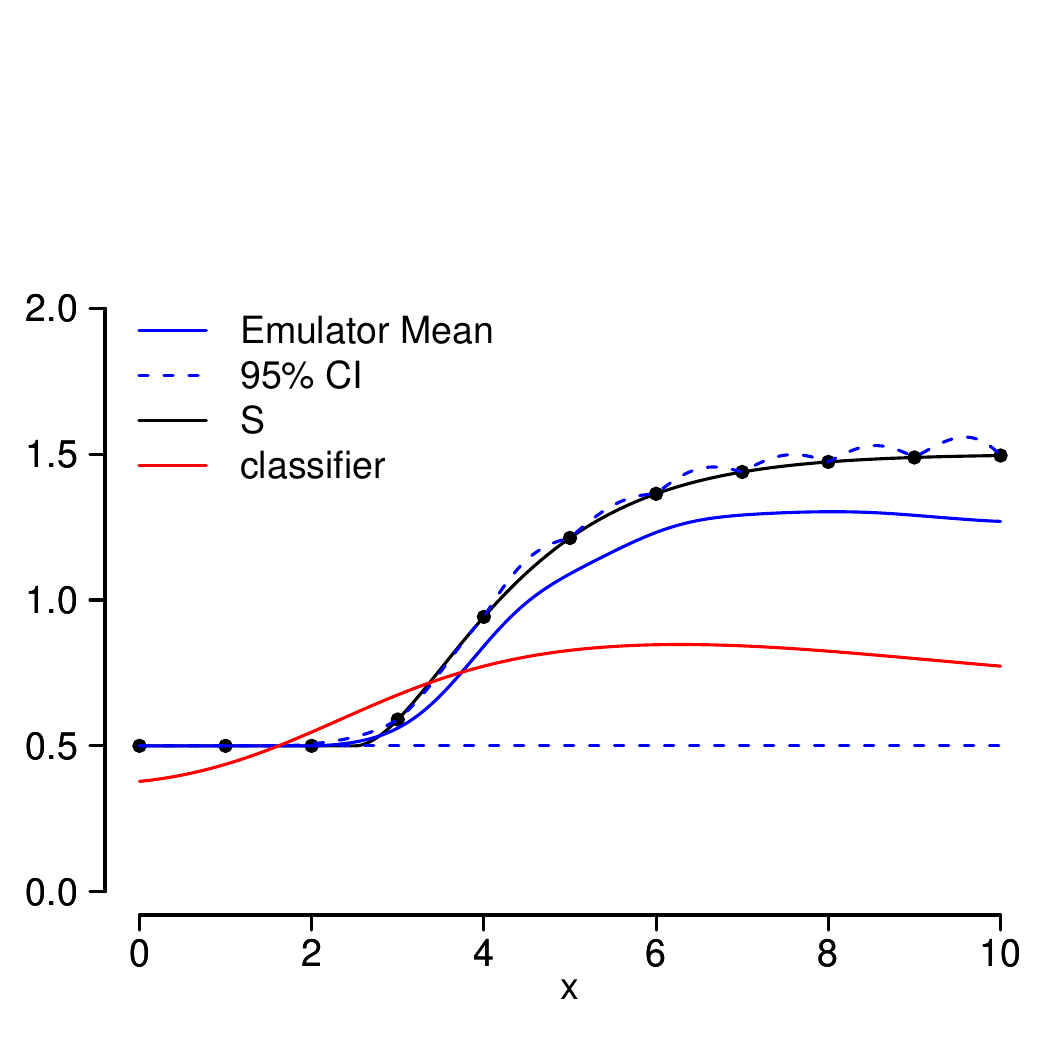}
    \end{subfigure}
    \begin{subfigure}{0.42\textwidth}
        \centering
        \includegraphics[width=\linewidth, trim={0 0 0 1.5cm},clip]{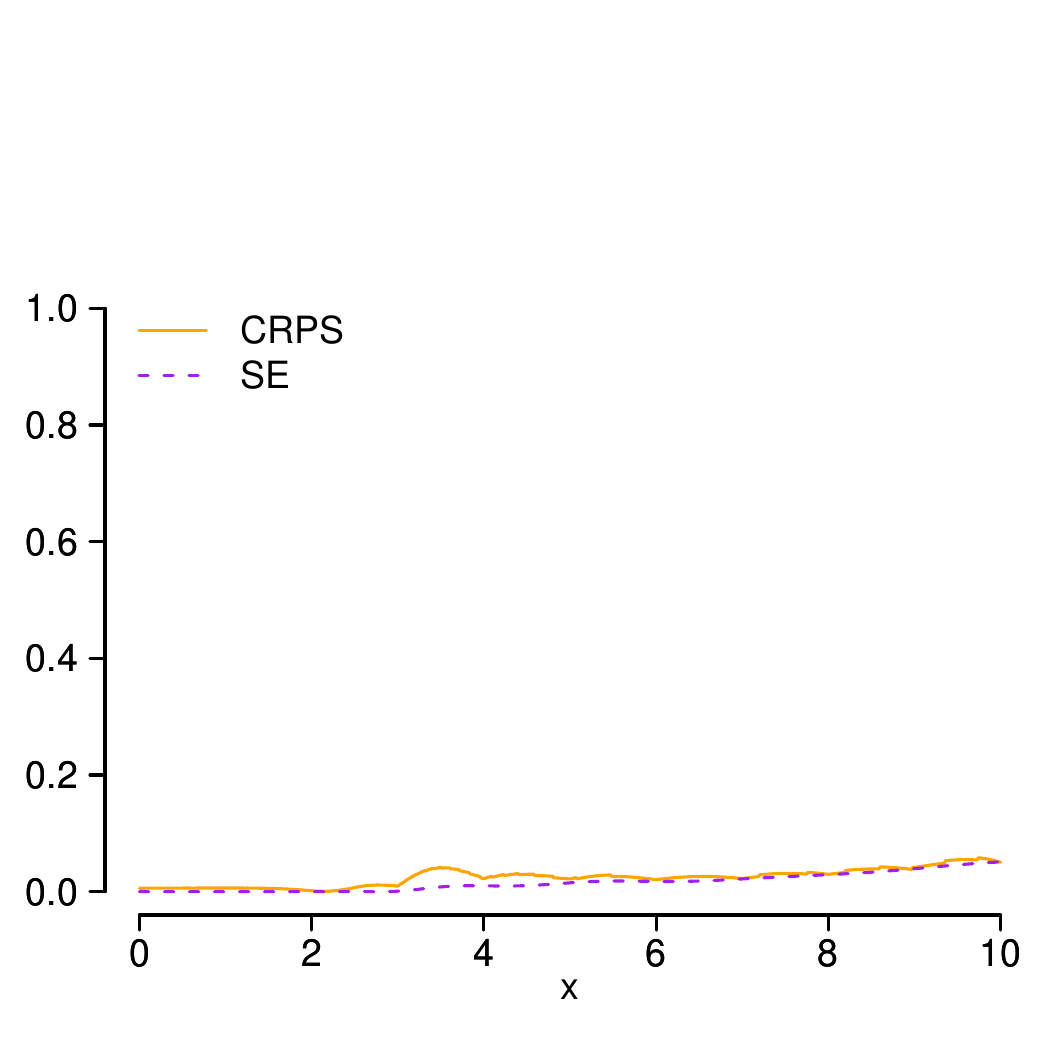}
    \end{subfigure}
    \caption{Comparing the GPE (top row) on the simulator in \eqref{1Dsimulator} with $\alpha = 2$ and $s = 2.5$ against two instances of the double emulator; the first fit uses a random forest (middle row) and the second a support vector machine (final row). Each classifier is denoted by the red line. The left column contains the fits with the right column containing the performance metrics; squared error (SE) and CRPS score as functions of $x$.}
    \label{fig:1DSoft}
\end{figure}

\begin{figure}[ht!]
    \centering
    \begin{subfigure}{0.42\textwidth}
        \centering
        \includegraphics[width=\linewidth, trim={0 0 0 0},clip]{Figures/1DSimulator/Hard/GPEHardLanding.pdf}
    \end{subfigure}
    \begin{subfigure}{0.42\textwidth}
        \centering
        \includegraphics[width=\linewidth, trim={0 0 0 0},clip]{Figures/1DSimulator/Hard/GPEHardLandingMetrics.pdf}
    \end{subfigure}
    \begin{subfigure}{0.42\textwidth}
        \centering
        \includegraphics[width=\linewidth, trim={0 0 0 1.5cm},clip]{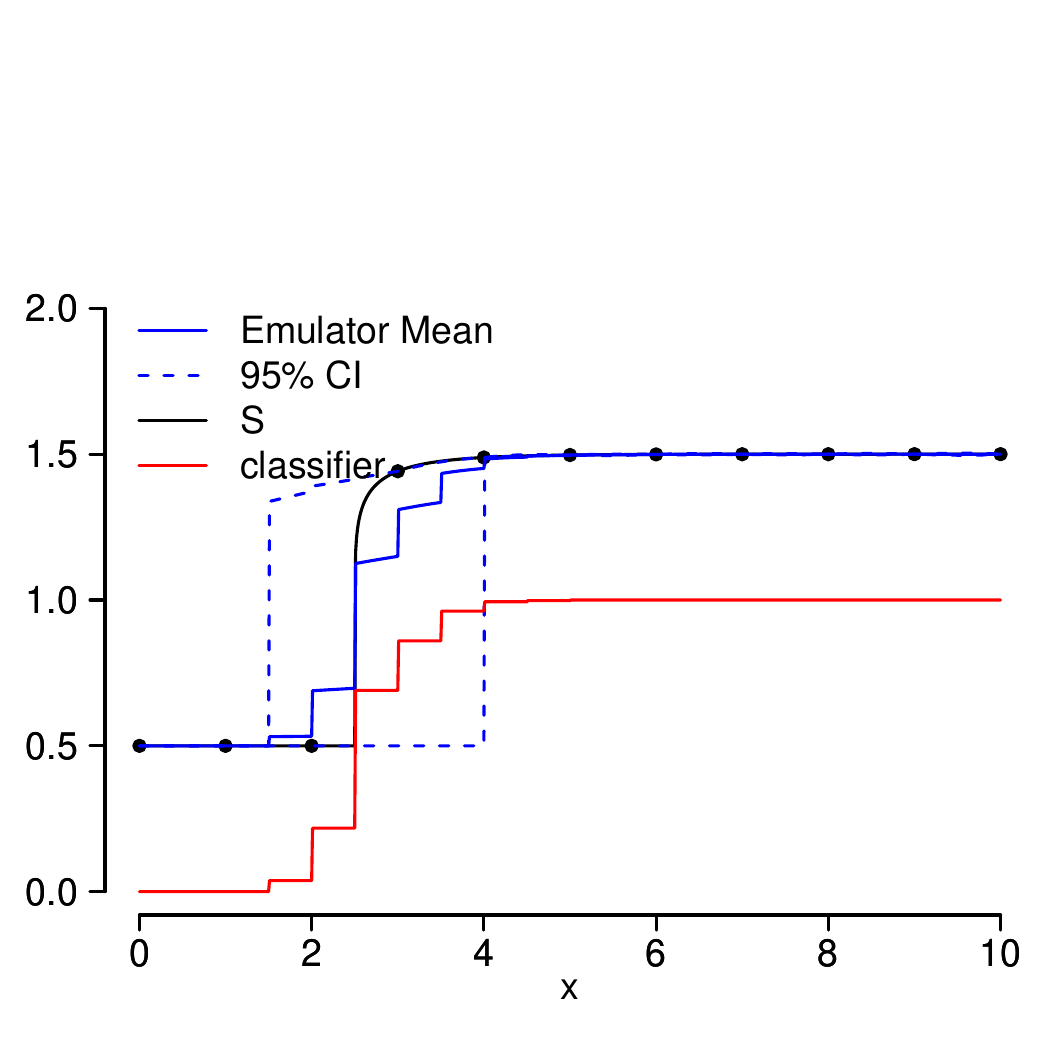}
    \end{subfigure}
    \begin{subfigure}{0.42\textwidth}
        \centering
        \includegraphics[width=\linewidth, trim={0 0 0 1.5cm},clip]{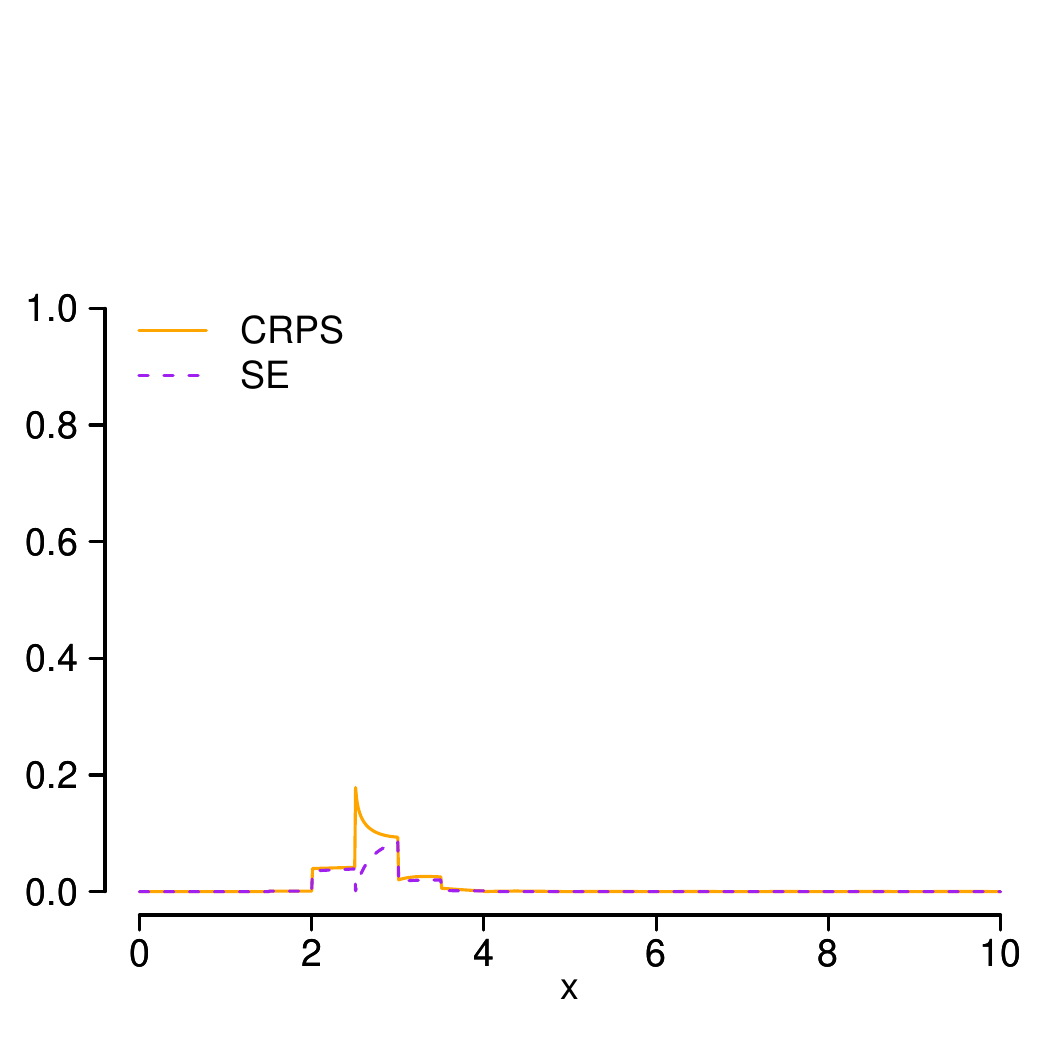}
    \end{subfigure}
    \begin{subfigure}{0.42\textwidth}
        \centering
        \includegraphics[width=\linewidth, trim={0 0 0 1.5cm},clip]{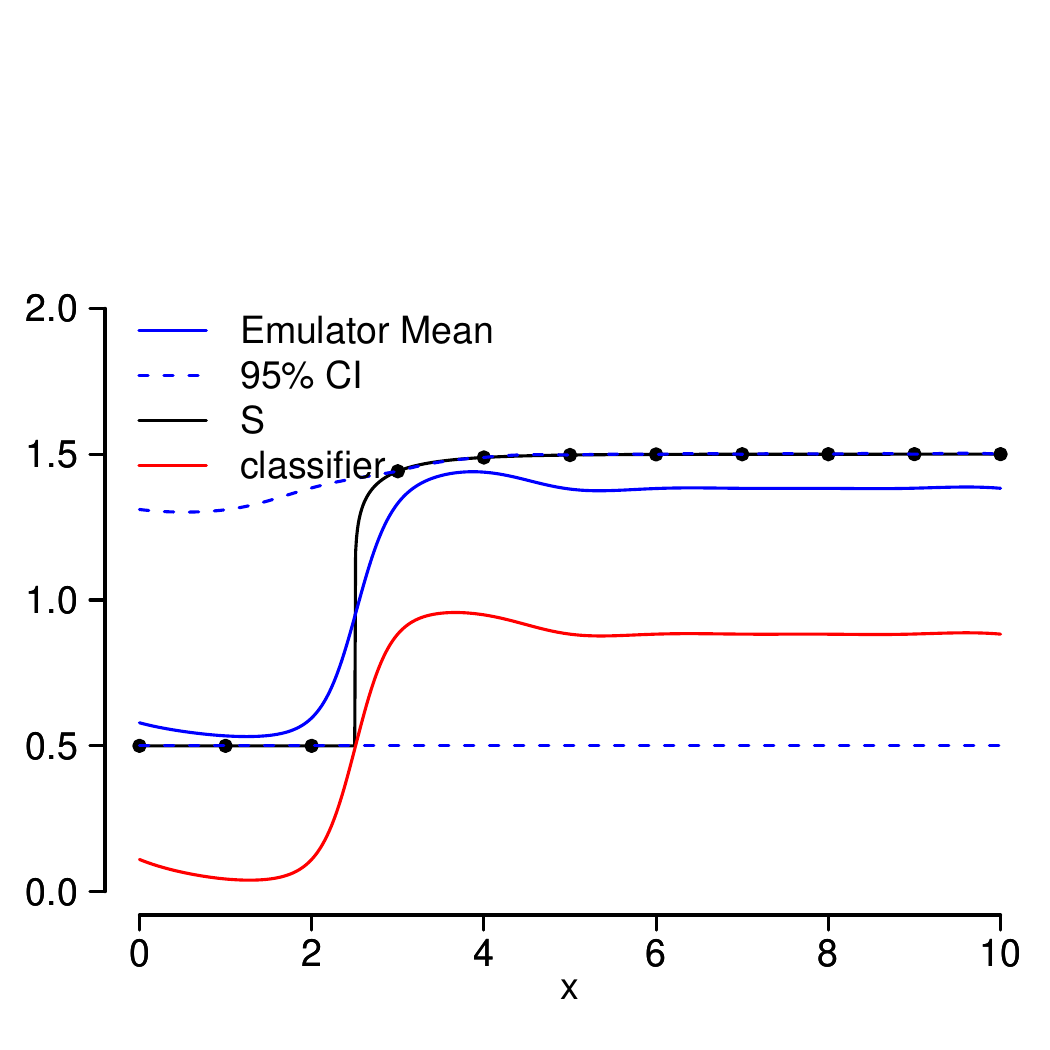}
    \end{subfigure}
    \begin{subfigure}{0.42\textwidth}
        \centering
        \includegraphics[width=\linewidth, trim={0 0 0 1.5cm},clip]{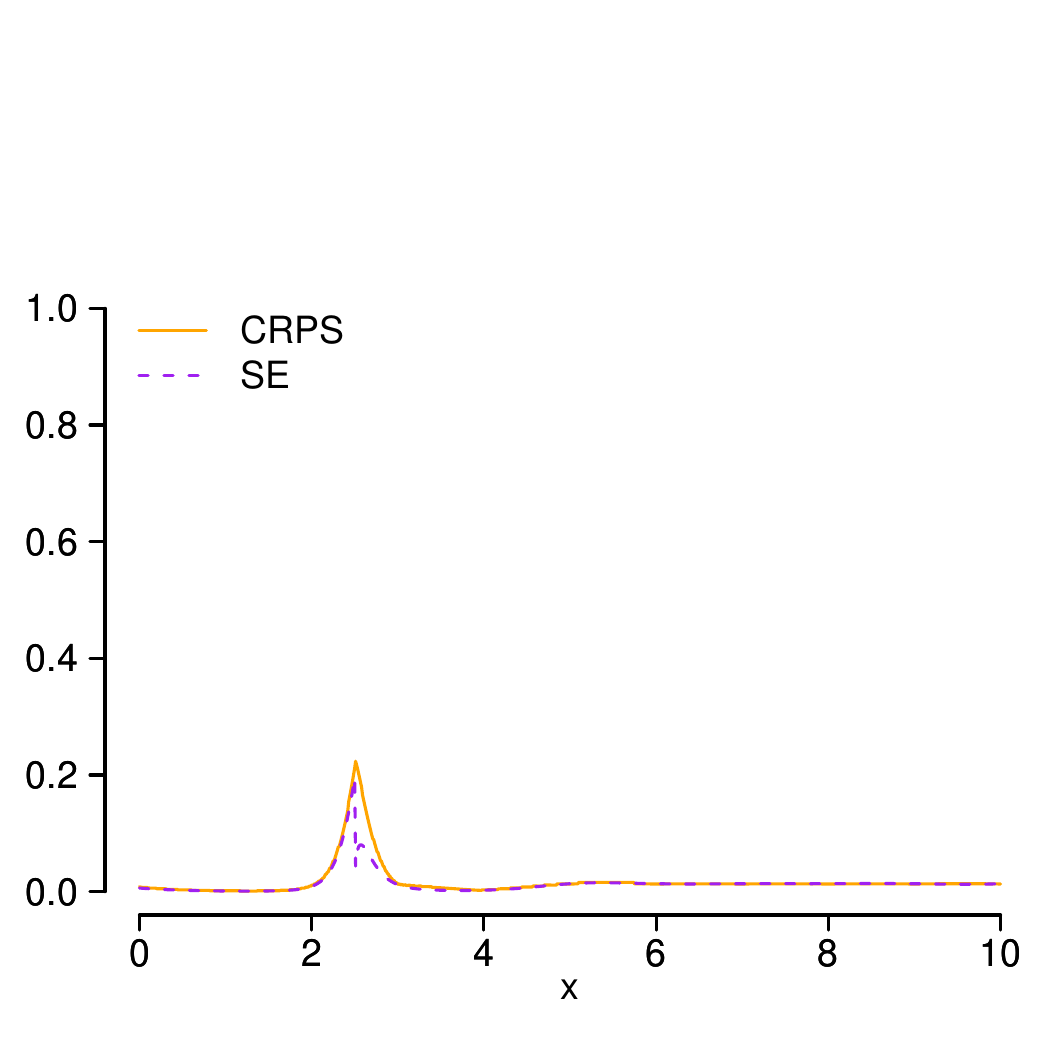}
    \end{subfigure}
    \caption{Comparing the GPE (top row) on the simulator in \eqref{1Dsimulator} with $\alpha = 0.1$ and $s = 2.5$ against two instances of the double emulator; the first fit uses a random forest (middle row) and the second a support vector machine (final row). Each classifier is denoted by the red line. The left column contains the fits with the right column containing the performance metrics; squared error (SE) and CRPS score as functions of $x$.}
    \label{fig:1DHard}
\end{figure}

Figures \ref{fig:1DSoft} and \ref{fig:1DHard} compare the fit of various double emulators with the GPE on the 1D Gamma example in \eqref{1Dsimulator} for the case of a soft landing ($\alpha = 2$ and $s = 2.5$) and hard landing ($\alpha = 0.1$ and $s = 2.5$), respectively. This is useful for visualizing the role of the classifier, and the associated CRPS and RMSE values.

\section{Higher Dimensional Synthetic Experiments}
\label{furtherresults}
Following on from Figure \ref{fig:results}, we present the corresponding RMSE scores for the series of experiments on the Banana simulator in \eqref{banana}, as well as the RMSE and CRPS scores for the simulator in \eqref{DPFunction}.

\begin{figure}[ht!]
    \centering
    \begin{subfigure}{0.3\textwidth}
        \centering
        \includegraphics[width=\linewidth]{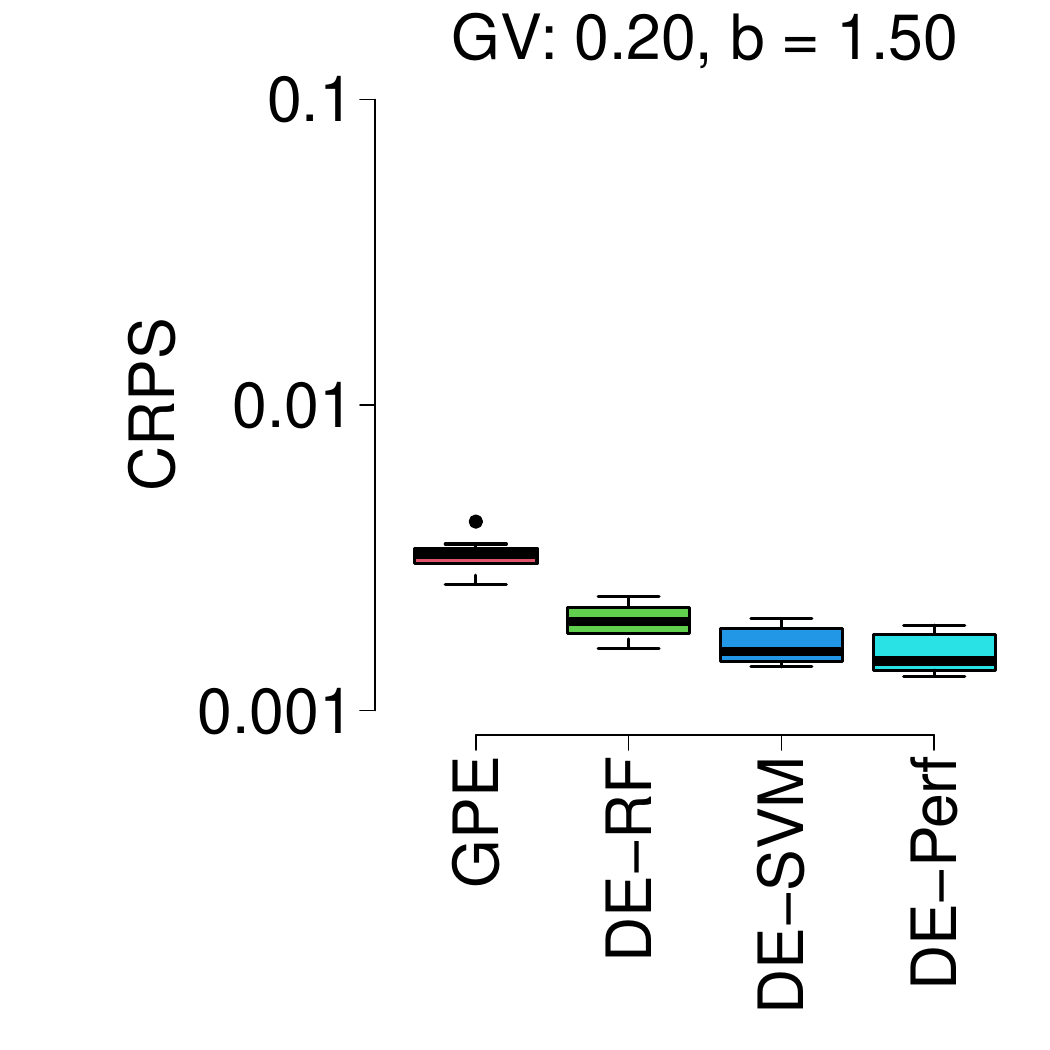}
    \end{subfigure}
    \begin{subfigure}{0.3\textwidth}
        \centering
        \includegraphics[width=\linewidth]{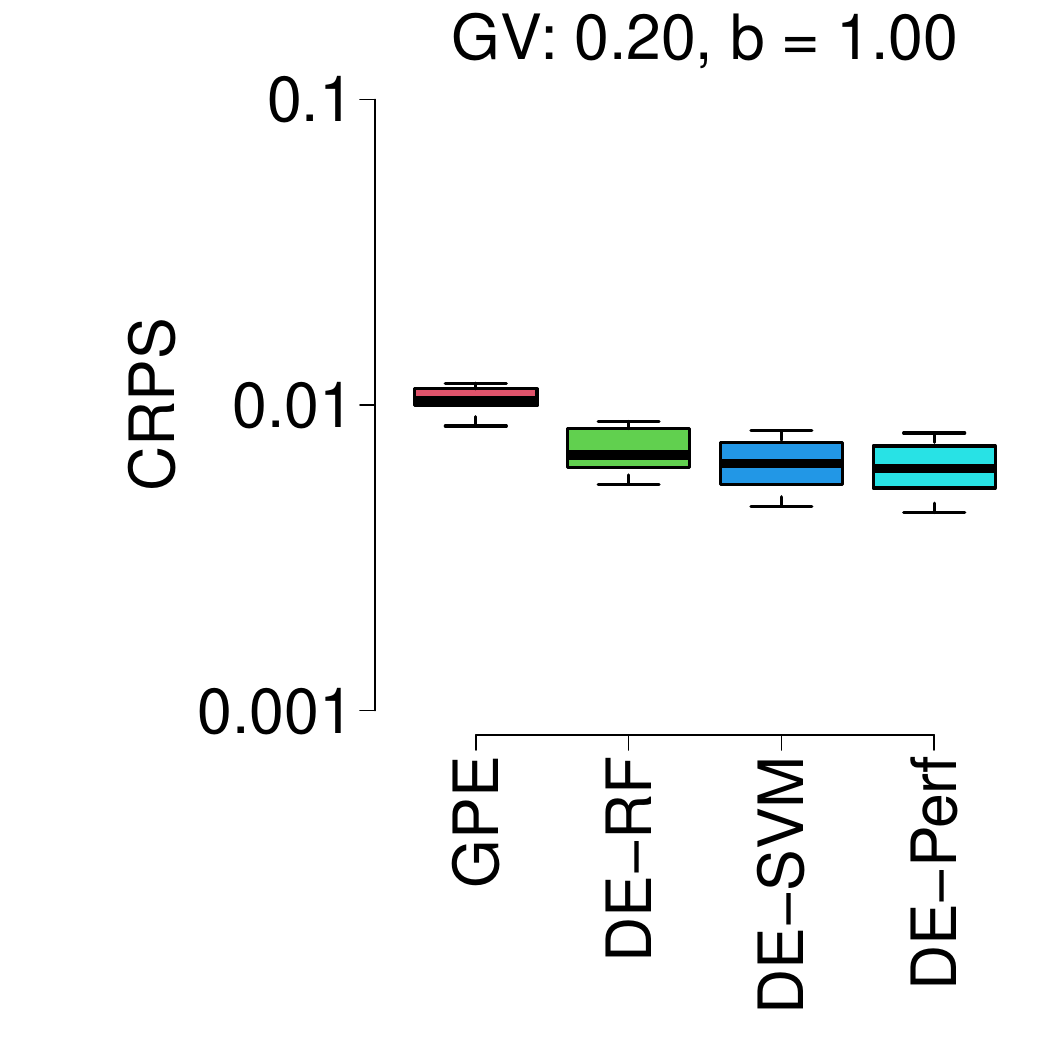}
    \end{subfigure}
    \begin{subfigure}{0.3\textwidth}
        \centering
        \includegraphics[width=\linewidth]{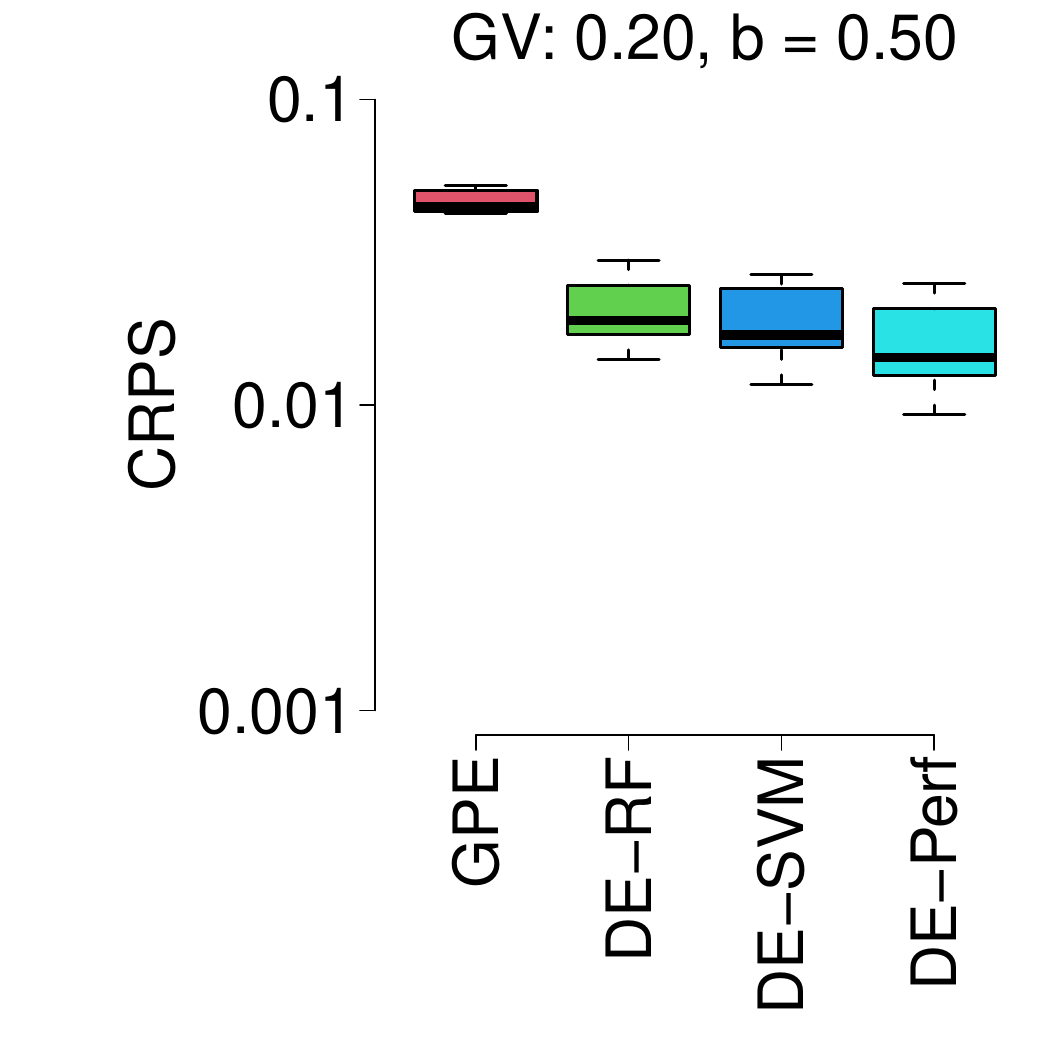}
    \end{subfigure}
    \begin{subfigure}{0.3\textwidth}
        \centering
        \includegraphics[width=\linewidth]{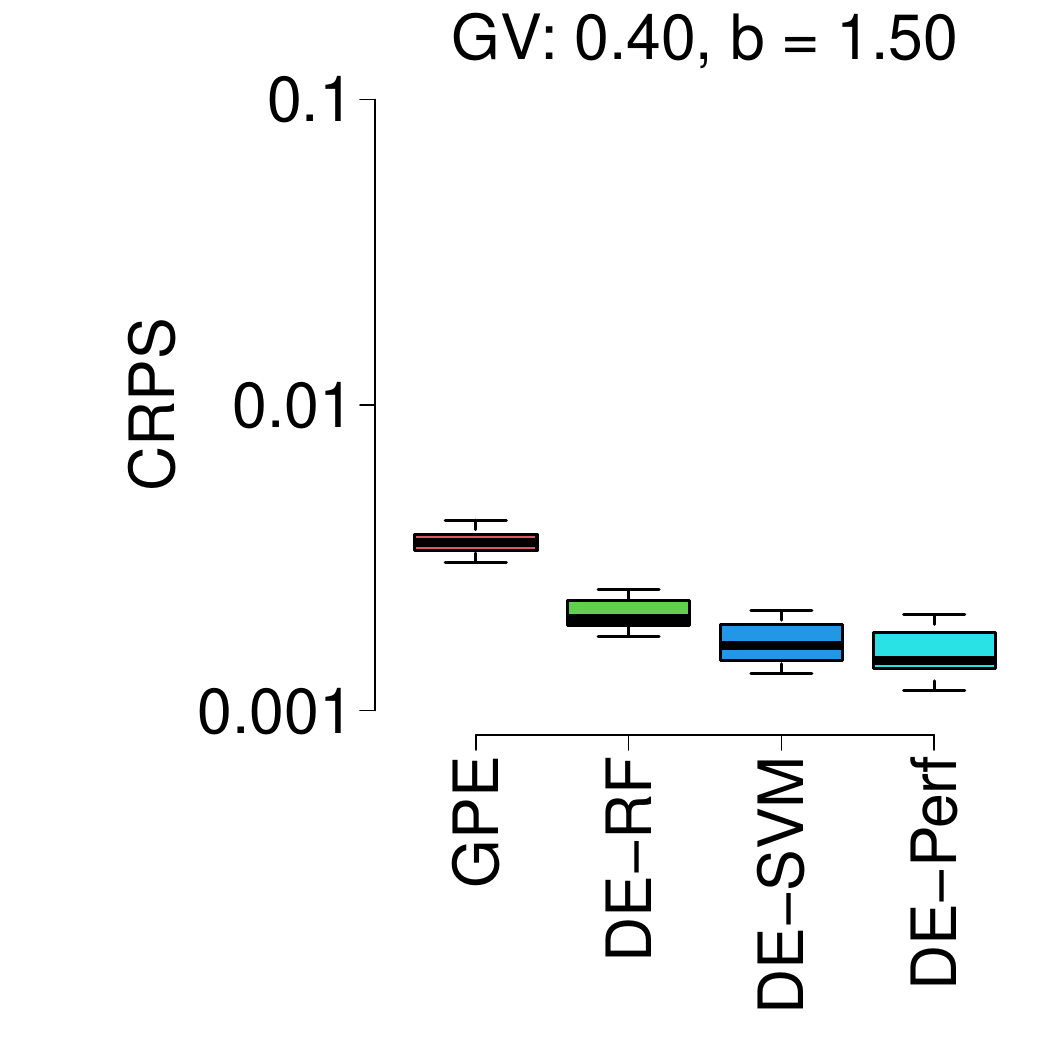}
    \end{subfigure}
    \begin{subfigure}{0.3\textwidth}
        \centering
        \includegraphics[width=\linewidth]{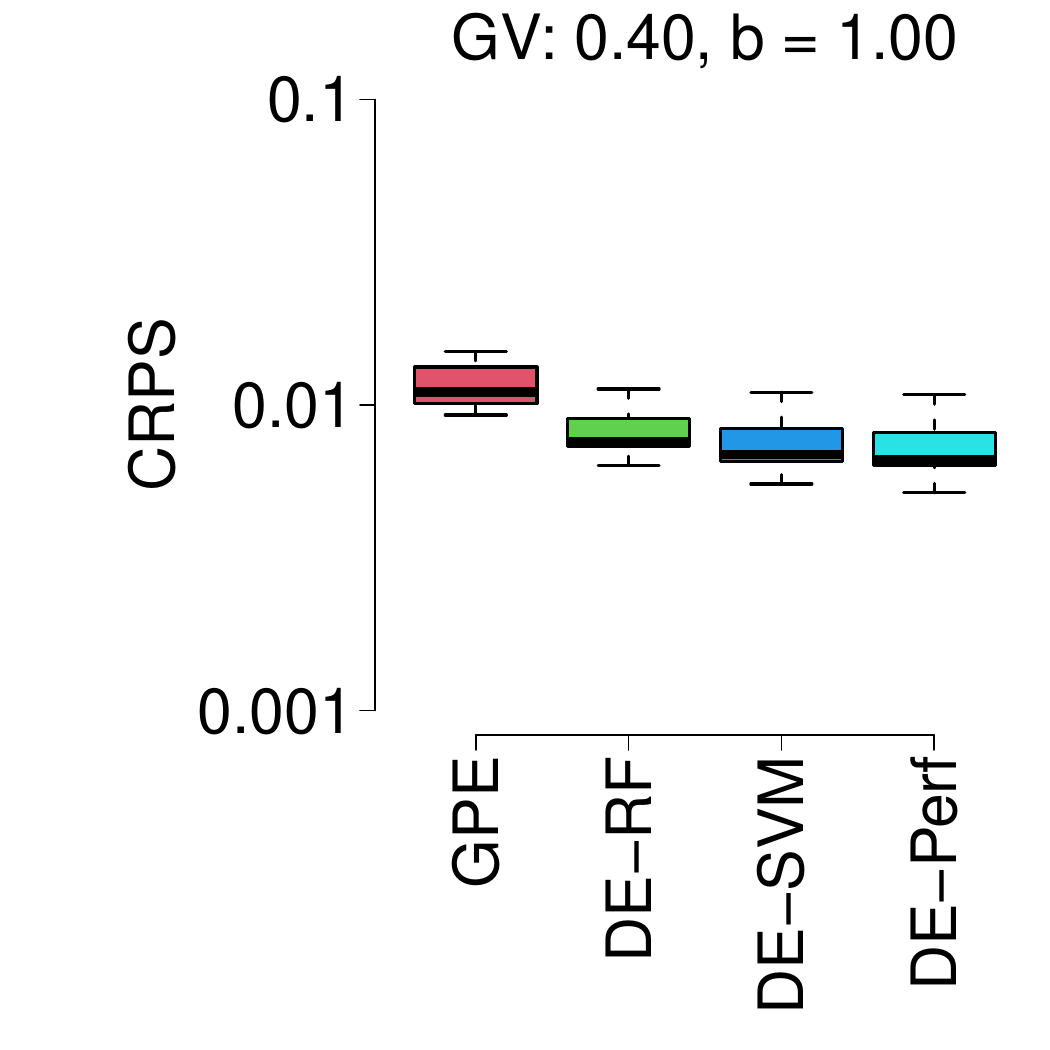}
    \end{subfigure}
    \begin{subfigure}{0.3\textwidth}
        \centering
        \includegraphics[width=\linewidth]{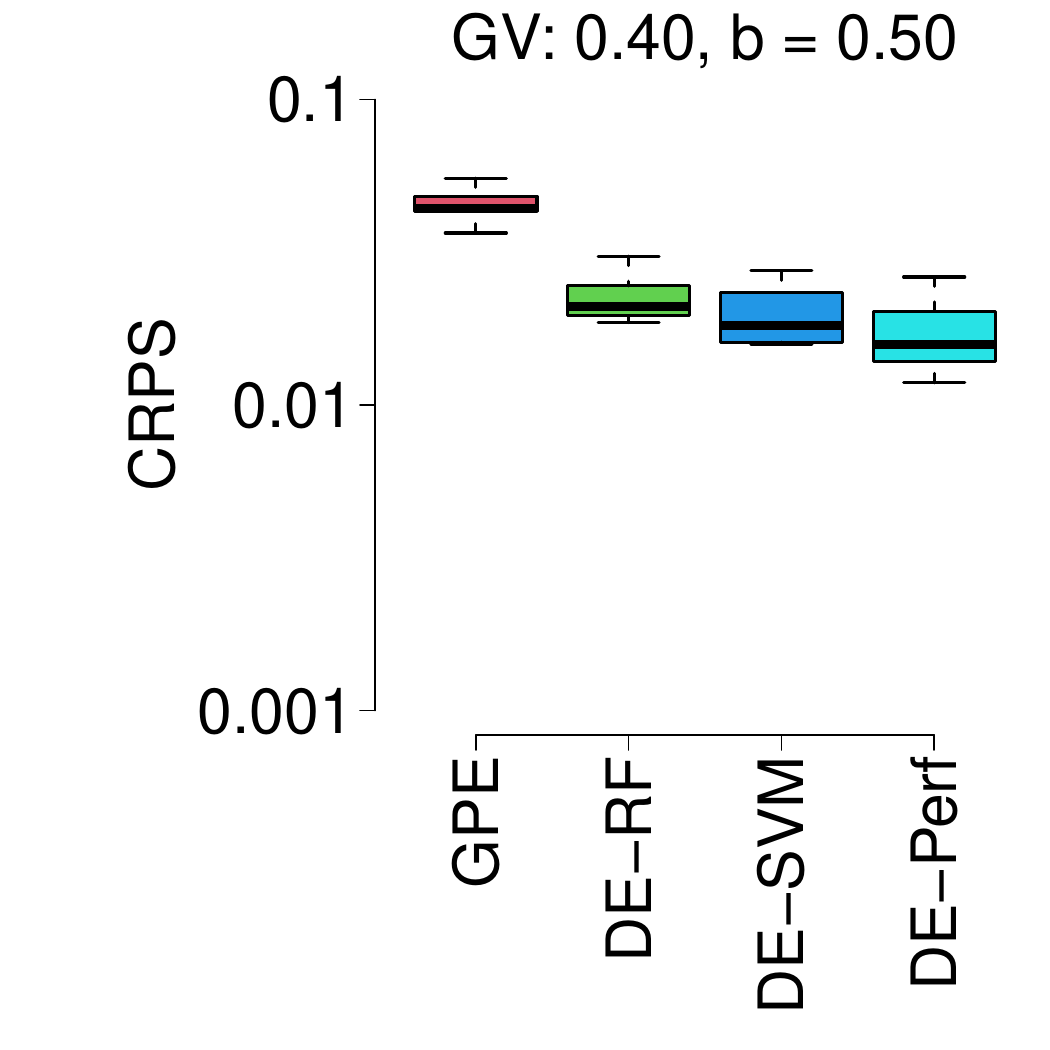}
    \end{subfigure}
        \begin{subfigure}{0.3\textwidth}
        \centering
        \includegraphics[width=\linewidth]{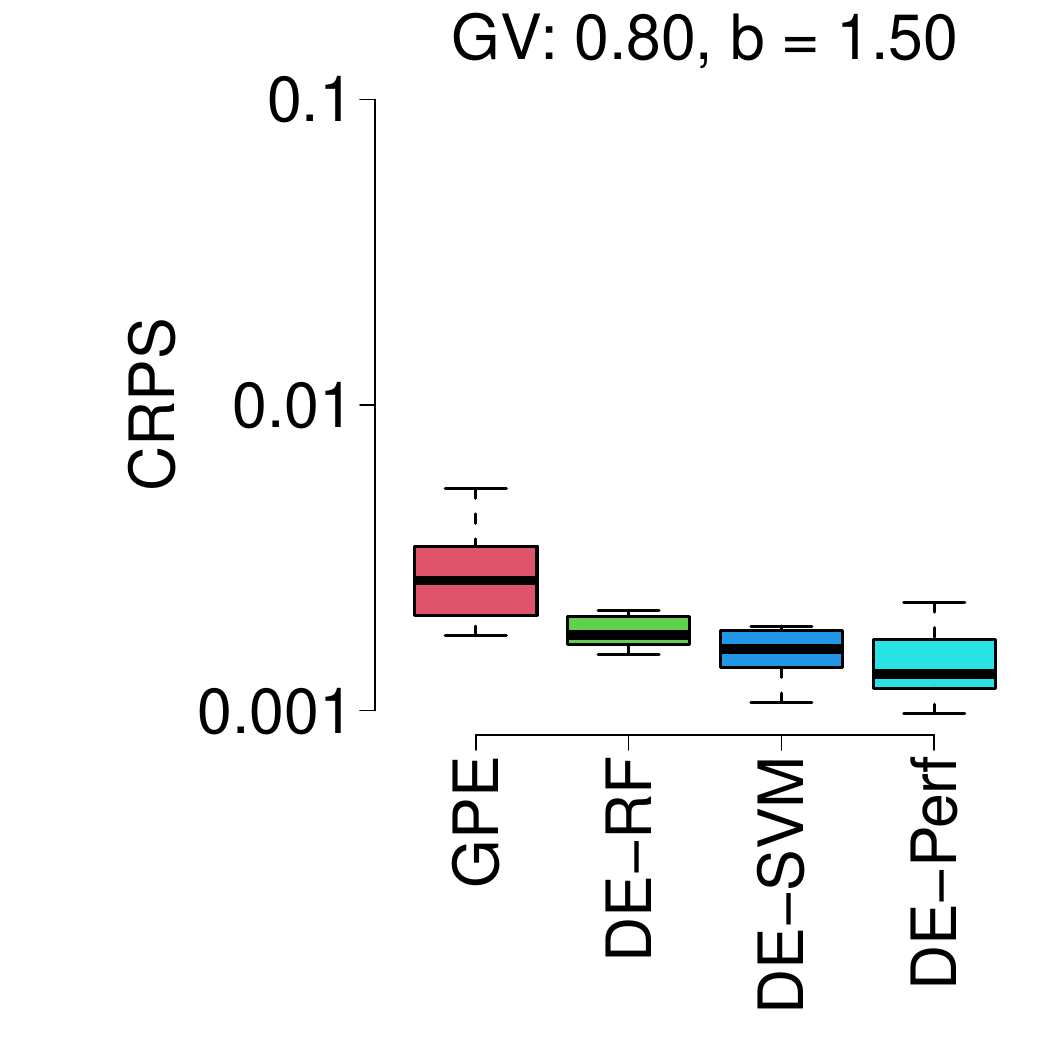}
    \end{subfigure}
    \begin{subfigure}{0.3\textwidth}
        \centering
        \includegraphics[width=\linewidth]{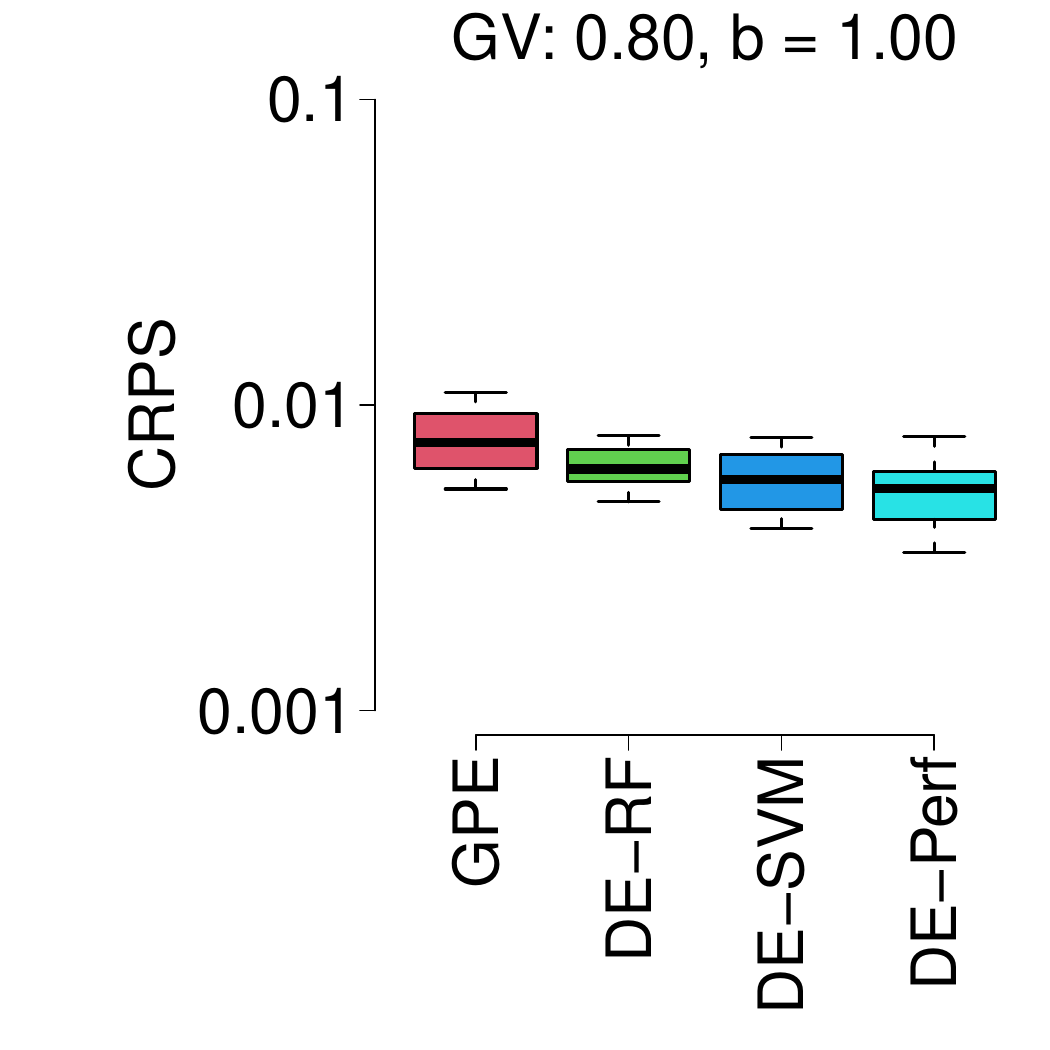}
    \end{subfigure}
    \begin{subfigure}{0.3\textwidth}
        \centering
        \includegraphics[width=\linewidth]{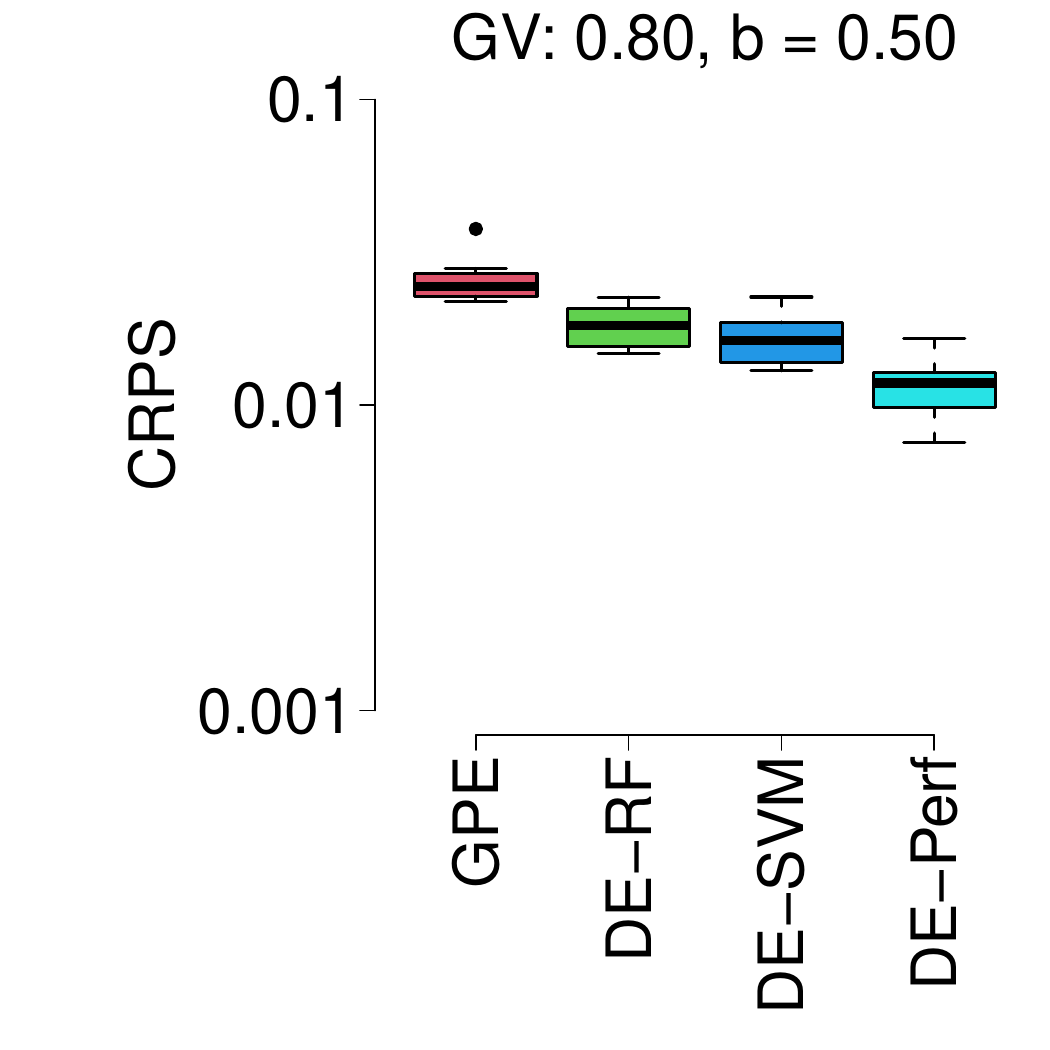}
    \end{subfigure}
    \caption{Comparing the performance of the GPE against three instances of the double emulator, fit using a RF (DE-RF), SVM (DE-SVM) and `perfect' classifier (DE-Perf), on the simulator in \eqref{DPFunction} using the CRPS score. We investigate the effect of the grounded volume (GV) via the offset, $a$, which is chosen so that the grounded volume increases from top to bottom from $20\%$ to $80\%$ of the input space. The hardness of the landing is increased from left to right via the exponent, $b$, in \eqref{bananasim}.}
    \label{fig:resultsdetpepcrps}
\end{figure}

\begin{figure}[ht!]
    \centering
    \begin{subfigure}{0.3\textwidth}
        \centering
        \includegraphics[width=\linewidth]{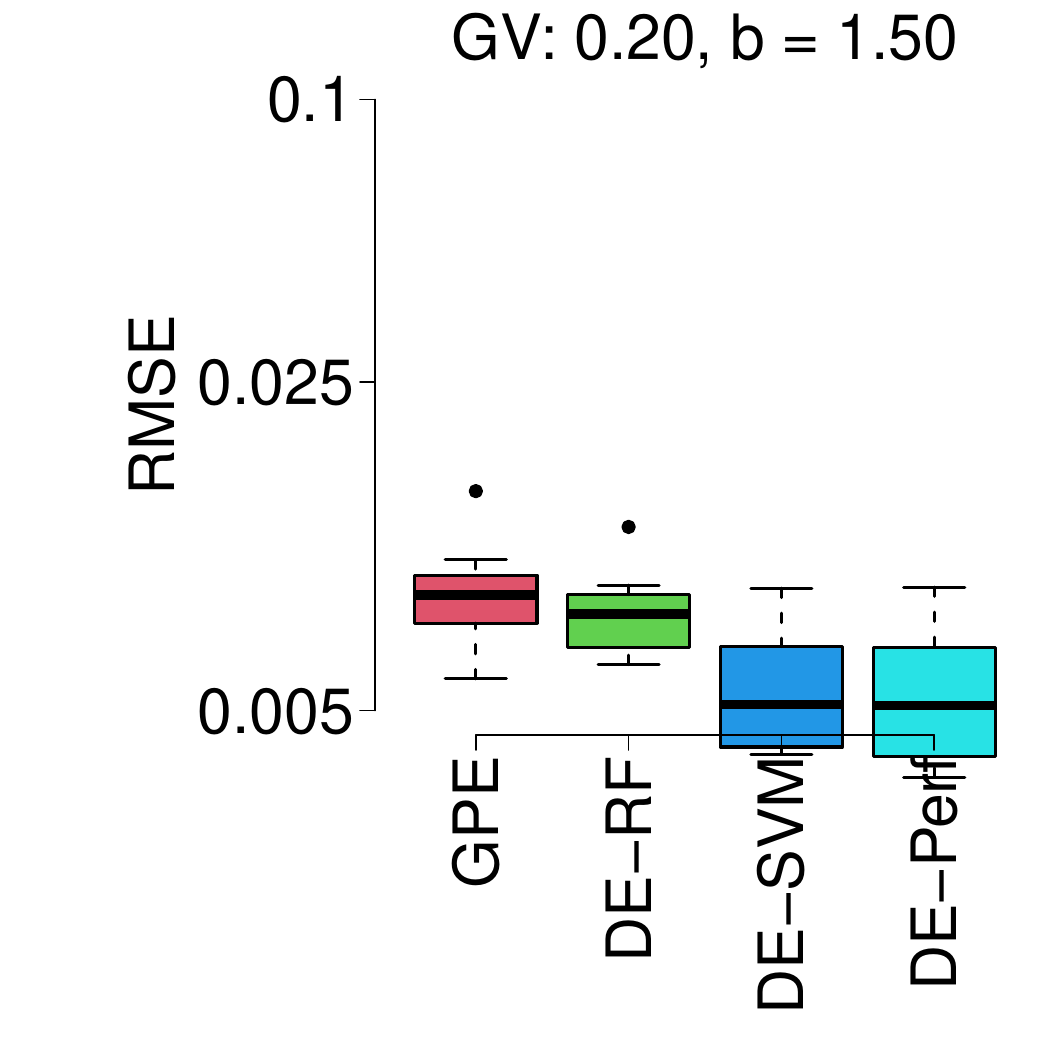}
    \end{subfigure}
    \begin{subfigure}{0.3\textwidth}
        \centering
        \includegraphics[width=\linewidth]{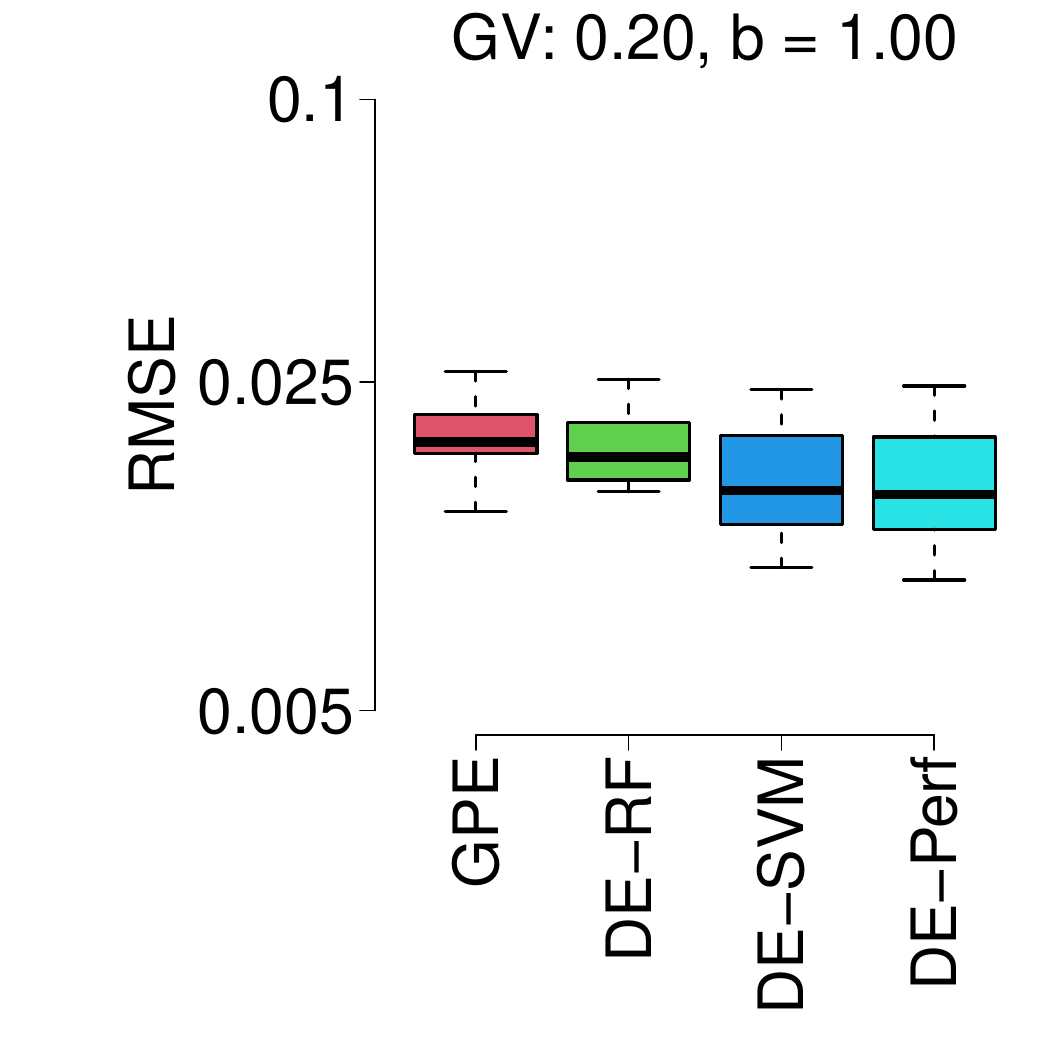}
    \end{subfigure}
    \begin{subfigure}{0.3\textwidth}
        \centering
        \includegraphics[width=\linewidth]{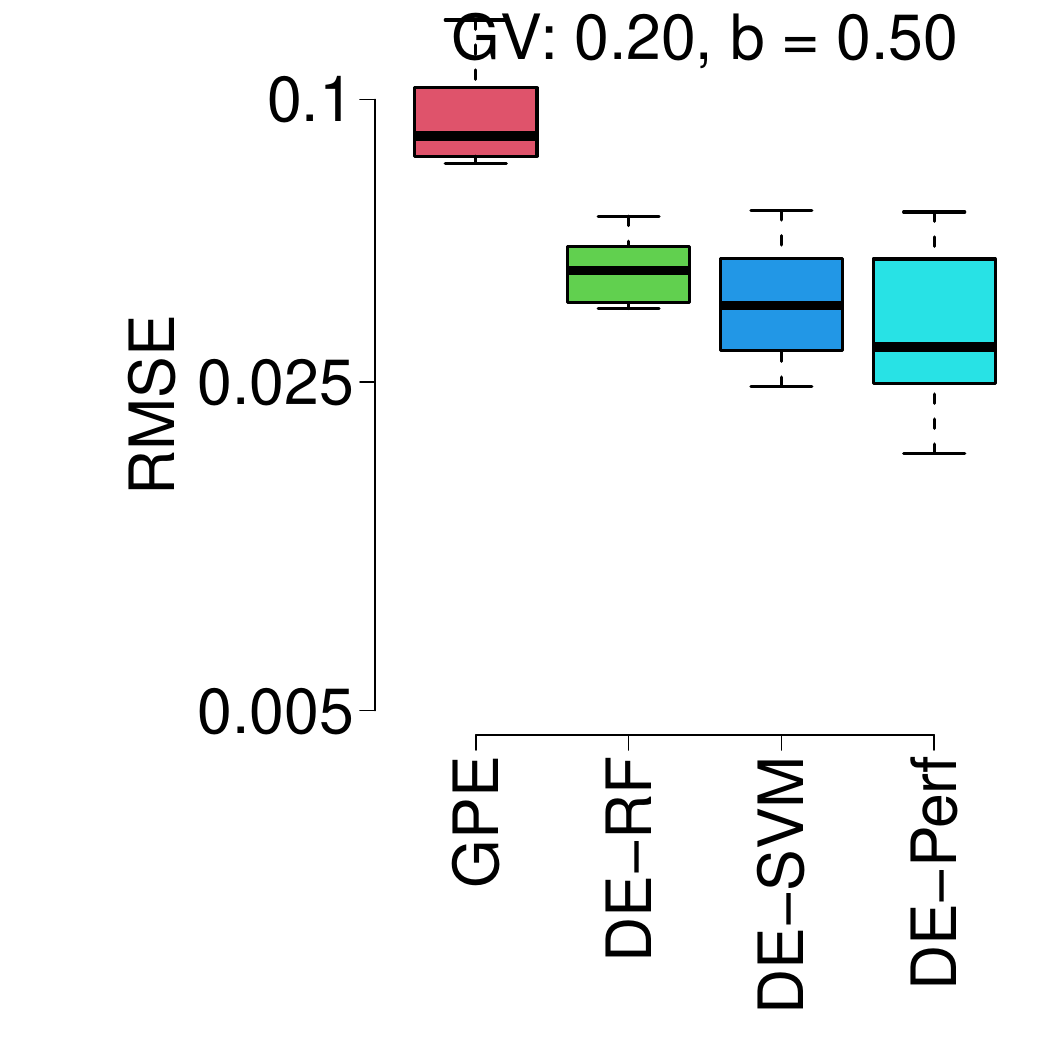}
    \end{subfigure}
    \begin{subfigure}{0.3\textwidth}
        \centering
        \includegraphics[width=\linewidth]{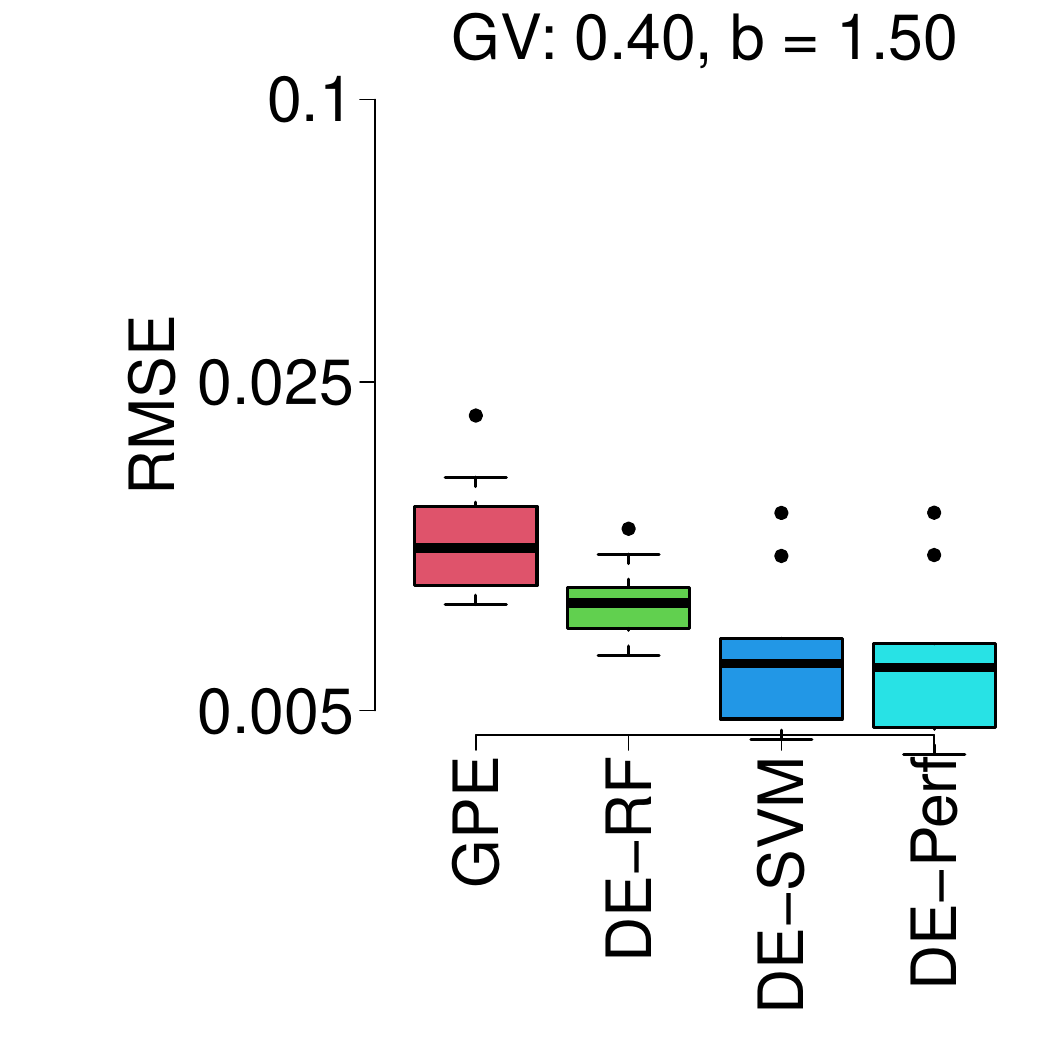}
    \end{subfigure}
    \begin{subfigure}{0.3\textwidth}
        \centering
        \includegraphics[width=\linewidth]{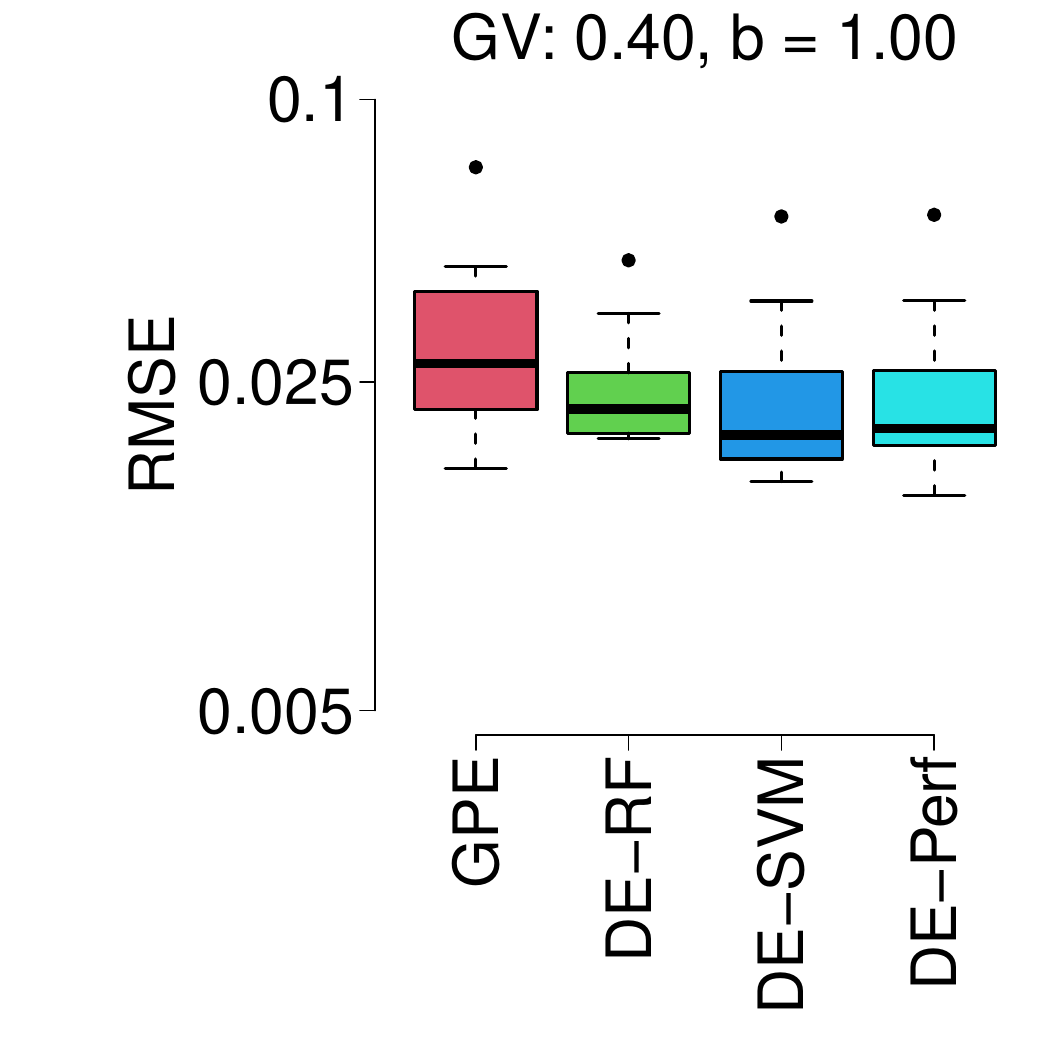}
    \end{subfigure}
    \begin{subfigure}{0.3\textwidth}
        \centering
        \includegraphics[width=\linewidth]{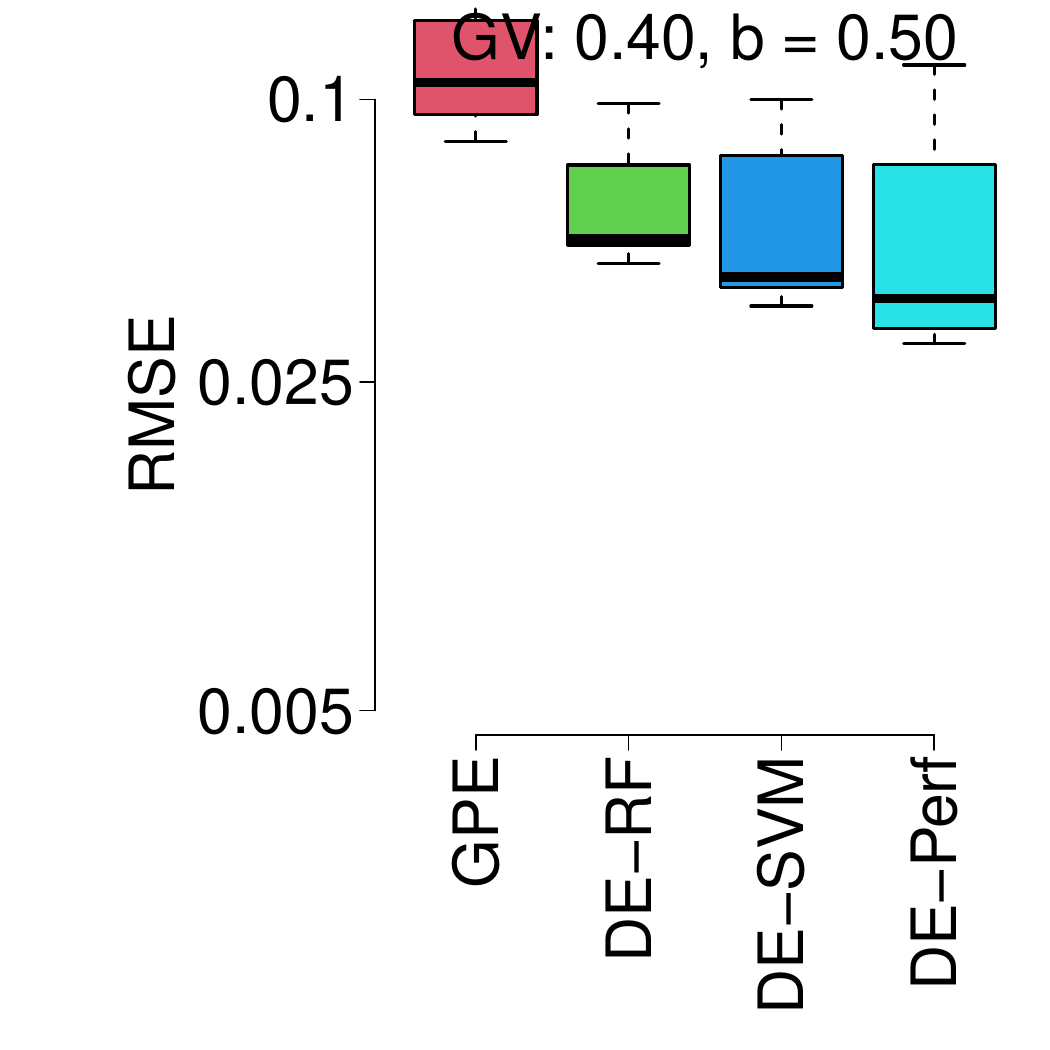}
    \end{subfigure}
        \begin{subfigure}{0.3\textwidth}
        \centering
        \includegraphics[width=\linewidth]{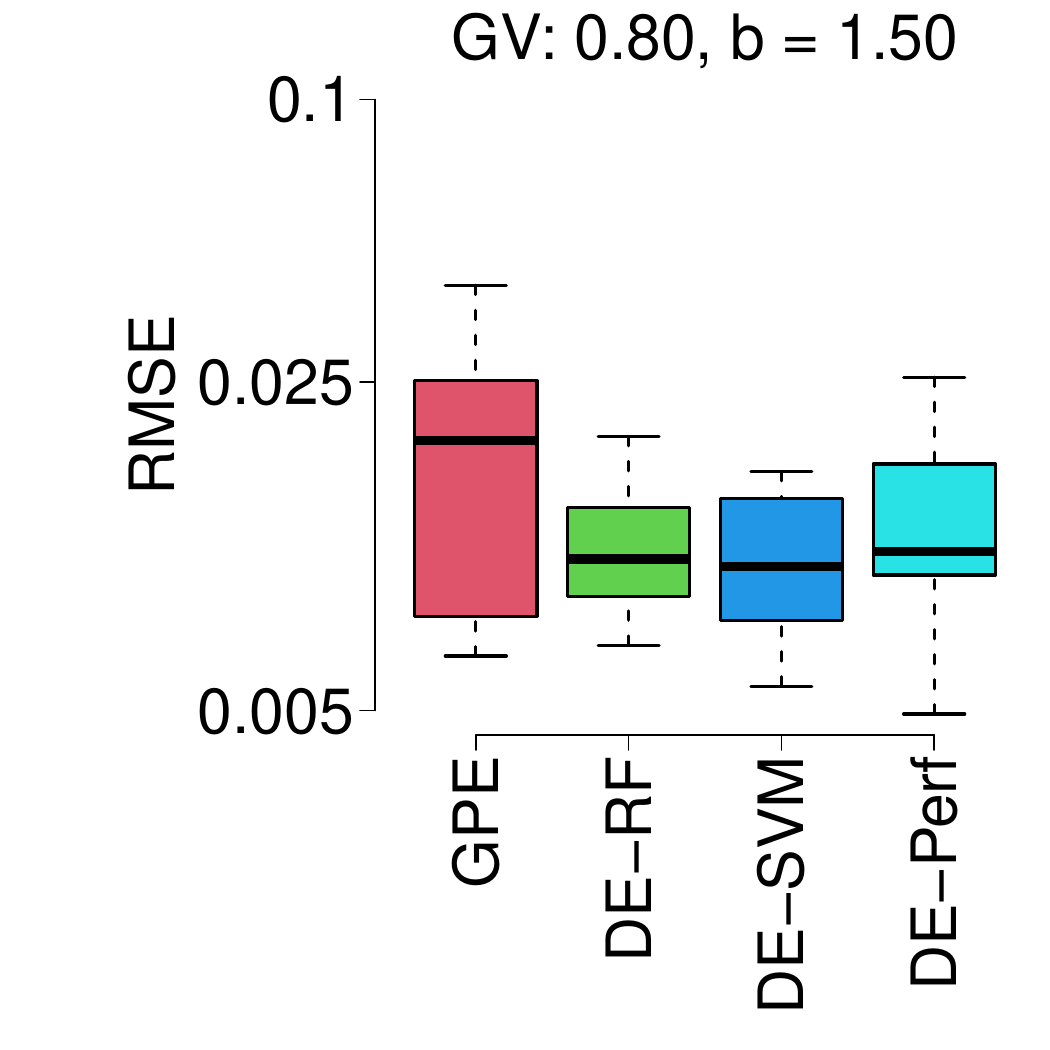}
    \end{subfigure}
    \begin{subfigure}{0.3\textwidth}
        \centering
        \includegraphics[width=\linewidth]{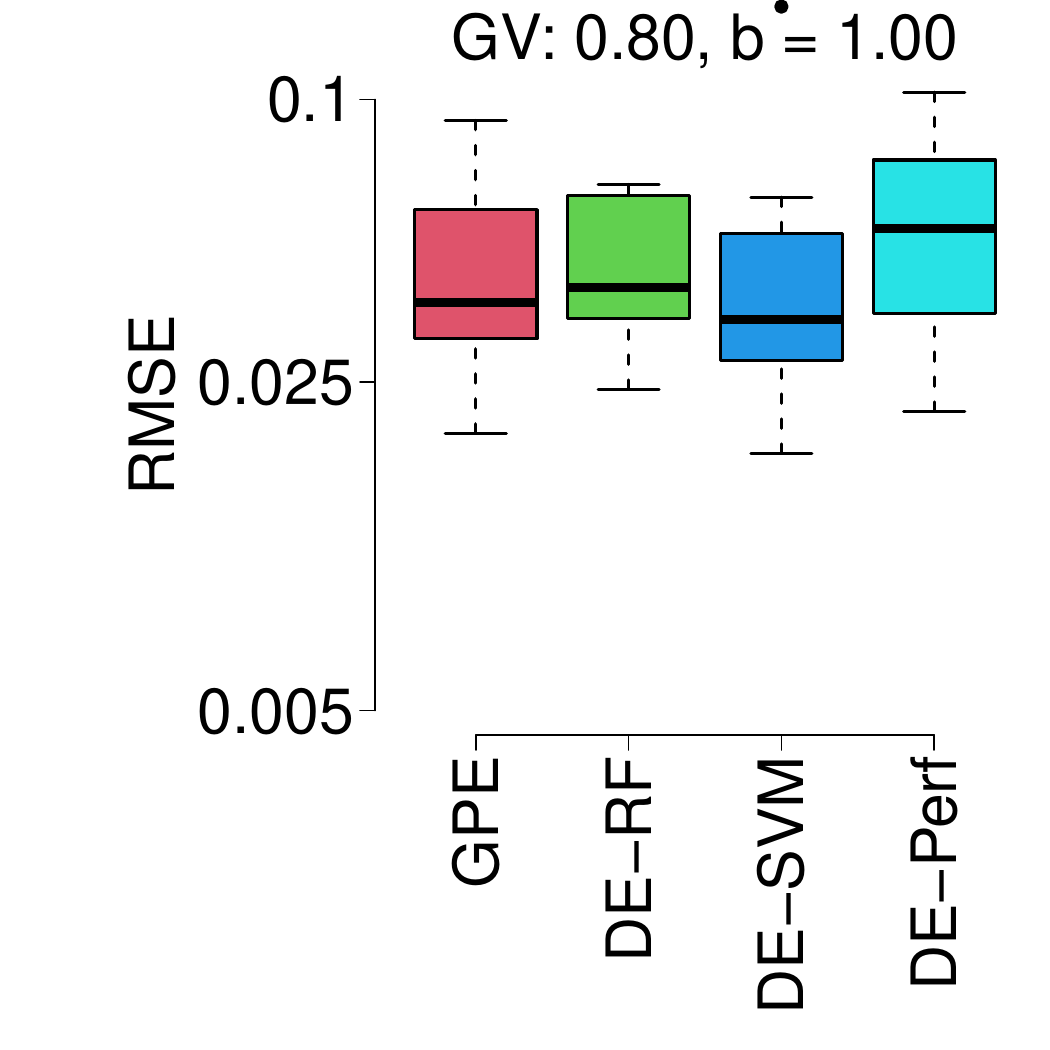}
    \end{subfigure}
    \begin{subfigure}{0.3\textwidth}
        \centering
        \includegraphics[width=\linewidth]{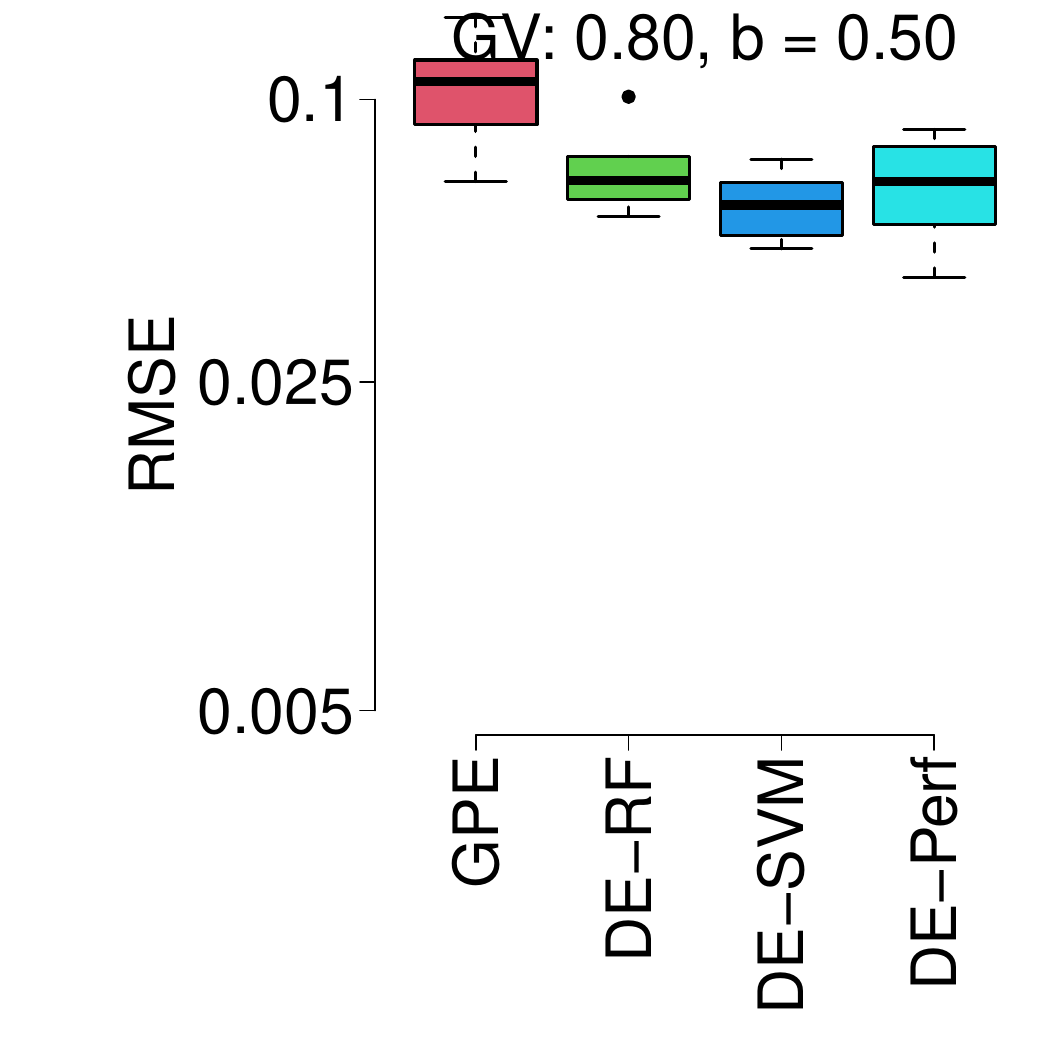}
    \end{subfigure}
    \caption{Comparing the performance of the GPE against three instances of the double emulator, fit using a RF (DE-RF), SVM (DE-SVM) and `perfect' classifier (DE-Perf), on the simulator in \eqref{DPFunction} using the RMSE. We investigate the effect of the grounded volume (GV) via the offset, $a$, which is chosen so that the grounded volume increases from top to bottom from $20\%$ to $80\%$ of the input space. The hardness of the landing is increased from left to right via the exponent, $b$, in \eqref{bananasim}.}
    \label{fig:resultsdetpeprmse}
\end{figure}

\begin{figure}[ht!]
    \centering
    \begin{subfigure}{0.3\textwidth}
        \centering
        \includegraphics[width=\linewidth]{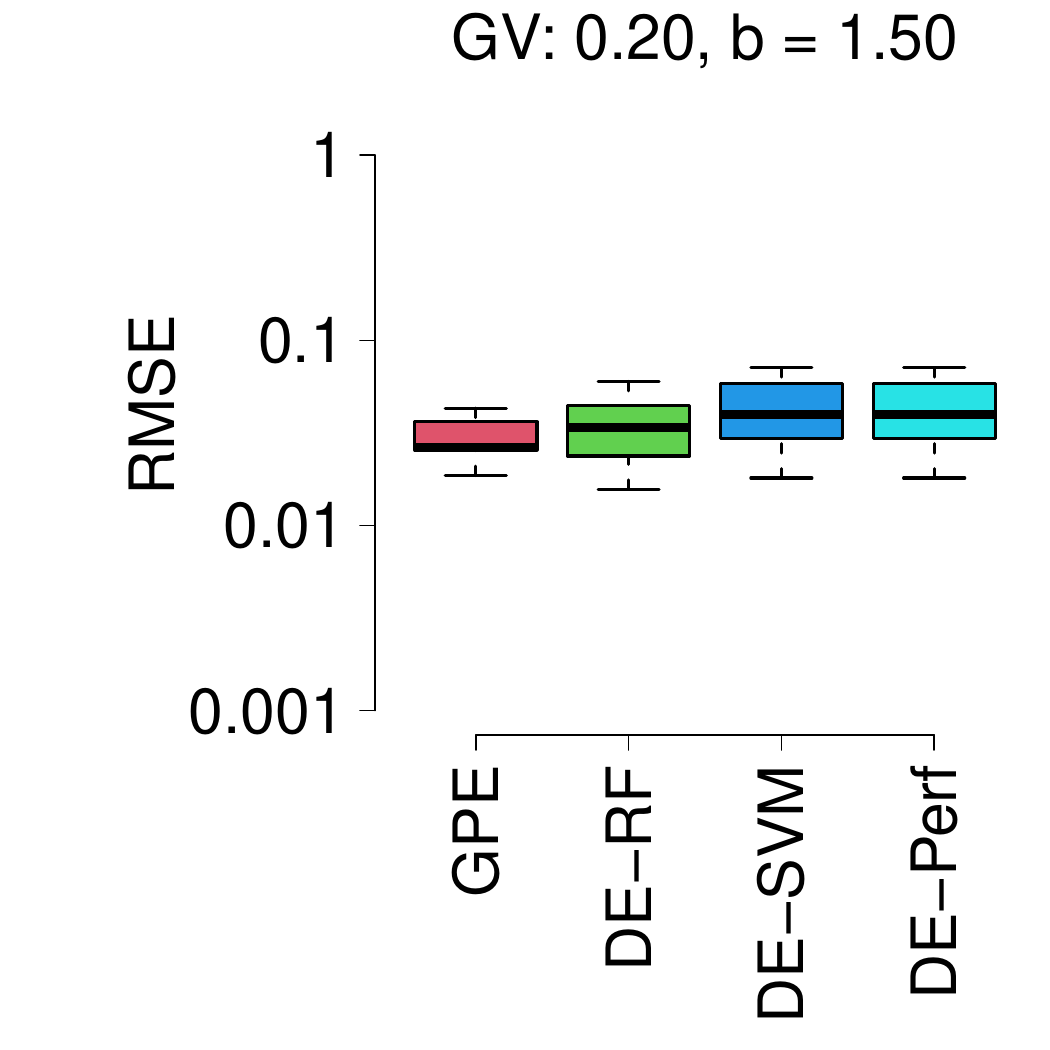}
    \end{subfigure}
    \begin{subfigure}{0.3\textwidth}
        \centering
        \includegraphics[width=\linewidth]{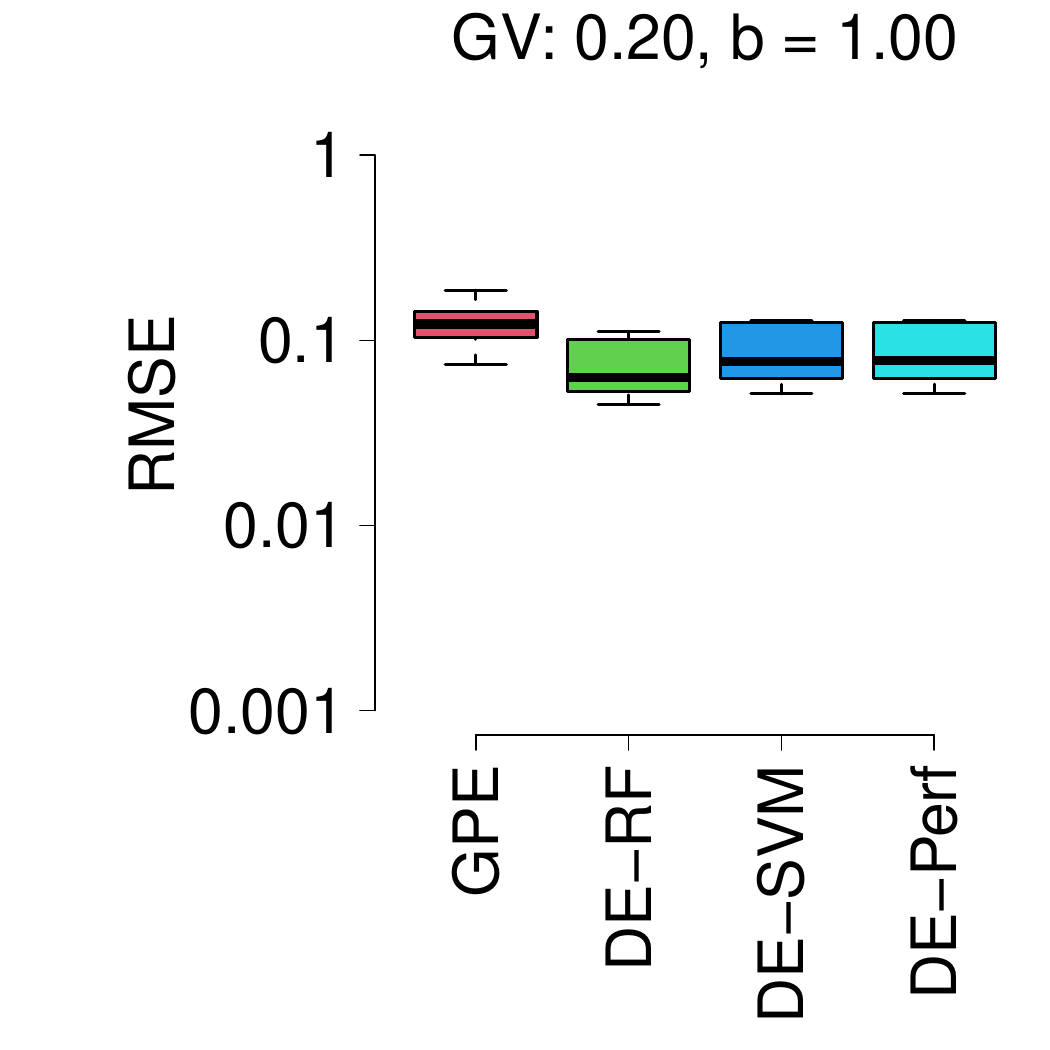}
    \end{subfigure}
    \begin{subfigure}{0.3\textwidth}
        \centering
        \includegraphics[width=\linewidth]{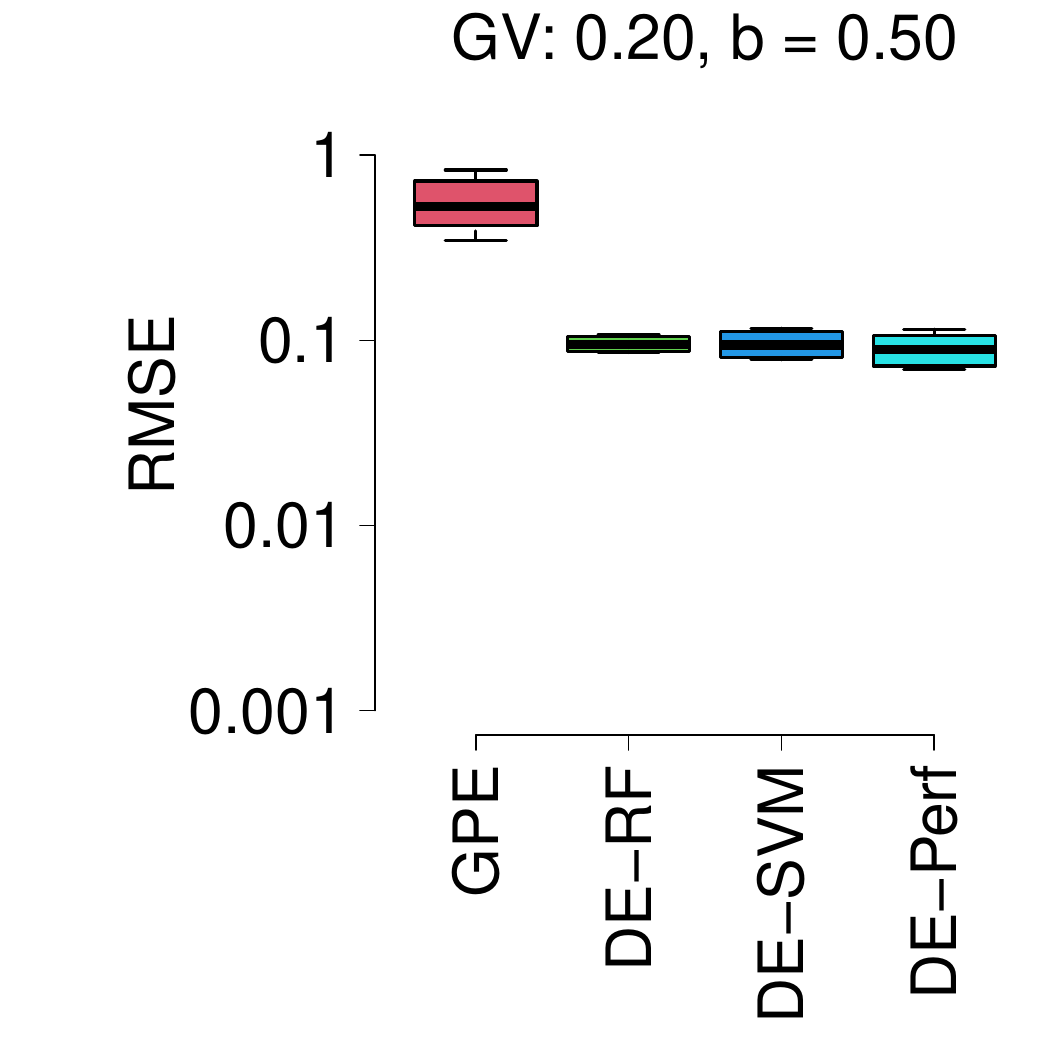}
    \end{subfigure}
    \begin{subfigure}{0.3\textwidth}
        \centering
        \includegraphics[width=\linewidth]{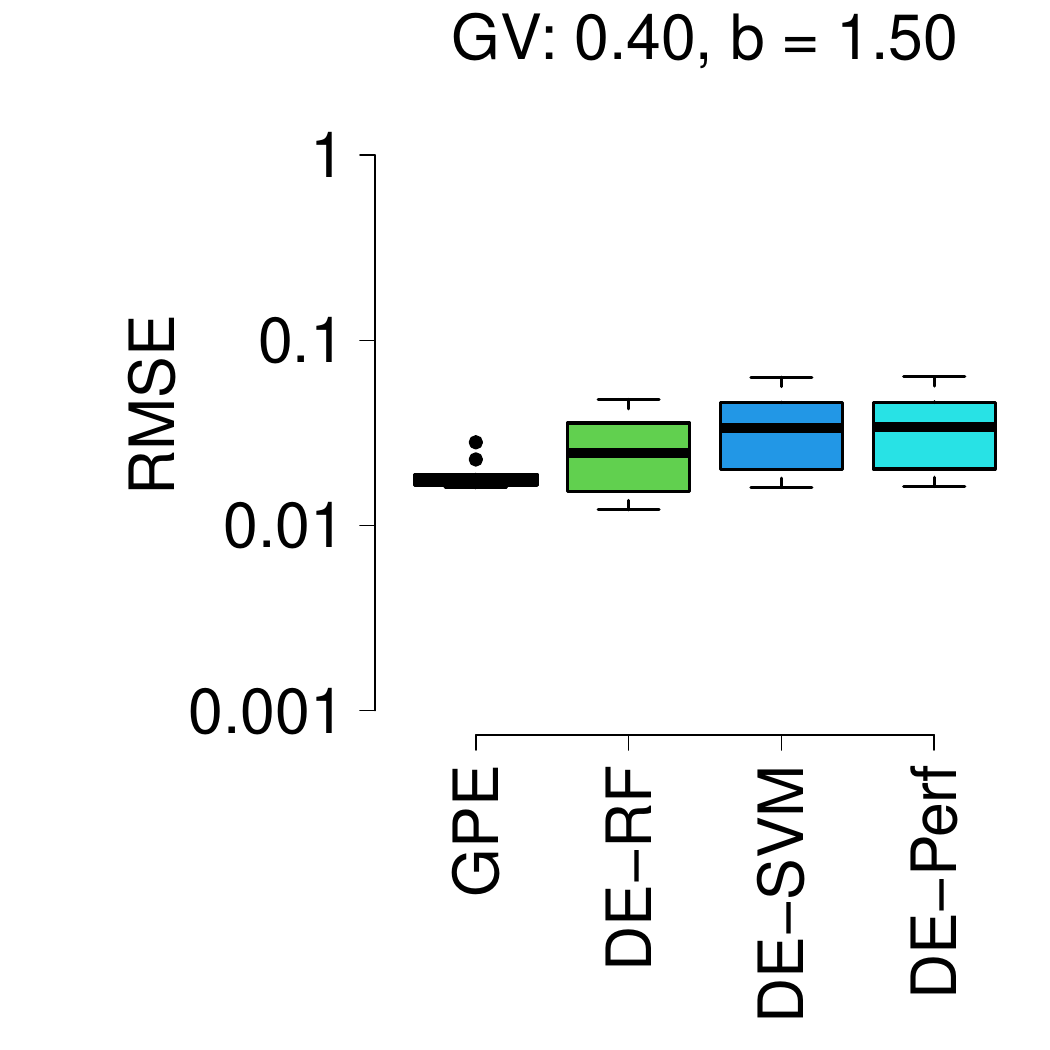}
    \end{subfigure}
    \begin{subfigure}{0.3\textwidth}
        \centering
        \includegraphics[width=\linewidth]{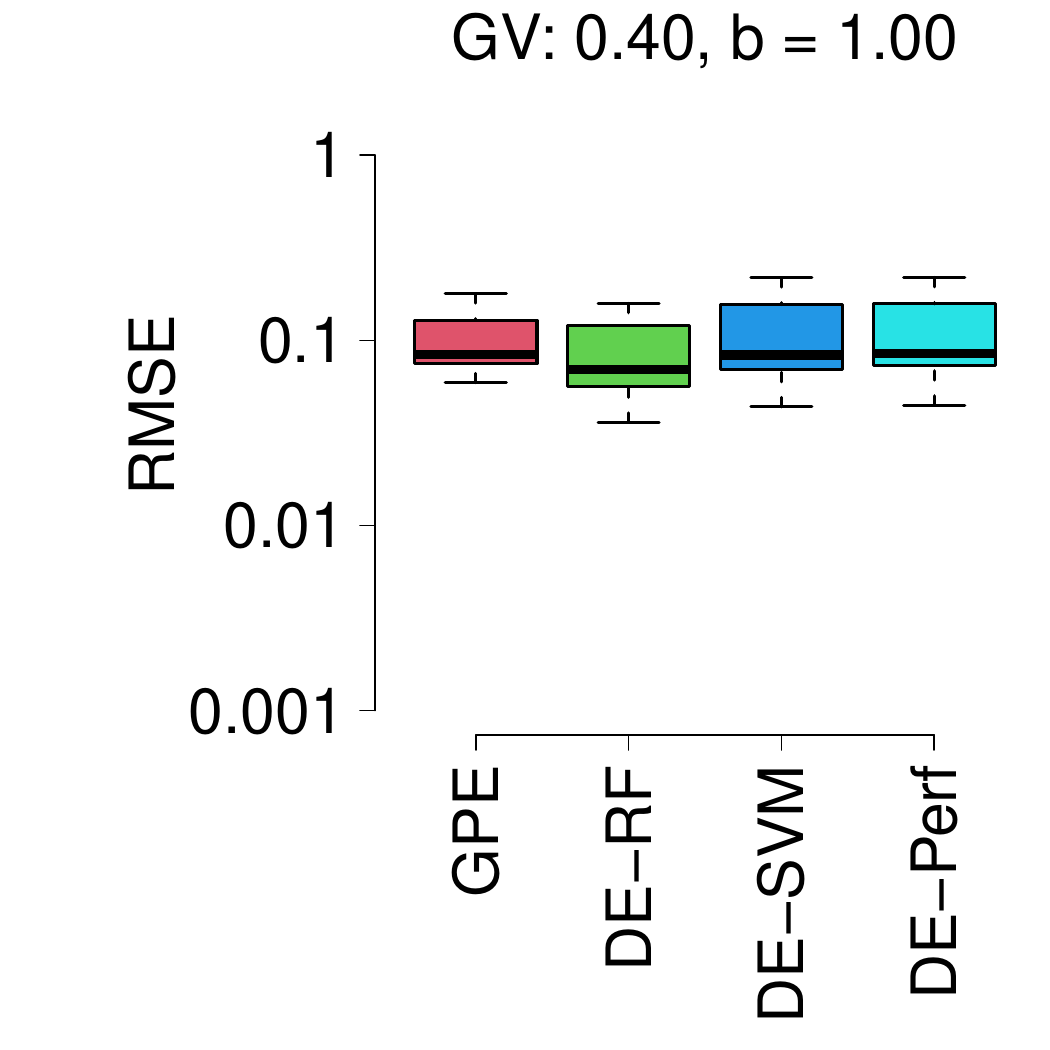}
    \end{subfigure}
    \begin{subfigure}{0.3\textwidth}
        \centering
        \includegraphics[width=\linewidth]{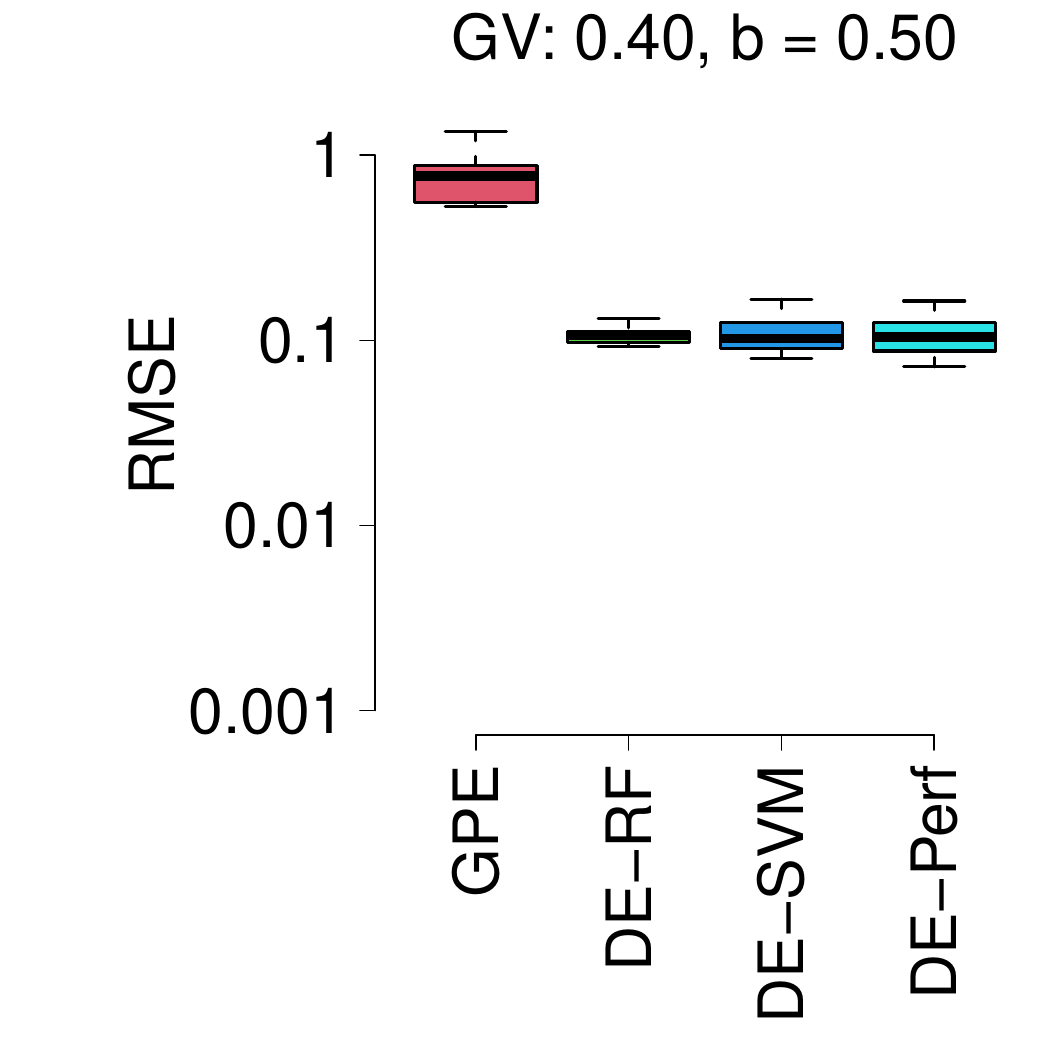}
    \end{subfigure}
        \begin{subfigure}{0.3\textwidth}
        \centering
        \includegraphics[width=\linewidth]{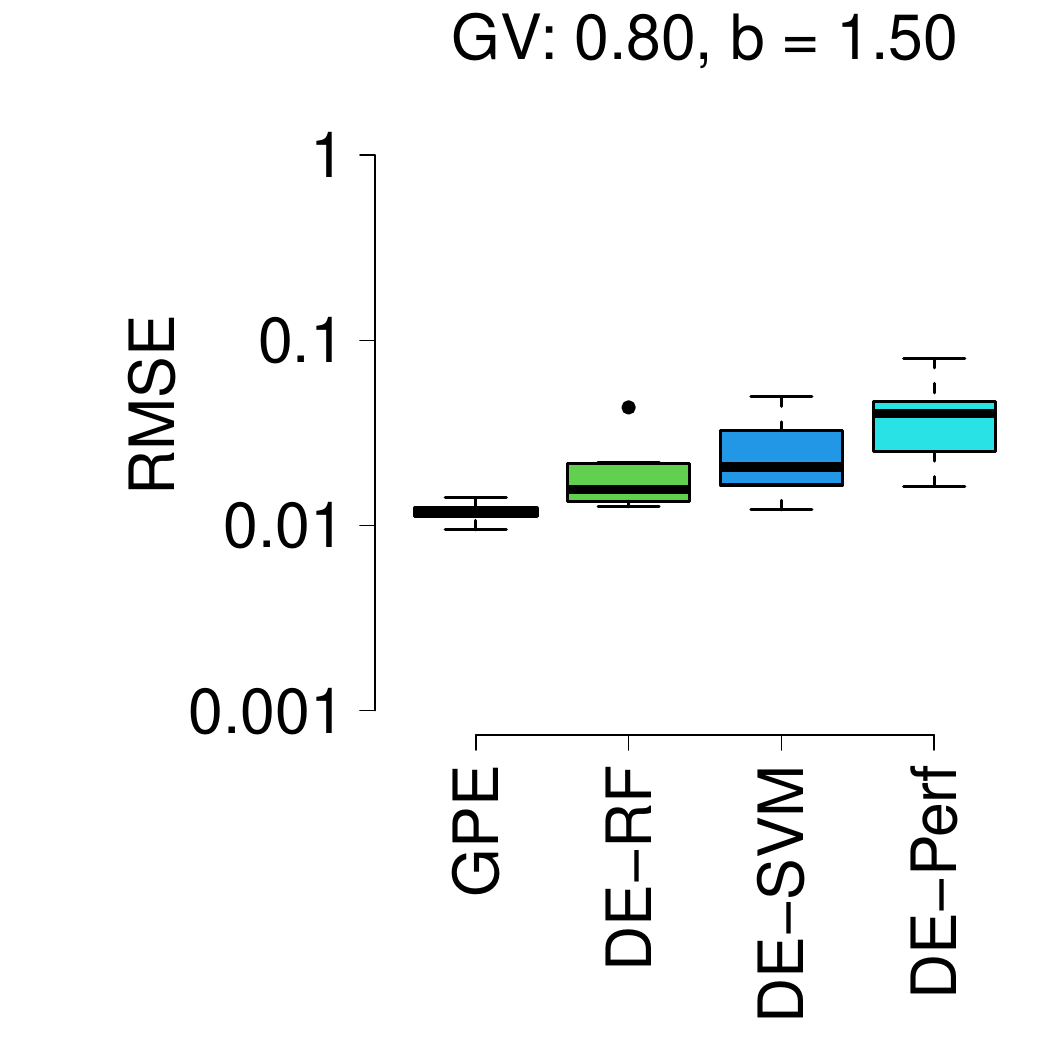}
    \end{subfigure}
    \begin{subfigure}{0.3\textwidth}
        \centering
        \includegraphics[width=\linewidth]{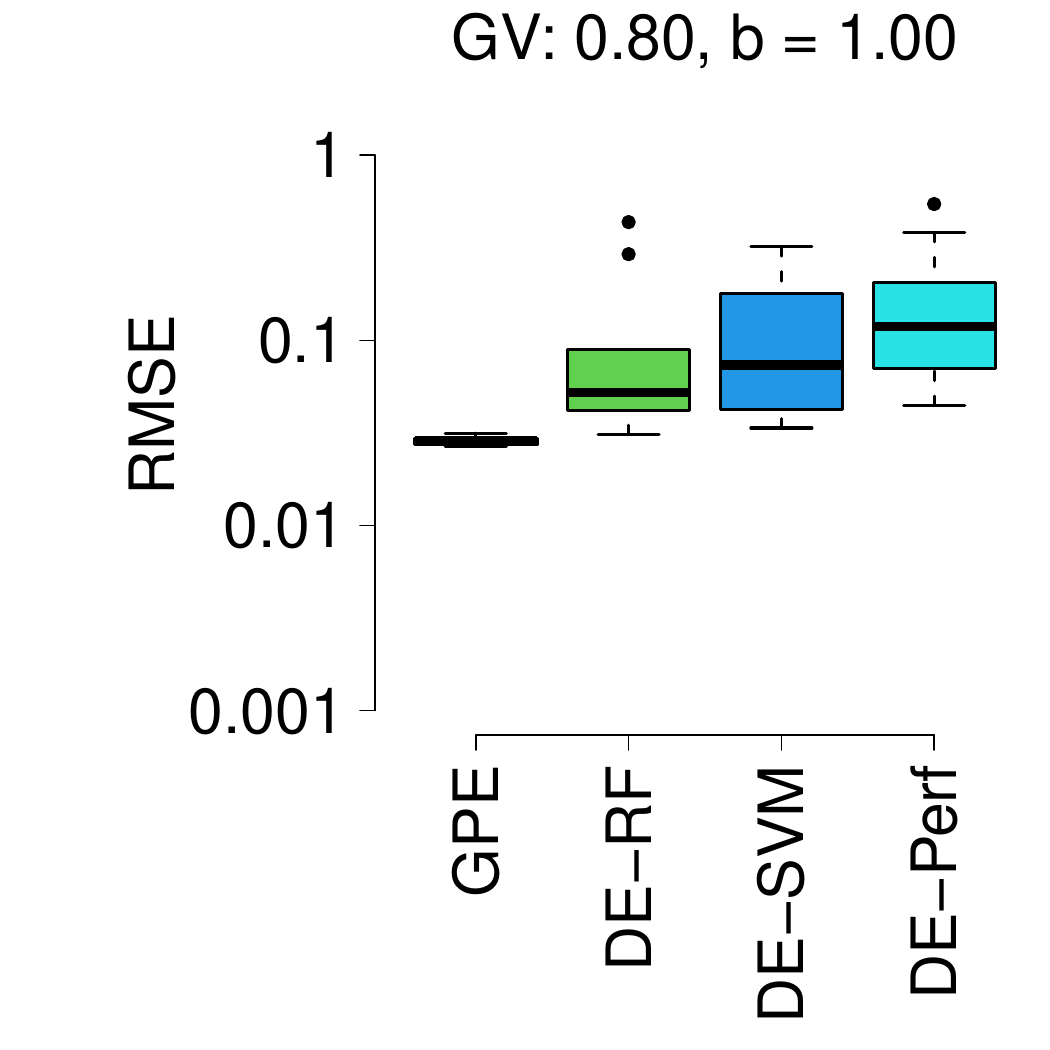}
    \end{subfigure}
    \begin{subfigure}{0.3\textwidth}
        \centering
        \includegraphics[width=\linewidth]{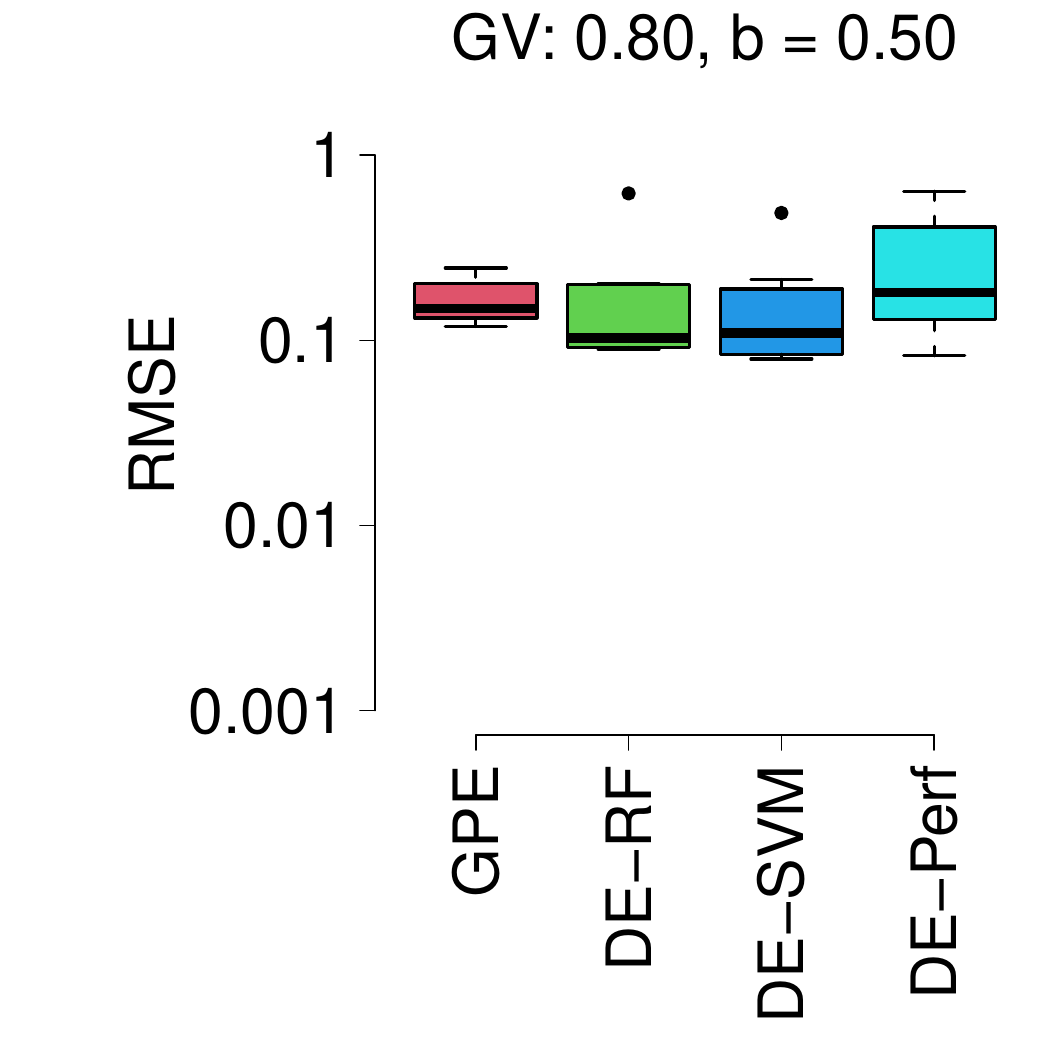}
    \end{subfigure}
    \caption{Comparing the performance of the GPE against three instances of the double emulator, fit using a RF (DE-RF), SVM (DE-SVM) and `perfect' classifier (DE-Perf), on the simulator in \eqref{banana} using the CRPS. We investigate the effect of the grounded volume (GV) via the offset, $a$, which is chosen so that the grounded volume increases from top to bottom from $20\%$ to $80\%$ of the input space. The hardness of the landing is increased from left to right via the exponent, $b$, in \eqref{bananasim}.}
    \label{fig:resultsbananarmse}
\end{figure}

\end{document}